\documentclass[onecolumn,authoryear]{els-mrw} 

\usepackage{amsmath,amssymb,amsfonts,amsthm,makeidx,graphicx}
\usepackage{txfonts}
\usepackage{helvet}
\usepackage{multicol}
\usepackage{hyperref}

\begin{document}

\chapter{Compact Objects in Globular Clusters}\label{chap1}

\author[1]{Kyle Kremer}%

\address[1]{\orgname{University of California, San Diego}, \orgdiv{Department of Astronomy \& Astrophysics}, \orgaddress{9500 Gilman Dr. La Jolla, CA 92093}}


\maketitle

\begin{abstract}[Abstract]
It is now widely established that globular clusters host robust populations of white dwarfs, neutron stars, and black holes throughout their lifetimes. Within clusters, dynamical processes enabled by stellar densities thousands to millions of times larger than typical galactic environments facilitate interactions involving these stellar remnants that give rise to an array of astrophysical phenomena. In particular, stellar clusters have emerged as an important formation site for X-ray sources, radio pulsars, and merging black hole binaries similar to those recently detected as gravitational wave sources by the LIGO/Virgo/KAGRA detectors. This article reviews our current understanding of compact objects in globular clusters, discussing current observational evidence, ways these objects influence the dynamical evolution of their hosts, and future prospects.
\end{abstract}

\begin{BoxTypeA}[chap1:box1]{\large{Key points}}
\begin{itemize}
\item{\term{Globular clusters:} Old and dense stellar populations of roughly $10^5-10^6$ stars and half-light radii of roughly a few parsec. Found in essentially all galaxy types. More than 150 globular clusters are observed in the Milky Way today.}
\item{\term{Relaxation time:} The characteristic timescale for a stellar system to reach thermal equilibrium via cumulative effect of gravitational encounters between objects. Globular clusters have relaxation times comparable to or less than their typical ages.}
\item{\term{Natal kick:} A velocity kick imparted to a compact object at birth as a result of asymmetries in the formation process. Plays critical role in retention of compact objects in stellar clusters upon formation.}
\item{\term{Mass segregation:} Dynamical process through which heavier members of a stellar system tend to move toward the system's center, while lighter members tend to move farther away. Important for formation of black hole subsystems in the centers of globular clusters.}
\item{\term{Cluster core collapse:} Process closely linked to dynamical relaxation that leads to a high concentration of stars at the center of a stellar cluster. Ultimate fate of all globular clusters, if given sufficient time to evolve. Roughly 20\% of Galactic globular clusters are core-collapsed today.}
\item{\term{Black hole burning:} Process through which stellar-mass black holes inject energy into the stellar bulk of their host cluster through dynamical interactions. Key to delaying onset of core collapse in most present-day globular clusters.}
\item{\term{Low-mass X-ray binary (LMXB):} A compact binary system where a neutron star or stellar-mass black hole accretes material from a low-mass stellar companion. Dozens of LMXBs are observed in globular clusters.}
\item{\term{Radio millisecond pulsar (MSP):} A highly magnetized neutron star that emits beams of observable electromagnetic radiation from its poles. Millisecond pulsars are distinguished by exceptionally fast spin periods, of order ten milliseconds or less. More than 300 MSPs are currently observed in Galactic globular clusters.}
\item{\term{Gravitational wave source:} For the purposes of this article, a double compact object binary that emits gravitational radiation observable by detectors like LIGO/Virgo/KAGRA as it inspirals and ultimately merges. Expected to form efficiently via dynamical processes in globular clusters.}
\item{\term{Intermediate-mass black hole (IMBH):} A black hole in the mass range $\sim10^2-10^5\,M_{\odot}$, more massive than stellar-mass black holes formed via massive star collapse and less massive than classic supermassive black holes in galactic centers. Expected to be found in some globular clusters, for example $\omega$Centauri.}
\item{\term{Fast radio burst (FRB):} Bright flares of coherent radio emission with roughly millisecond durations. Some are observed to repeat. Origin and burst mechanisms remain unknown. A repeating FRB is observed in a globular cluster in M81.}
\end{itemize}
\end{BoxTypeA}



\section{Introduction}

\subsection{Stellar cluster basics}

Dense conglomerates of stars called stellar clusters are ubiquitous throughout the universe. Stellar clusters contain populations of tens to tens of millions of stars bound together by their mutual gravitational attraction. They come in a variety of types distinguished by their varying ages, masses, central densities, stellar metallicities, and formation histories. The most massive star clusters ($M \gtrsim 10^7\,M_{\odot}$) called nuclear star clusters \citep{Neumayer2020} reside at the centers of galaxies and are closely connected to supermassive black holes that power active galactic nuclei and regulate galaxy evolution. Low-mass young star clusters and stellar associations ($M \lesssim 10^3\,M_{\odot}$) are the birth places for the majority of stars, making these systems fundamental building blocks of galaxies \citep[e.g.,][]{LadaLada2003}. This article focuses specifically upon \textbf{globular clusters}, ancient stellar populations with masses roughly $10^5-10^6\,M_{\odot}$ that host some of the oldest stars in the universe. 

\begin{table}[t]
\TBL{\caption{Typical properties of stellar clusters and other astrophysical environments}\label{table1}}
{\begin{tabular*}{\textwidth}{@{\extracolsep{\fill}}lcccccc}
\toprule
\multicolumn{1}{c}{} &
\multicolumn{1}{c}{Stellar number} &
\multicolumn{1}{c}{Radius} &
\multicolumn{1}{c}{Stellar density} &
\multicolumn{1}{c}{Velocity dispersion} &
\multicolumn{1}{c}{Crossing time, $R/v$} &
\multicolumn{1}{l}{Relaxation time, $t_{\rm rl}$}\\
\multicolumn{1}{c}{} &
\multicolumn{1}{c}{} &
\multicolumn{1}{c}{($\rm{pc}$)} &
\multicolumn{1}{c}{($\rm{pc}^{-3}$)} &
\multicolumn{1}{c}{($\rm{km/s}$)} &
\multicolumn{1}{c}{($\rm{yr}$)} &
\multicolumn{1}{c}{($\rm{yr}$)}\\
\colrule
Local Galactic disk & $10^{9}$ & $10^3$ & $.1$ & $30$ & $10^7$ & $10^{14}$ \\
Open cluster & $10^{2}-10^4$ & $2$ & $10$ & $1$ & $10^6$ & $10^{7}$ \\
Globular cluster & $10^5-10^{6}$ & $3$ & $10^4$ & $10$ & $10^5$ & $10^{9}$ \\
Nuclear cluster & $>10^{7}$ & $10$ & $10^5$ & $>30$ & $10^5$ & $10^{10}$ \\
\botrule
\end{tabular*}}{%
\begin{tablenotes}
\footnotetext{}
\end{tablenotes}
}%
\end{table}

Perhaps the defining feature of globular clusters (indeed all stellar clusters) is their high stellar densities, which can extend many orders of magnitude above typical galactic field environments (see Table~\ref{table1} for summary of typical cluster properties compared to the Galactic disk). As a consequence of their high stellar densities, the stars that populate stellar clusters undergo frequent close gravitational fly-by encounters with one another. Individually, these encounters can lead to a variety of exotic dynamical processes including the formation of binary systems, stellar collisions and tidal encounters, hypervelocity stars flung from their host clusters via gravitational slingshot, and a range of high-energy transient sources. In aggregate, these gravitational encounters govern the large-scale evolution of the cluster as a whole. A key timescale for globular clusters is the \textbf{relaxation time}, the timescale on which two-body gravitational encounters transfer energy between individual stars and cause the system to establish thermal equilibrium. The relaxation time can be expressed as:

\begin{equation}
    \label{eq:t_rl}
    t_{\rm rl} \approx \frac{0.1 N^{1/2} R^{3/2}}{G^{1/2} \langle m \rangle^{1/2} \ln N} \approx \frac{0.1 N}{\ln N} \frac{R}{v} 
\end{equation}
\citep{BinneyTremaine2008}, where $N$ is the total number of stars in the system, $R$ is the half-mass radius, and $\langle m \rangle$ is the average stellar mass. To attain the right-hand side of Equation~\ref{eq:t_rl}, we have assumed virial equilibrium so that $GM_{\rm cl}/R \sim v^2$, where $M_{\rm cl} = N \langle m \rangle$ is the total cluster mass and $v$ is the system's velocity dispersion. For typical globular clusters -- $N\approx 10^6$, $R \approx 3\,$pc, and $v\approx 10\,$km/s -- we find typical relaxation times of $\lesssim 10^9\,$yr. Globular clusters, like other ``collisional'' stellar populations, have relaxation times comparable to or less than their typical ages. In this case, interactions between particles are efficient with respect to the lifetime of the system and play a fundamental role in the system's evolution. This is in contrast to ``collisionless'' systems where such interactions are negligible and the constituent particles move under the influence of the smoothed-out gravitational field of the system as a whole (the disk of the Milky Way is a classic example; see Table~\ref{table1}).

\begin{figure}[t]
\centering
\includegraphics[width=\textwidth]{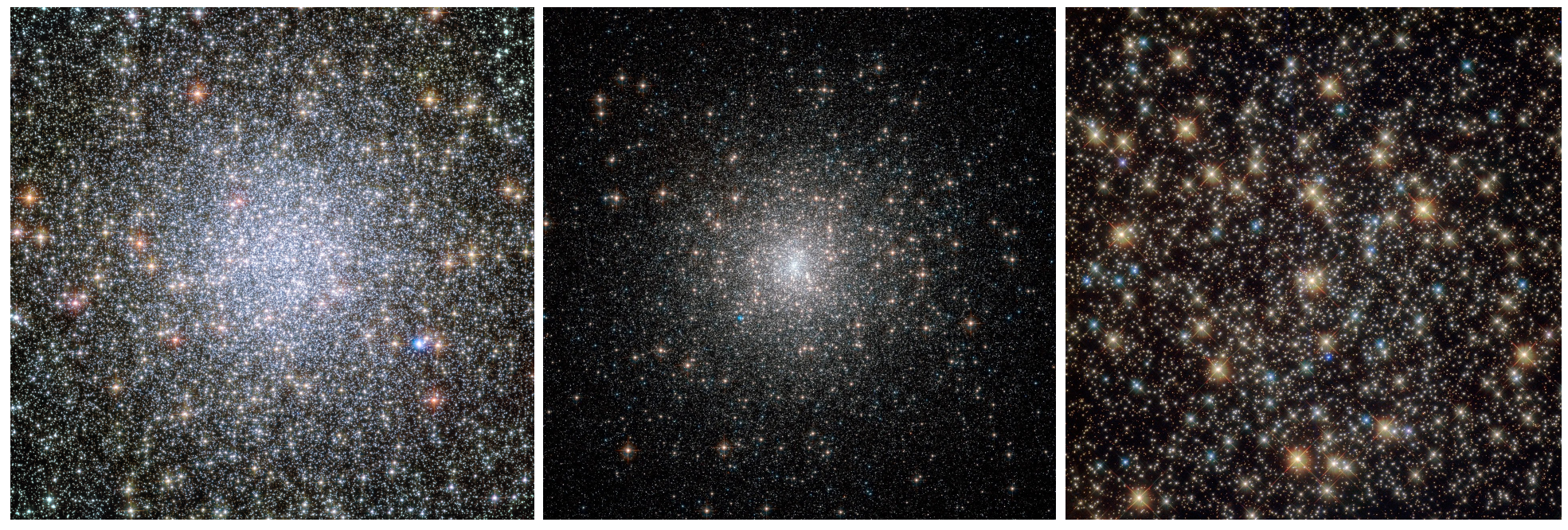}
\caption{Hubble Space Telescope HST) images of three Galactic globular clusters. From left to right: 47 Tucanae, a massive cluster with many observed radio millisecond pulsars \citep{Manchester1991} and white dwarfs directly observed via HST photometry \citep{Kalirai2012}; Messier 15, a core-collapsed globular cluster home to two young radio pulsars \citep{Zhou2024} and a double neutron star binary \citep{Jacoby2006}, and the inner region of NGC~3201, a large core-radius cluster host to three known stellar-mass black hole candidates \citep{Giesers2019}.
\textit{Image credits:} NASA, ESA, and the Hubble Heritage (STScI/AURA)-ESA/Hubble Collaboration; Acknowledgment: J. Mack (STScI) and G. Piotto (University of Padova, Italy)
}
\label{fig:47tuc}
\end{figure}

More than a hundred globular clusters have been observed and studied in the Milky Way for decades \citep{Harris1996}. See Figure~\ref{fig:47tuc} for a few examples from observations with the Hubble Space Telescope. More recently, astronomers have uncovered populations of extragalactic globular clusters and it is now understood globular clusters are common features of essentially all galaxy types \citep{BrodieStrader2006}. Both Galactic and extragalactic globular clusters have long been leveraged as laboratories for studying astrophysical dynamics and exotic sources that result from these processes \citep[e.g.,][]{HeggieHut2003}.

\subsection{Compact objects in globular clusters}

In the past couple of decades, it has become clear that globular clusters are host to stellar compact objects -- white dwarfs, neutron stars, and stellar-mass black holes -- setting the stage for a range of exciting astrophysical implications.


Globular clusters are old populations with ages of roughly $10\,$Gyr or more. As such, the luminous stellar content of present-day globular clusters consists only of low-mass stars less than roughly $1\,M_{\odot}$. Although old globular clusters are devoid of massive stars today, they almost certainly contained robust populations of massive stars at the time of their formation. For a typical globular cluster with roughly $10^6$ stars at birth, a standard initial stellar mass function \citep[e.g.,][]{Kroupa2001} suggests of order $10^3$ black holes and neutron stars and in excess of $10^4$ white dwarfs form via evolution of the most massive stars in the system at earlier times.

Formation from a massive stellar population is a necessary but not sufficient condition for the presence of compact objects in clusters today. During formation, many compact objects -- especially neutron stars and black holes -- receive large \textbf{natal kicks} caused by asymmetries in the final stages of collapse of their progenitor star. The typical escape velocity of a globular cluster can be written as

\begin{equation}
    \label{eq:v_esc}
    v_{\rm esc} \approx \sqrt{\frac{4 G M_{\rm cl}}{R}} \approx 100 \, \Bigg( \frac{M_{\rm cl}}{10^6\,M_{\odot}} \Bigg)^{1/2} \Bigg( \frac{R}{3\,\rm{pc}} \Bigg)^{-1/2}\,\rm{km/s}
\end{equation}
\citep{BinneyTremaine2008}. Any compact object formed with a natal kick in excess of this escape velocity will be ejected from its host shortly after formation. This is a non-issue for white dwarfs, which are generally expected to receive kicks of at most a few km/s at birth \citep[e.g.,][]{Fellhauer2003}. Many neutron stars, meanwhile, are known to receive large kicks up to several hundred km/s or more \citep{Hobbs2005}, suggesting many neutron stars (but certainly not all; see Section~\ref{sec:pulsars}) born at early times in clusters are immediately lost. Black hole kicks are more uncertain \citep[e.g.,][]{Mirabel2001,JonkerNelemans2004,Smartt2009,AndrewsKalogera2022}, but it is generally thought that at least some black holes \citep[particularly more massive ones formed at low metallicities;][]{Fryer2012} receive natal kicks less than roughly $100\,$km/s. Thus, retention of at least a fraction of the initial black hole population in globular clusters is expected.

Any compact objects retained after formation can potentially remain in their host for billions of years, lurking at present day amidst the luminous stellar bulk of their host cluster. From the perspective of stellar dynamics, white dwarfs and neutron stars are by-and-large passive members of their hosts. With typical masses of roughly $1-2\,M_{\odot}$, white dwarfs and neutron stars behave similar dynamically to the $\lesssim 1\,M_{\odot}$ luminous stars that are by far most numerous in a cluster. As such, these low-mass compact objects are merely along for the ride, undergoing dynamical encounters that lead to a range of exciting astrophysics but, in general, never having a major effect upon the host cluster itself. This is not the case for black holes. With masses of up to $40\,M_{\odot}$ or more, stellar-mass black holes are significantly more massive than typical cluster stars. As a consequence, once formed in the first few to tens of Myr of their host cluster's evolution, black holes promptly \textbf{mass segregate} to their host's center. 
The mass segregation timescale for objects of mass $M$ can be approximated as:

\begin{equation}
    \label{eq:t_ms}
    t_{\rm ms} \approx \frac{\langle m \rangle}{M} t_{\rm rl} \approx 30\,\Bigg( \frac{ \langle m \rangle}{M_{\odot}} \Bigg) \Bigg( \frac{ M}{30\,M_{\odot}} \Bigg)^{-1} \Bigg( \frac{ t_{\rm rl}}{1\,\rm{Gyr}} \Bigg)\,\rm{Myr}
\end{equation}
Thus, for a typical globular cluster, the stellar black holes sink and form a dense central subsystem on sub-Gyr timescales, leaving ample time throughout the subsequent evolution of the cluster for a range of black hole dynamics.

The long-term retention up to the present day of these compact objects -- particularly the black holes -- has long been a source of debate. It was traditionally argued that such a black hole subsystem would dynamically decouple from the rest of its host cluster \citep[e.g.,][]{Spitzer1969,Kulkarni1993}. In this case, the black holes would then undergo series of strong dynamical encounters and ultimately eject all but a few black holes from the cluster on sub-Gyr timescales. However, in the past decade, a number of analytic analyses and modern numerical simulations have shown that this argument of rapid black hole evaporation is incorrect and in fact, many black holes are likely retained up to the present day in most clusters \citep[e.g.,][]{Mackey2008, BreenHeggie2013, Morscher2015, Kremer2020_catalog}. These advances in theory have also been complemented by various observations (described further in Section 2), confirming that indeed most old globular clusters contain populations of black holes and other compact objects today.

\subsection{How black holes shape globular clusters}

As self-gravitating systems with negative heat capacities, globular clusters naturally feature a runaway flow of energy from their strongly self-gravitating cores to their relatively sparse halos that inevitably leads to core contraction and ultimately \textbf{core collapse} \citep{Henon1961,Antonov1962,Lynden-Bell1968}. However, observationally Milky Way globular clusters exhibit a clear bimodal distribution in core radii separating core-collapsed and non-core-collapsed clusters \citep{Harris1996}. Explaining this bimodality, and specifically explaining why, in spite of their short relaxation times, most globular clusters are \textit{not} core collapsed at present (as the above argument may imply), has been a major puzzle in stellar dynamics for decades.

Recent work has begun to show that stellar-mass black holes may provide the answer. Through series of strong dynamical encounters in the inner, black hole subsystem region of a typical globular cluster, black holes are frequently ejected to higher orbits in their host cluster potential, leading to interactions with luminous stars in the outer parts of the cluster. Through these interactions, the black holes deposit energy into the globular cluster's stellar bulk and influence the large-scale structural properties of their host cluster \citep{Mackey2008, BreenHeggie2013, Wang2016, Kremer2019a}. Thus, while present in globular clusters, black holes naturally produce dynamical energy through so-called \textbf{``black hole burning”} at a rate sufficient to delay the collapse of the cluster’s core. Because most (roughly 80\%) of Galactic globular clusters have large well-resolved cores, we expect most globular clusters at present still retain dynamically-important populations of black holes at their centers. Only when a globular cluster's black hole population has been almost fully depleted can a cluster attain a core-collapsed architecture \citep{Kremer2018b}. In core-collapsed clusters, neutron stars and massive white dwarfs become the most dynamically-active comapct object populations \citep{Kremer2021_wd, Vitral2023}.

\section{Observations of stellar remnants in clusters}

A variety of observational evidence across the electromagnetic spectrum accumulated over the past several decades now confirms the theoretical arguments for compact objects in clusters discussed in the previous section. Here we summarize some of the key observational advances.

\subsection{X-ray binaries}

The first pieces of observational evidence for compact objects in globular clusters came via the \textit{Uhuru} and OSO-7 satellites
in the 1970s. These revealed a population of luminous ($L_X \gtrsim 10^{36}\,$erg/s) \textbf{low-mass X-ray binaries} (LMXBs) in the Milky Way globular clusters. In these systems, a neutron star accretes material from a low-mass ($\lesssim M_{\odot}$) stellar (or possible white dwarf) companion. These initial observations revealed that LMXBs are roughly 100 times more common (per unit stellar mass) in globular clusters relative to isolated binaries in the Galactic field \citep{Clark1975,Katz1975}. Shortly thereafter, a variety of theoretical arguments demonstrated this overabundance arises via unique formation channels in globular clusters enabled by their high central densities \citep{Fabian1975,Hills1975,Heggie1975}.

\begin{figure}[t]
\centering
\includegraphics[width=\textwidth]{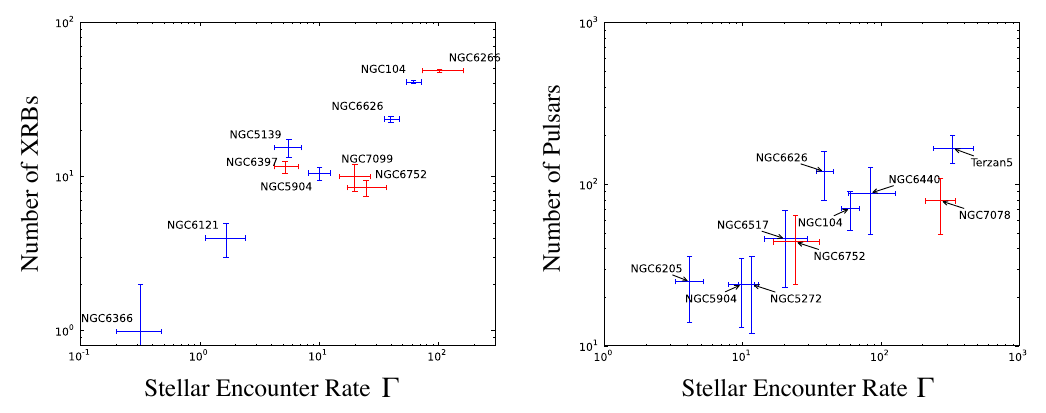}
\caption{Adapted from \citet{Bahramian2013}. \textit{Left panel:} Number of XRBs \citep{Pooley2003,Lugger2007} per cluster versus stellar encounter rate, $\Gamma$, as defined in \citet{Bahramian2013}. \textit{Right panel:} Number of radio pulsars per cluster \citep[from][]{Bagchi2011} versus $\Gamma$. In both panels, core-collapsed clusters are shown in red, non-core-collapsed clusters in blue. In both cases, a clear correlation exists between source count and stellar encounter rate, demonstrating that stellar dynamics plays a critical role in the formation of these systems in globular clusters.}
\label{fig:gamma}
\end{figure}

More recently, the \textit{Chandra} X-ray Observatory has enabled further breakthrough in our understanding of X-ray sources in globular clusters \citep[e.g.,][]{Grindlay2001,Rutledge2002,Pooley2002}. In addition to bright ($L_X \gtrsim 10^{36}\,$erg/s) LXMBs with neutron star accretors, \textit{Chandra} has also revealed large numbers of fainter ($L_X \lesssim 10^{34}\,$erg/s)  sources: quiescent LXMBs, cataclysmic variables (where a white dwarf accretes from a stellar companion), and active main-sequence or subgiant binaries. The robust X-ray source catalog from \textit{Chandra} has also enabled exploration of X-ray source populations with cluster properties \citep[e.g.,][]{Pooley2003,Bahramian2013}. It has been demonstrated by numerous studies that the number of X-ray sources in globular clusters correlates with the stellar encounter rate of their host, often parameterized with some variation of the so-called ``Gamma'' parameter: $\Gamma \propto \rho^2/v$ where $\rho$ is the mass density of stars in the cluster's center and $v$ is the central velocity dispersion. Clusters with the highest interaction rates generally feature the highest number of X-ray sources (see Figure~\ref{fig:gamma}), indicating that stellar dynamical processes play a crucial role in the formation of these sources.


Although neutron stars and white dwarfs have been observed in clusters for decades, the first observational evidence of stellar-mass black holes in globular clusters has come only recently. In extragalactic globular clusters, ultraluminous ($L_X \gtrsim 10^{39}\,$erg/s) X-ray sources hint at the presence of black hole accretors \citep[e.g.,][]{Maccarone2007}.
In the Milky Way globular clusters, detections of black hole LMXBs in quiescence have been enabled by deep radio continuum observations. Identifying quiescent black hole LMXBs is challenging because black holes do not produce distinct X-ray signatures like thermal surface emission produced by quiescent neutron star LMXBs \citep[e.g.,][]{Heinke2003}. However, accreting black holes do emit faint radio jets with higher radio luminosities than neutron star systems with similar X-ray luminosities \citep{MigliariFender2006}. Since the early 2010s, several studies have identified stellar-mass black hole candidates in Galactic globular clusters via detection of this faint radio emission \citep{Strader2012,Chomiuk2013, Miller-Jones2015,Shishkovsky2018}. These detections, along with the aforementioned ultraluminous X-ray sources in extragalactic clusters around this same time, provided the first observational evidence for stellar-mass black holes in globular clusters and provided the first hints that some globular clusters may indeed retain black holes up to the present day. 

\subsection{Radio pulsars}
\label{sec:pulsars}

Neutron stars are well-known sources of radio emission powered by spin-down luminosity enabled by magnetic dipole radiation \citep{Manchester2005}. Once formed, a neutron star evolving via magnetic dipole radiation will spin down over time, following tracks of roughly constant characteristic field strength

\begin{equation}
    \label{eq:B}
    B \approx \Bigg( \frac{3c^3I}{8\pi^2R_{\rm ns}^6} \Bigg)^{1/2} (P \dot{P} )^{1/2} \approx 10^{12} \Bigg( \frac{P}{100\,\rm{ms}}\Bigg)^{1/2} \Bigg( \frac{\dot{P}}{10^{-14}\rm{s/s}}\Bigg)^{1/2}\,\rm{G}
\end{equation}
(here $I\approx 0.4 M R_{\rm ns}^2$ is the neutron star moment of inertia, with $M\approx 1.4M_{\odot}$ and $R_{\rm ns}\approx10\,$km) and crossing lines of constant characteristic age

\begin{equation}
    \label{eq:tau}
    \tau \approx \frac{P}{2\dot{P}} \approx 10^5\, \Bigg( \frac{\it{P}}{50\,\rm{ms}}\Bigg)^2 \Bigg( \frac{\it{B}}{10^{12}\,\rm{G}}\Bigg)^{-2}\,\rm{yr}
\end{equation}
\citep[e.g.,][]{ShapiroTeukolsky1983}. Eventually, pulsars spin down sufficiently to fall below the so-called ``death line'', an empirical boundary near $B/P^2 = 1.7\times10^{11}\,\rm{G\,s}^{-2}$ \citep[e.g.,][]{RudermanSutherland1975} below which they no longer emit observable radiation. For a given initial magnetic field, the characteristic spin-down time to reach this death line is roughly $\tau_{\rm sd} \approx 10^8 (B/10^{12}\,\rm{G})^{-1}\,\rm{yr}$. This suggests that in typical old globular clusters, with stellar populations with ages $10^{10}\,$yr or more, any pulsars formed from the initial population of massive stars will have long ago spun down and will no longer be detectable as radio sources.

However, if a neutron star accretes material in a LMXB system, it can be spun up \citep{PhinneyKulkarni1994}. This process enables a previously undetectable neutron star to be ``recycled'' and reactivated as a bright radio source. Such recycled neutron stars are observed as radio \textbf{millisecond pulsars} (MSPs) characterized by spin periods as short as a few milliseconds. The presence of neutron stars accreting as LMXBs in globular clusters naturally implies a population of MSPs in these systems and indeed in the 1980s, the first globular cluster MSPs were identified \citep{Lyne1987}. Since these initial detections, more than 300 MSPs have been found in Galactic globular clusters.\footnote{For an up-to-date list, see Paulo Freire's \hyperlink{https://www3.mpifr-bonn.mpg.de/staff/pfreire/GCpsr.html}{``Pulsars in Globular Clusters''} catalog.}

Analogous to LMXBs, the specific abundance of MSPs in globular clusters is roughly $100$ times higher relative to MSPs in the Galactic field, suggesting again that dynamical processes unique to these dense environment provide efficient pathways for forming MSPs not accessible for isolated stellar binaries. Furthermore, (also akin to LMXBs) pulsars exhibit a clear correlation with stellar encounter rate (see right panel of Figure~\ref{fig:gamma}). Indeed, modern N-body simulations \citep[e.g.,][]{Ye2018}, confirm that dynamical processes play a critical role in formation of pulsar and pulsar binaries similar to those observed in globular clusters. 

Amidst the full population of several hundred radio pulsars known today in globular clusters, a smaller number of specific sources that hint at exciting phenomena in clusters are worth further comment:

\textit{Slow pulsars: formed via white dwarf mergers? --} Although the vast majority of pulsars in globular clusters are millisecond pulsars (spin periods $\lesssim 10\,$ms), a subset of six \textit{slowly spinning} pulsars (spin periods of roughly $0.3-1\,$s) have also been revealed \citep{Boyles2011,Zhou2024}. These slow pulsars have inferred ages of roughly $10^7-10^8\,$yr, remarkable because any massive stars typically associated with neutron star formation have been depleted in globular clusters for $10^{10}\,$yr. These objects have therefore been touted as evidence for alternative neutron star formation scenarios in old stellar populations, in particular involving collapse of massive white dwarfs via accretion from a binary companion \citep{Tauris2013} or white dwarf binary merger \citep{Kremer2023_psr}. Furthermore, all six of these young pulsars are observed in dense clusters that have undergone or are near core collapse, consistent with expectations from a white dwarf collapse scenario \citep{Kremer2023_psr}.

\textit{Pulsar+planet system --} One of the first pulsars discovered in a globular cluster -- PSR B1620-26 in the nearby cluster M4 -- is a recycled millisecond pulsar with a roughly $0.3\,M_{\odot}$ binary companion on a $191\,$d orbit \citep{Lyne1988}. Measurements of the higher time derivatives of the pulsar's spin period \citep{Thorsett1993} and direct optical observations of the pulsar's companion \citep{Sigurdsson2003} demonstrate the companion is a roughly $10^8\,$yr old undermassive white dwarf and further reveal the presence of an outer tertiary companion of Jupiter mass. At the time of its discovery, this object marked one of the first detections of an exoplanet and offers tantalizing hints that globular clusters may harbor large populations of planets. However, note that subsequent attempts to identify additional planets in globular clusters via transit observations have proven difficult \citep[e.g.,][]{Gilliland2000}.

\textit{Double neutron star binaries --} In the dense centers of core-collapsed clusters bereft of stellar-mass black holes, the dynamical formation of double neutron binaries seems inevitable. Indeed, a handful of such binaries are known. The first such system was identified in the core-collapsed globular cluster M15 --  a millisecond pulsar with a roughly $1.35\,M_{\odot}$ neutron star companion \citep{Anderson1990,Jacoby2006}. This system has a gravitational wave inspiral time of roughly $200\,$Myr implying a nonzero binary neutron star merger rate in globular clusters (we return to this in Section 3). Additional double neutron star binary candidates have been found in the clusters NGC 1851 \citep[companion mass $1.22\,M_{\odot}$;][]{Freire2004,Ridolfi2019}, NGC 6544 \citep[companion mass $1.2\,M_{\odot}$;][]{Lynch2012}, and NGC 7099 \citep[companion mass $\gtrsim 1.1\,M_{\odot}$;][]{Ransom2004, Balakrishnan2023}, however a massive white dwarf companion cannot yet be ruled out for these systems. More recently, the MeerKat radio telescope revealed a binary pulsar in NGC 1851 with a companion of mass $2.1-2.7\,M_{\odot}$ at 95\% confidence \citep{Barr2024}. This companion lies within the ``mass gap'' between neutron stars and black holes and is consistent with the masses measured for merger products of known binary neutron star mergers detected by LIGO/Virgo \citep[e.g.,][]{GW170817}.



\subsection{Detached black hole binaries}

\begin{figure}[t]
\centering
\includegraphics[width=\textwidth]{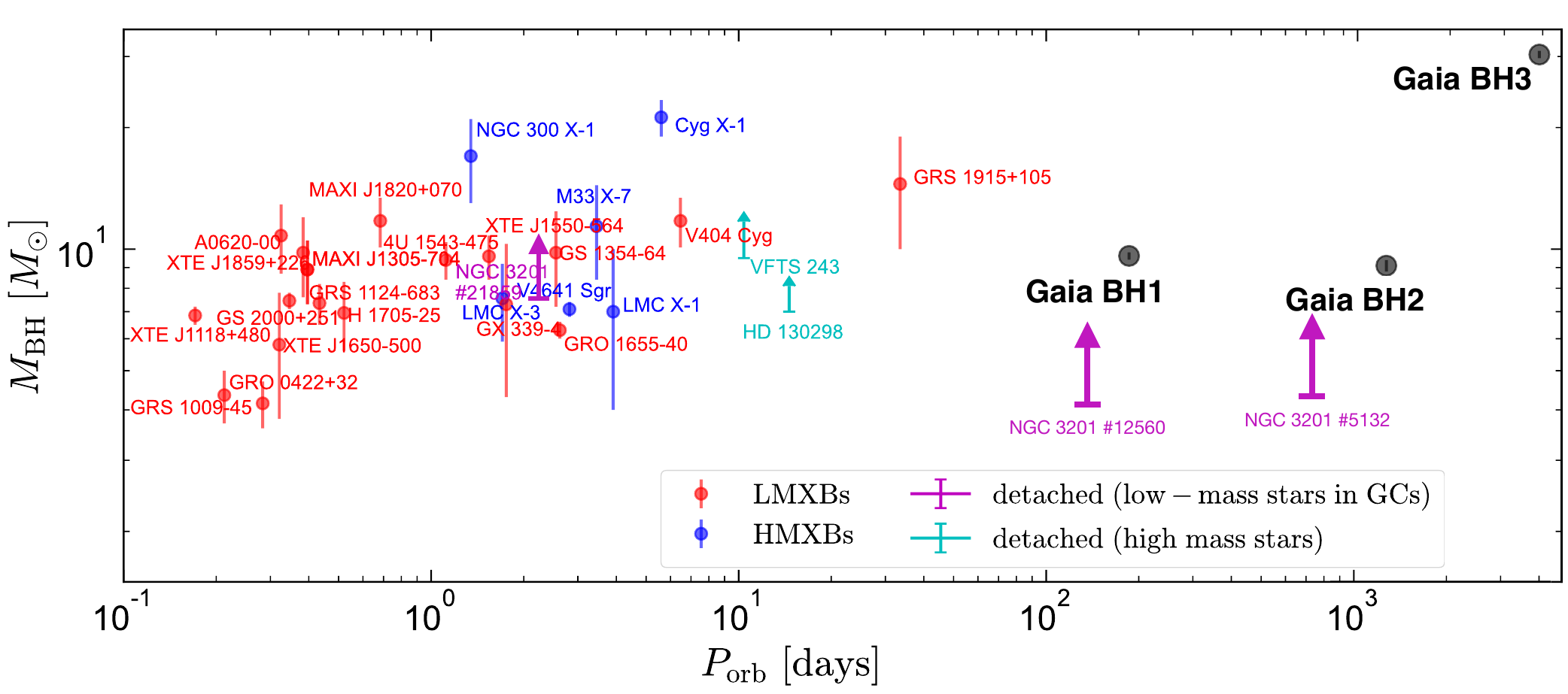}
\caption{Adapted from \citet{El-Badry2023b}. Known stellar-mass black hole binaries with black hole mass measurements. In red and blue, we show X-ray binaries from \citet{RemillardMcClintock2006,Corral-Santana2016}. In black, we show the three \textit{Gaia} black holes detected to date, and in magenta we show the minimum masses for the radial velocity black holes in the globular cluster NGC~3201.}
\label{fig:bh_masses}
\end{figure}

Perhaps the most robust observational evidence for stellar-mass black holes in globular clusters has emerged in just the past few years. The latest generation of integral field spectrograph instruments, namely the Multi Unit Spectroscopic Explorer (MUSE) instrument on the Very Large Telescope \citep{Bacon2014} enable star-by-star \textbf{radial velocity measurements} even in the crowded cores of globular clusters \citep{Kamann2016}. A recent radial velocity survey of Galactic globular clusters with MUSE revealed a $0.81\,M_{\odot}$ main sequence star in the cluster NGC~3201 with radial velocity variations consistent with the presence of a compact object companion \citep{Giesers2018}. Orbital solutions reveal a $180\,$d orbital period and a minimum companion mass of $4.3\,M_{\odot}$, almost certainly a black hole. This object marked the first stellar-mass black hole ever identified via a blind radial velocity survey and the first mass measurement for a black hole in a globular cluster.

Follow up numerical efforts \citep[e.g.,][]{Kremer2018b} demonstrated NGC~3201 likely contains a much larger underlying population of roughly 100 black holes including additional black hole+star binaries.
Indeed, subsequent observations via MUSE revealed two additional detached black hole+star binary candidates in this cluster with minimum companion masses of $4.4\,M_{\odot}$ and $7.7\,M_{\odot}$ and orbital periods of $760\,$d and $2.2\,$d, respectively \citep{Giesers2019}. Current work is underway to survey additional Galactic globular clusters to uncover more detached black hole binary systems.

These three detached black hole binaries found in NGC~3201 via radial velocity measurements complement analogous sources recently discovered in the field of the Milky Way via \textbf{astrometric measurements} enabled by the \textit{Gaia} mission \citep{Gaia2023}. To date, three black hole binaries have been identified within the \textit{Gaia} data: Gaia BH1, a $10\,M_{\odot}$ black hole in a $180\,$d orbit with a $0.9\,M_{\odot}$ near-solar-metallicity G dwarf \citep{El-Badry2023a}; Gaia BH2, a $9\,M_{\odot}$ black hole in a $1200\,$d orbit with a $1\,M_{\odot}$ red giant \citep{El-Badry2023b}; and Gaia BH3, a $33\,M_{\odot}$ black hole in a $4200\,$d orbit with a metal-poor ([Fe/H]$\approx -2.6$) $0.76\,M_{\odot}$ giant \citep{GaiaBH3_2024}. Both Gaia BH1 and BH2 have Galactocentric orbits consistent with origin in the Galactic plane \citep[for discussion of possible formation scenarios, see][]{Rastello2023, DiCarlo2024}. Gaia BH3 meanwhile is found in the Galactic halo and is chemically and kinematically associated with the metal-poor ED-2 stream \citep{Balbinot2024}. The progenitor of the ED-2 stream was most likely a disrupted globular cluster, suggesting that Gaia BH3 likely formed dynamically \citep{MarinPina2024}, perhaps in a manner similar to the three sources currently known in NGC~3201. Furthermore, Gaia BH3's low metallicity stellar companion provides the first clear observational evidence that low-metallicity stars \citep[similar to metallicities of typical globular clusters;][]{Harris1996} enable formation of more massive black holes, presumably due to reduced stellar wind mass loss at low metallicity \citep[e.g.,][]{Vink2001}. In Figure~\ref{fig:bh_masses}, we show all known black hole binaries with mass measurements, including X-ray binaries, the three current \textit{Gaia} black holes, and the three black hole candidates in the globular cluster NGC~3201.

\subsection{Indirect observational evidence of compact objects from dynamical modelling}

X-ray, radio, and optical radial velocity measurements all present methods for directly observing individual compact object sources in globular clusters. However, populations of compact objects can also be constrained via \textit{indirect} measurements, by leveraging the expected dynamical influence of the compact objects upon the luminous stellar bulk of their host cluster. This is particularly effective for identifying black holes, which owing to their large masses exhibit a pronounced influence, actively shaping the long-term evolution and present-day structure of their host globular cluster. When many black holes are retained in a cluster, energy flow between black holes and stars enabled by frequent dynamical encounters effectively ``heats'' the cluster leading to cluster core expansion \citep{Mackey2008,BreenHeggie2013,Kremer2020_bhburning}. As a cluster's black hole population is inevitably depleted on $10\,$Gyr timescales through dynamical ejections (see Section~1), the cluster's core radius slowly shrinks. In the rare cases where \textit{all} black holes are ejected within a Hubble time, the cluster undergoes core collapse \citep{Kremer2019a}. In the absence of black holes, neutron stars and white dwarfs become the dynamically-dominant stellar remnants in core-collapsed clusters \citep{Kremer2021_wd}.

This underlying connection between compact objects and cluster dynamics makes structural parameters like core radii powerful diagnostics of clusters' stellar remnant content. For example, NGC~3201, which has three known black hole binary candidates, features an anomalously large core radius of $2\,$pc, which has been linked to the influence of a large central black hole population \citep{Giesers2019}.
More generally, \citet{ArcaSedda2018} and \citet{Kremer2019a} used N-body simulations to predict the populations of black holes in a number of Milky Way globular clusters, based on their observed density profiles. \citet{Weatherford2020} used measurements of mass segregation of luminous stars to provide an independent prediction of the sizes of black hole subsystems in several dozen Galactic globular clusters. \citet{Vitral2022,Vitral2023} used a combination of proper motion data attained with \textit{Gaia} and HST and N-body simulations to constrain the mass and radial extent of black hole subsystems in NGC~3201 and M4 and of a white dwarf subsystem in the core-collapsed globular cluster NGC~6397. As a final example, \citet{Gieles2021} used the tidal tail features of Palomar~5 to argue this cluster features a supra-massive subsystem of black holes that is currently driving the cluster to dissolution.

\section{Globular clusters as factories of gravitational wave sources}

\begin{figure}[t]
\centering
\includegraphics[width=0.6\textwidth]{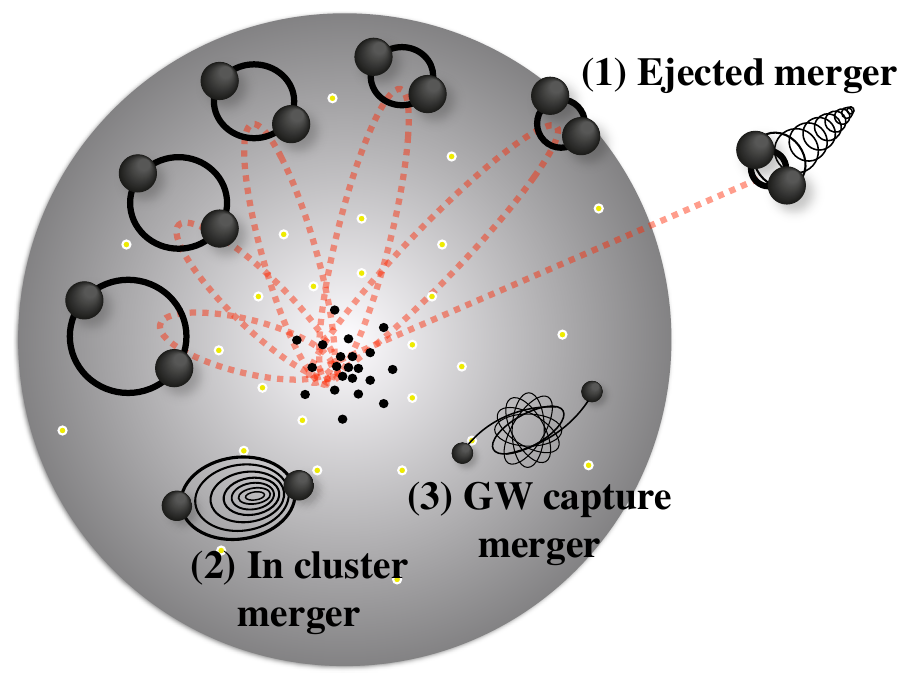}
\caption{Illustration of the three key dynamical channels for formation of merging black hole binaries in globular clusters discussed in the text. \textit{Ejected mergers} (roughly 50\% of all mergers) are dynamically hardened until ejection from their host cluster. These systems inspiral and merge outside of their original host. \textit{In-cluster mergers} (roughly 40\% of all mergers) feature relatively high eccentrities, enabling them to inspiral and merge within their host cluster before they can be dynamically ejected. \textit{Gravitational wave capture mergers} ($\lesssim 10\%$ of all mergers) form via close fly-by encounters between pairs of black holes. These sources have the highest eccentricities of all cluster mergers and may even yield sources with measurable eccentrities as they enter the LIGO frequency band \citep{Rodriguez2018b}.}
\label{fig:merger_cartoon}
\end{figure}

\subsection{Dynamics of black hole binary mergers}
\label{sec:bh_mergers}

A key consequence of the presence of compact objects in globular clusters is the dynamical formation, and ultimate merger, of compact object binary pairs that are prominent sources of \textbf{gravitational waves}. 
Over the past few years, the groundbreaking detections of gravitational wave signals from merging binary black holes and neutron stars by the LIGO/Virgo/KAGRA collaboration have opened a new window to the cosmos \citep{Abbott2016a}. 
Despite all we have learned already, the basic question of \textit{how} these gravitational wave sources formed remains unanswered. Dynamical formation in dense environments like globular clusters has emerged as an important formation channel for gravitational wave sources, particularly binary black hole mergers \citep[e.g.,][]{Rodriguez2016a}.\footnote{For the sake of brevity, we focus here only on formation in globular clusters. However, a number of other plausible formation channels have been proposed including isolated stellar binaries \citep[e.g.,][]{Belczynski2016a}, young massive clusters \citep[e.g.,][]{Banerjee2017}, nuclear star clusters \citep[e.g.,][]{Antonini2016}, hierarchical triples \citep[e.g.,][]{Silsbee2017}, disks of active galactic nuclei \citep[e.g.,][]{McKernan2018}, and primordial black holes \citep[e.g.,][]{Bird2016}. Some combination of these scenarios may in fact be most likely \citep[e.g.,][]{Zevin2021}.} In the dense core of a globular cluster, black holes undergo frequent dynamical encounters with one another and with cluster stars, ultimately leading to binary formation and eventually binary mergers. 

Dynamical encounters in a globular cluster can be loosely separated into two types -- strong encounters and weak encounters -- characterized by the distance of the interacting particles at closest approach. In a strong encounter between two objects of equal mass $m$, the total energy of the system at infinity, $m v^2$, is comparable to the total energy of the pair at the distance of closest approach, $Gm^2/r_p$. We can then define a minimum pericenter distance, $r_s$, for strong encounters as

\begin{equation}
    r_{s} \lesssim \frac{G m}{v^2} \approx 200 \, \Bigg( \frac{m}{30\,M_{\odot}} \Bigg) \Bigg( \frac{v}{10\,\rm{km/s}} \Bigg)^{-2}\,\rm{au}
\end{equation}
where $v \approx 10\,$km/s is the typical velocity dispersion for globular clusters. For strong encounters with $r_p < r_s$, the trajectories of the interacting particles are perturbed significantly. For individual weak encounters with $r_p \gg r_s$, the trajectories are changed negligibly, however weak encounters are far more common and their cumulative effect plays the dominant role in the relaxation process of the system as a whole (Section 1).

The characteristic time scale for black holes within the center of a globular cluster to undergo strong encounters can be estimated as

\begin{equation}
    \label{eq:t_enc}
    t_{\rm enc} \approx (n \Sigma v)^{-1} \approx 10^6 \Bigg( \frac{n}{10^5\,\rm{pc}^{-3}} \Bigg)^{-1} \Bigg( \frac{r_s}{200\,\rm{au}} \Bigg)^{-2} \Bigg( \frac{v}{10\,\rm{km/s}} \Bigg)^{-1}\,\rm{yr}
\end{equation}
Here $n\approx 10^5\,\rm{pc}^{-3}$ is the typical density of black holes in the center of globular cluster \citep[obtained from simulations;][]{Wang2016,Kremer2020_catalog} and $\Sigma \sim r_s^2$ is the interaction cross for strong encounters. Thus, a typical black hole in the center of a globular cluster will undergo thousands of dynamical encounters over the $10^{10}\,$yr  lifetime of the system.

The essential first step in forming a gravitational wave source is formation of a binary system. The presence of stellar binaries in globular clusters is now well-established observationally \citep{Hut1992} and observed black hole+star binaries (identified via both radial velocity and X-ray/radio measurements; see Section 2) demonstrate clearly that black hole binaries are formed in clusters. There are likely two primary mechanisms through which binaries can form in clusters. The first scenario is \textit{primordial cluster binaries.} When a globular cluster initially forms, a fraction of its massive stars are likely born in binary systems. Observations of O- and B-type stars in the Galactic field reveal binary fractions of nearly 100\% \citep{Sana2012} and observations of young massive clusters, the likely progenitors of globular clusters, have similarly high massive star binary fractions \citep{Sana2009}. Primordial binaries aside, binaries can also be assembled dynamically through a process called \textit{three-body binary formation} \citep{AarsethHeggie1976, Atallah2024}. Here, three initially unbound bodies interact sufficiently closely to pair up two of the objects. During a strong encounter between two black holes with separation $r_s$, the probability for a third black hole being nearby is $r_s^3 n \approx 10^{-5}$. Assuming each of these three-body encounters leads to binary formation,\footnote{See \citet{Atallah2024} for a more modern discussion of the subtleties of the three-body binary formation process.} the rate of three-body binary formation for the entire black hole population a cluster is given by $t_{\rm enc}^{-1} r_s^3 n$ and the characteristic time scale for three-body binary formation is:

\begin{equation}
    t_{\rm 3BB} \approx (n^2 r_s^5 v N_{\rm bh})^{-1} \approx 10^{8} \, \Bigg( \frac{n}{10^5\,{\rm{pc}^{-3}}} \Bigg)^{-2} \Bigg( \frac{r_s}{200\,\rm{au}} \Bigg)^{-5} \Bigg( \frac{v}{10\,\rm{km/s}} \Bigg)^{-1} \Bigg( \frac{N_{\rm bh}}{100} \Bigg)^{-1} \, \rm{yr}
\end{equation}
\citep{BinneyTremaine2008}. Thus over the lifetime of a cluster, we expect $\mathcal{O}(100)$ binary black holes to form via three-body encounters, roughly comparable to the expected number of dynamically-hard black hole binaries formed in the primordial binary population.

Once formed, black hole binaries will naturally undergo subsequent dynamical encounters with other black holes (and binaries) in the cluster center on timescale $t_{\rm enc}$ (Equation~\ref{eq:t_enc}). Analytic work dating back to the 1970s \citep{Heggie1975} shows that ``dynamically hard'' binaries (i.e., binaries with semi-major axis $a \lesssim r_s$) will naturally harden further as a consequence of subsequent encounters with other objects of energy $mv^2$.
As a binary undergoes dynamical encounters, its center-of-mass motion is also altered. Conservation of energy suggests that the characteristic center-of-mass recoil velocity of a binary is comparable to its orbital velocity:

\begin{equation}
    \label{eq:v_rec}
    v_{\rm rec} \sim v_{\rm orb} \sim \sqrt{\frac{Gm}{a}} 
\end{equation}
Thus, as a binary hardens ($a$ decreases), it attains increasingly large dynamical recoil kicks, launching the binary increasingly further out into the cluster. Once kicked, the binary mass segregates back to the center (see Equation~\ref{eq:t_ms}) and the process repeats: encounter, harden, kick, mass segregate. However, the binary cannot harden indefinitely, because globular clusters are governed by a natural speed limit set by the system's escape velocity (Equation~\ref{eq:v_esc}). In this case, the characteristic minimum separation, $a_{\rm ej}$, a binary can achieve via dynamical hardening before ejection is set by the criterion $v_{\rm rec} \sim v_{\rm esc}$, which when combined with Equations \ref{eq:v_esc} and \ref{eq:v_rec} gives

\begin{equation}
    \label{eq:a_ej}
    a_{\rm ej} \approx \frac{G m}{v_{\rm esc}^2} \approx  1\,\Bigg( \frac{m}{30\,M_{\odot}} \Bigg) \Bigg( \frac{v_{\rm esc}}{100\,\rm{km/s}} \Bigg)^{-2} \,\rm{au}.
\end{equation}
As a result, the fate of most black hole binaries in a typical globular cluster is to ultimately be ejected from their host with characteristic orbital separation roughly $a_{\rm ej}$. Because $v_{\rm esc} \propto M_{\rm cl}^{1/2}$, higher (lower) mass clusters tend to produce more compact (wider) ejected black hole binaries.

Once launched from their cluster out into the halo of their host galaxy, these binary black holes will continue to inspiral via emission of gravitational waves. For eccentric systems like those expected to form dynamically, the time to merger is given by:

\begin{equation}
    \label{eq:t_insp}
    t_{\rm insp} = \frac{3}{85} \frac{a^4 c^5}{G^3 m_1 m_2 (m_1+m_2)} (1-e^2)^{7/2} \approx 10^{10} \, \Bigg( \frac{a_{\rm ej}}{1\,\rm{au}} \Bigg)^{4} \Bigg( \frac{m}{30\,M_{\odot}} \Bigg)^{3} (1-e^2)^{7/2} \,\rm{yr}
\end{equation}
\citep{Peters1964} where we have assumed $e=0.9$ and $m=m_1=m_2$. For inspiral times less than a Hubble time, a successful black hole merger is ultimately expected. 

\begin{figure}[t]
\centering
\includegraphics[width=0.6\textwidth]{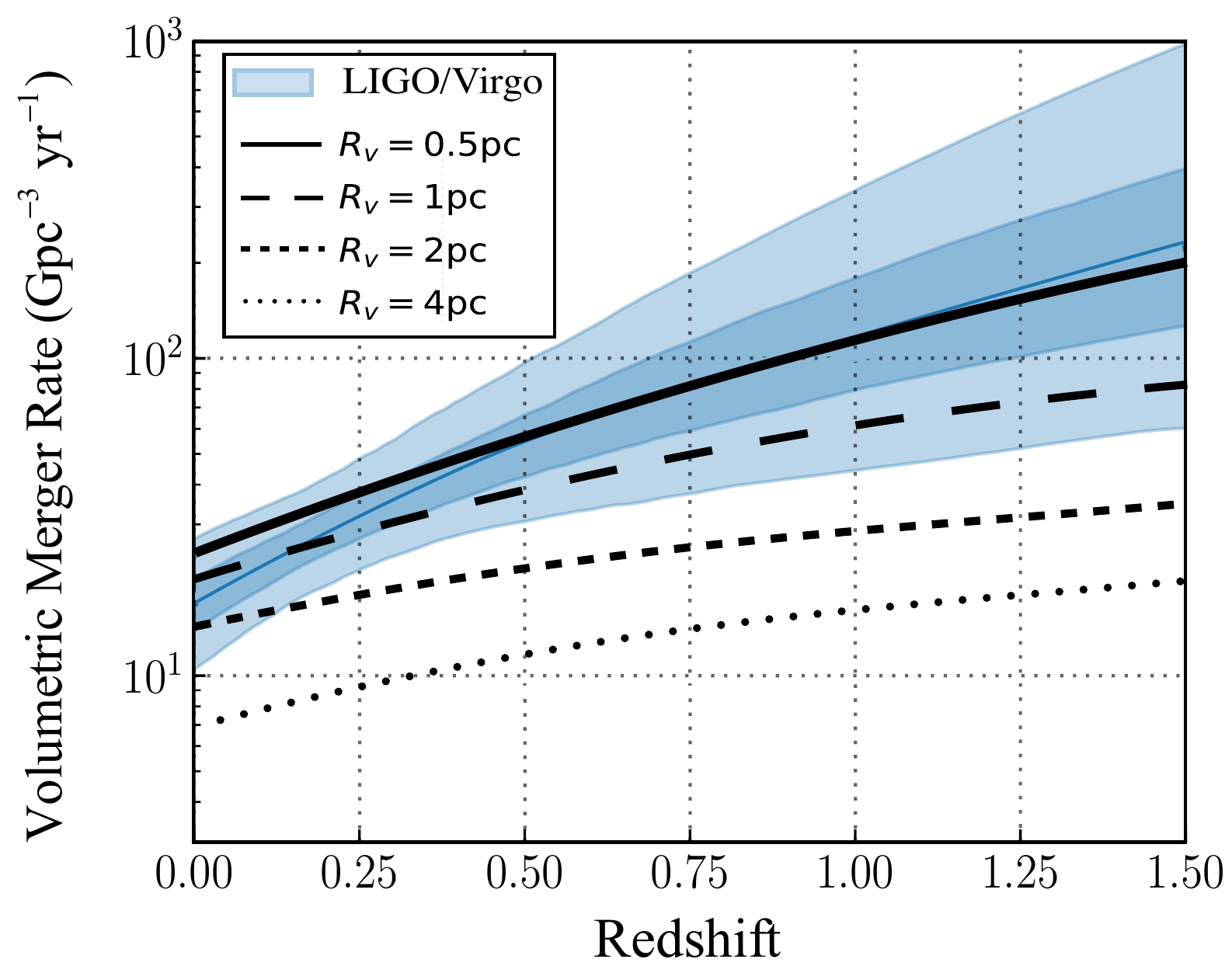}
\caption{Adapted from \citet{Rodriguez2021_RNAAS}. Volumetric rate of black hole binary mergers from globular clusters in the local universe, computed from N-body cluster simulations \citep{Kremer2020_catalog}. Different black curves assume different values for initial cluster virial radii, a key uncertain parameters in cluster birth properties. In blue, we show the volumetric rate inferred from gravitational wave detections up through the end of LIGO's third observing run \citep{LIGO2023}.
}
\label{fig:rates}
\end{figure}

Of course, this simple order-of-magnitude style calculation does not capture the full complexities of the dynamics at play. Numerical simulations of globular clusters show that in fact a range of orbital separations are produced, with average values roughly comparable to the $1\,$au limit derived above \citep[e.g., see Figure 1 of][]{Rodriguez2016a}. Furthermore, as black hole binaries undergo dynamical encounters they occasionally attain eccentrities much higher than the fiducial $0.9$ value adopted in Equation ~\ref{eq:t_insp}. As $e$ approaches unity, the gravitational wave inspiral time drops sharply, in some cases allowing mergers to occur before the binary can reach sufficiently small orbital separations to be ejected. In fact, cluster simulations show that roughly $40\%$ of all black hole mergers formed in typical globular clusters occur \textit{inside} their host \citep{Rodriguez2018a,Kremer2020_catalog}. In-cluster black hole mergers can also occur through a separate process called gravitational wave capture. In this case, a pair of interacting black holes come close enough for gravitational wave emission to remove sufficient energy from the orbit to result in a bound pair. The minimum distance required for a capture of two black holes of mass $m$ is

\begin{equation}
    r_{\rm cap} = \Bigg( \frac{85 \pi}{6} \Bigg)^{2/7} \frac{Gm}{c^{10/7} v^{4/7}} \approx 5\times 10^4 \Bigg( \frac{m}{30\,M_{\odot}} \Bigg) \Bigg( \frac{v}{10\,\rm{km/s}} \Bigg)^{-4/7}\,\rm{km}
\end{equation}
\citep{QuinlanShapiro1987}. Such events can occur between a pair of single black holes \citep{Samsing2020} or via close fly-bys during binary-mediated resonant encounters \citep{Samsing2014}. Relative to the typical strong encounter (Equation~\ref{eq:t_enc}), the dynamical cross section for these gravitational wave captures is very small, thus these events are rare. Simulations show captures constitute at most $10\%$ of all mergers expected in globular clusters \citep{Rodriguez2019}. Nonetheless, these captures may produce unique high-eccentricity mergers that may be used by detectors like LIGO to point toward dynamical formation \citep[][see Section~\ref{sec:ecc}]{Zevin2021_ecc}. In Figure~\ref{fig:merger_cartoon}, we illustrate the three key processes leading to black hole binary mergers in clusters that have been summarized here.

Finally, we can estimate the rate of black hole mergers expected from globular clusters. Assuming each of the $\mathcal{O}(100)$ black hole binaries formed over the $10^{10}\,$yr lifetime of a typical cluster ultimately merges, we can compute an approximate local universe merger rate of:

\begin{equation}
    \mathcal{R} \approx \Bigg( \frac{10^2\,\rm{mergers}}{10^{10}\,\rm{yr}} \Bigg) \Bigg( \frac{n_{\rm GC}}{1\,\rm{Mpc}^{-3}} \Bigg) \approx 10\,\rm{Gpc^{-3}\rm{yr}^{-1}}
\end{equation}
where $n_{\rm GC}$ is the number density of globular clusters in the local universe \citep{Rodriguez2015a}. This is comparable to within a small factor of the black hole merger rate of $17.9-44\,\rm{Gpc}^{-3}\,\rm{yr}^{-1}$ inferred from the most recent catalog of LIGO/Virgo/KAGRA events \citep{LIGO2023}, suggesting globular clusters may indeed be a significant channel for forming gravitational wave sources. Results from state-of-the-art $N$-body simulations coupled to realistic globular cluster formation histories corroborate this simple estimate. In Figure~\ref{fig:rates}, we show the volumetric merger rate versus redshift computed from simulations for a few different values of initial cluster virial radii, a key uncertainty in cluster birth properties \citep{Kremer2019a}.

\subsection{Black hole binary masses, eccentricities, and spins}
\label{sec:ecc}

The populations of black hole binary mergers assembled dynamically in stellar clusters are expected to exhibit a number of characteristic properties that, if measured with detectors like LIGO, may be leveraged as tell-tale features of the cluster formation channel. We summarize three of these key features below.

\textit{Massive black holes --} The maximum mass of black holes formed via evolution of massive stars is an open question. It is generally expected that stars that build helium cores of masses in the range $\sim40-130\,M_{\odot}$ are subject to the pair instability \citep{Fowler1964}, which may drive significant mass loss via violent pulsations or destroy the star completely prior to collapse \citep[e.g.,][]{Woosley2002}. It has long been speculated that the pair instability process may lead to a dearth of black holes with masses in the range $\sim 40\,M_{\odot}-130\,M_{\odot}$, the so-called ``upper mass gap'' \citep[e.g.,][]{Woosley2017}. However, the details of this gap (e.g., the width, boundaries, and dependence on stellar properties like metallicity) are highly uncertain \citep[e.g.,][]{Farmer2019}.

A surprising feature of the gravitational wave detections made to-date has been the presence of merging binaries with masses residing in the purported upper mass gap. The canonical event of this class is GW190521 \citep{GW190521}, with component masses of roughly $85\,M_{\odot}$ and $65\,M_{\odot}$. The direct detections of GW190521 and other upper-mass-gap candidate events \citep{LIGO2023} spark questions concerning our understanding of the stellar black hole mass spectrum.

Dense star clusters provide several avenues through which massive black hole binary mergers may form, consistent with our current understanding of the pair-instability process. First, heavy black holes can form via earlier black hole mergers. Merging binary black holes experience an impulsive velocity kick from anisotropic emission of gravitational waves \citep[e.g.,][]{Favata2004}. In a cluster, if the recoil velocities of such merger products are less than their host cluster's escape speed, the newly-formed black holes will be retained by the cluster, creating a new generation of black holes that may merge again \citep[e.g.,][]{AntoniniRasio2016,Rodriguez2019}. These ``second-generation'' mergers have unique masses, mass ratios, and spins which may be difficult or impossible to produce from first-generation black hole binaries formed from collapsing stars \citep[e.g.,][]{GerosaBerti2017}. Second, heavy black holes in, or beyond, the upper mass gap may arise from the collapse of anomalously massive progenitor stars \citep[e.g.,][]{DiCarlo2019,Gonzalez2021}. These can form via massive stellar collisions, which occur frequently at early times in dense clusters
\citep[e.g.,][]{PortegiesZwart2004} and have long been connected to formation of intermediate-mass black holes with masses in the range $\sim10^2-10^4\,M_{\odot}$ (Section~\ref{sec:imbh}).

\textit{Isotropic spins --} In general, the two best-measured spin terms for a merging black hole binary are the effective spin $\chi_{\rm eff}$, which encodes a mass-weighted projection of the spin vectors on the orbital angular momentum axis, and the precessing spin $\chi_{\rm p}$, which is related to the strength of precession of the orbital angular momentum about the total angular momentum and is determined by the projection of the component spin vectors onto the plane of the orbit \citep[e.g.,][]{Damour2001,Schmidt2015}. Black hole binaries assembled dynamically in globular clusters (or other gas-poor environments) are expected to have random (isotropically distributed) spin orientations as a result of the chaotic encounters that lead to their formation \citep[e.g.,][]{HeggieHut2003}. Thus, if indeed cluster dynamics is a significant contributor to the population of binary black hole mergers detectable in the local universe, subsets of mergers with negative $\chi_{\rm eff}$ and/or high values of $\chi_{\rm p}$ (both challenging to produce via isolated binary evolution) are naturally expected. Detection of these features (in the full GW populations or in individual events) may then place constraints on the contribution of dynamical formation in dense stellar clusters \citep[e.g.,][]{Rodriguez2016c,Fishbach2022,Payne2024}.



\textit{High eccentricity binaries --} 
Systems assembled in globular clusters are expected to form with high eccentricities \citep[on average, drawn from a thermal distribution, $f(e) = 2e de$;][]{Heggie1975}. This is in contrast to merging binaries formed via isolated stellar binaries, which are generally expected to feature relatively circular orbits \citep[e.g.,][]{Postnov2014}. In this case, observation (or lack thereof) of eccentric black hole binary systems may constrain the fraction of systems that originate from dynamical environments like globular clusters \citep[e.g.,][]{Zevin2021_ecc}.\footnote{Note that additional proposed formation channels for gravitational wave sources may also yield highly eccentric sources, most notably secular evolution of hierarchical triple systems \citep[e.g.,][]{AntoniniPerets2012,SilsbeeTremaine2017}.}

By the time most merging binaries (including those formed dynamically) enter the frequency band detectable by current ground-based detectors like LIGO ($f \gtrsim 10\,$Hz), any residual eccentricity will have mostly been erased \citep[e.g.,][]{Peters1964}. The exception is the subset of binaries assembled via gravitational wave capture that form with $e \sim 1$ \citep{QuinlanShapiro1987} and may still have measurable eccentricities even in the LIGO band \citep[e.g.,][]{O'Leary2009,Samsing2014,Rodriguez2018a}. However, these highly-eccentric sources are intrinsically rare, expected to constitute at most $\sim10\%$ of all mergers in globular clusters (see Section~\ref{sec:bh_mergers}). Furthermore, unambiguous detection of eccentricity of LIGO/Virgo/KAGRA sources is a challenge at present \citep[for discussion, see e.g.,][]{Zevin2021_ecc}. Future low-frequency detectors like LISA, which will observe binaries closer to the frequencies at which they formed, may provide a much more efficient way to use eccentricity as a means to distinguish formation scenarios \citep{Breivik2016}. For example, as a many as half of all black hole binaries formed in globular clusters that are resolvable by LISA at millihertz frequencies are expected to have measurable eccentricities \citep{D'OrazioSamsing2018,Kremer2019b}.

\subsection{White dwarf and neutron star mergers}

The dynamical processes outlined in this basic picture are not limited to black hole binary mergers. In core-collapsed clusters that have already ejected all of their black holes, an analogous process operates that leads to formation and merger of white dwarf and neutron star binaries. Although the total neutron star merger rate in clusters is likely a small fraction of the total neutron star merger rate inferred by LIGO \citep{Ye2020_BNS}, neutron star mergers may still provide a viable pathway for forming objects like the mass-gap companion of pulsar NGC~1851E discussed in Section 2.2. White dwarf mergers in clusters may be observable as gravitational wave sources by next generation detectors like LISA and DECIGO \citep{Kremer2021_wd} and may ultimately power events such as Type Ia supernovae \citep[e.g.,][]{Bregman2024} and Calcium-strong transients \citep[e.g.,][]{Shen2019} or collapse into young neutron stars \citep{Kremer2023_psr}.

\section{Recent advances and the future}

\subsection{Intermediate-mass black holes}
\label{sec:imbh}

We have focused here primarily on stellar-mass black holes. However, globular clusters have also long been linked to so-called \textbf{intermediate mass black holes} \citep[IMBHs;][]{Greene2020}. With masses in the range $\sim 10^2-10^5\,M_{\odot}$, IMBHs would provide the ``missing link'' between stellar-mass black holes formed from the evolution of massive stars and supermassive black holes found at the centers of most galaxies. Extension of the $M_{\rm bh}-\sigma_{\star}$ relation \citep{FerrareseMerritt2000} to low masses suggests globular cluster as a natural place to search for IMBHs. Furthermore, several formation scenarios for IMBHs in globular clusters have been theorized for decades, including runaway mergers of massive stars \citep[e.g.,][]{PortegiesZwart2004} or stellar-mass black holes \citep[e.g.,][]{MillerHamilton2002} or accretion of gas in the nascent giant molecular by seed black holes \citep[e.g.,][]{Shi2023}.

In spite of this strong theoretical motivation, the search for observational evidence of IMBHs in globular clusters has been a challenge. A number of IMBH candidates have been put forth from X-ray and radio observations \citep[e.g.,][]{Tremou2018} and from dynamical measurements \citep[e.g.,][]{Feldmeier2013,Lutz2011,Noyola2010, Perera2017}. Nonetheless, most of these candidates remain controversial as unambiguous confirmation of IMBHs \citep[e.g.,][]{Gieles2018,Zocchi2019}.
Recent work, however, has revealed perhaps the most compelling evidence yet for an IMBH in a globular cluster. Astrometric measurements enabled by HST reveal seven fast-moving stars in the central 3 arcseconds of the Galactic globular cluster $\omega$Centauri \citep{Haberle2024},  analogous to the S-stars in the Galactic center orbiting the supermassive black hole Sgr A$^\star$ \citep{Ghez2008}.
These seven stars can only be explained if they are bound to a central black hole, as the velocities of the stars are significantly higher than the expected central escape velocity of their host cluster. From these velocity measurements, \citet{Haberle2024} infers a lower mass limit of roughly $8000\,M_{\odot}$ for the central black hole in $\omega$Centauri, the strongest candidate yet for an IMBH in the local universe.

\subsection{Fast radio bursts}

As summarized in Section 2.2, globular clusters have a rich history in radio astronomy. A new chapter was added to this story in 2020 when a repeating \textbf{fast radio burst} (FRB) was localized to a globular cluster in M81 \citep{Kirsten2022}. The origin of FRBs in general remains unclear, however a Galactic FRB coincident with a known magnetar \citep{Bochenek2020} provides clear evidence that some FRBs are magnetar-powered. However, in old ($\gtrsim10\,$Gyr) globular clusters, magnetars formed through massive star evolution have been inactive for billions of years. Thus, alternative models are necessary to explain the M81 FRB. Recent studies have postulated collapse of massive white dwarfs \citep{Kremer2021_frb, Lu2022} and black hole X-ray binaries \citep{Sridhar2021} as possible explanations.

Remarkably, the M81 FRB is both the \textit{closest} known extragalactic FRB and the \textit{only} known extragalactic FRB where a host cluster identification was possible, hinting that FRB association with globular clusters may in fact be common \citep{Kremer2023_frb}. Ongoing efforts to identify FRB host galaxy properties \citep[e.g.,][]{Gordon2023, Law2024} will provide crucial insight over the coming years into the origin of FRBs and the possible role of cluster dynamics beyond just the M81 source.

\subsection{Electromagnetic transients from stellar collisions}

The high densities in stellar clusters facilitate frequent close-fly by encounters that can lead to tidal disruptions or even direct collisions between pairs of objects. These disruptive interactions lead to an incredibly diverse range of outcomes, depending on the types of stars involved.
Mergers of pairs of main sequence stars are one pathway to form blue straggler stars, which are observed in nearly every Galactic globular cluster \citep{Bailyn1995}. Mergers of giant-branch stars lead to common envelope-like events that are observable as luminous red novae \citep{MetzgerPejcha2017}. Series of multiple collisions in young clusters can create very massive stars $\mathcal{O}(100)\,M_{\odot}$ or more, which may ultimately lead to pair-instability supernovae or collapse into IMBHs \citep{Kremer2020_imbh}. Close interactions between stars and compact objects may produce ultra-compact X-ray binaries \citep[e.g., as observed in 47 Tuc;][]{Bahramian2017} or collisions that create powerful electromagnetic transients \citep{Perets2016,Kremer2022_sph}.

Current all-sky electromagnetic surveys like the Zwicky Transient Facility and the Sloan Digital Sky Survey (SDSS) have unveiled whole zoos of new astronomical transients, laying the foundation for the Vera Rubin Observatory which will detect \textit{millions} of transient events each night \citep{LSST2019}. Attaining a better understanding of the array of transients expected to form dynamically in dense stellar clusters will be paramount to interpretation of future all-sky transient survey data.

\subsection{Conclusions \& looking forward}
It is now clear from a variety of observational evidence that globular clusters contain robust and dynamically-active populations of white dwarfs, neutron stars, and black holes throughout their lifetimes. It is also clear we have only scratched the surface observationally of the full populations present. Current state-of-the-art radio telescopes like FAST and MeerKat promise to increase the number of known radio pulsars by \textit{up to ten times}, including in Galactic globular clusters \citep[e.g.,][]{FAST2021,TRAPUM2021}. Radial velocity measurements enabled by integral field spectrographs like the Multi Unit Spectroscopic Explorer, MUSE, have only recently enabled the first direct mass measurements of black holes in globular clusters and are being complemented by astrometric measurements of stellar-mass black holes by \textit{Gaia} and HST. Furthermore, in the coming years and decades, the field of gravitational wave astrophysics is also destined to blossom. At design sensitivity, LIGO and its partners, Virgo and KAGRA, will detect hundreds of gravitational wave events per year. The recent detection of nanohertz gravitational waves by pulsar timing arrays \citep{NANOGrav2023} has launched the era of multiband gravitational wave astronomy. In the 2030s, the Laser Interferometer Space Antenna \citep[LISA;][]{LISA2023} will unveil millihertz frequencies,
and by 2040, new ground-based detectors like Cosmic Explorer and Einstein Telescope will enable detection of nearly every black hole binary merger in the observable universe \citep[e.g.,][]{Baibhav2019}. As the gravitational wave universe continues to be unveiled, the details on compact object populations in globular clusters (and their possible role in the formation of gravitational wave sources), will inevitably become more apparent.


\begin{multicols}{2}
\bibliographystyle{apj}
\bibliography{reference}

\begin{thebibliography}{}
\expandafter\ifx\csname natexlab\endcsname\relax\def\natexlab#1{#1}\fi

\bibitem[{{Aarseth} \& {Heggie}(1976)}]{AarsethHeggie1976}
{Aarseth}, S.~J., \& {Heggie}, D.~C. 1976, \aap, 53, 259

\bibitem[{{Abbott} {et~al.}(2016){Abbott}, {Abbott}, {Abbott}, {Abernathy}, {Acernese}, {Ackley}, {Adams}, {Adams}, {Addesso}, {Adhikari}, \& et~al.}]{Abbott2016a}
{Abbott}, B.~P., {Abbott}, R., {et~al.} 2016, \apjl, 818, L22

\bibitem[{{Abbott} {et~al.}(2017){Abbott}, {Abbott}, {Abbott}, {Acernese}, {Ackley}, {Adams}, {Adams}, {Addesso}, {Adhikari}, {Adya}, {Affeldt}, {Afrough}, {Agarwal}, {Agathos}, {Agatsuma}, {Aggarwal}, {Aguiar}, {Aiello}, {Ain}, {Ajith}, {Allen}, {Allen}, {Allocca}, {Altin}, {Amato}, {Ananyeva}, {Anderson}, {Anderson}, {Angelova}, {Antier}, {Appert}, {Arai}, {Araya}, {Areeda}, {Arnaud}, {Arun}, {Ascenzi}, {Ashton}, {Ast}, {Aston}, {Astone}, {Atallah}, {Aufmuth}, {Aulbert}, {AultONeal}, {Austin}, {Avila-Alvarez}, {Babak}, {Bacon}, {Bader}, {Bae}, {Bailes}, {Baker}, {Baldaccini}, {Ballardin}, {Ballmer}, {Banagiri}, {Barayoga}, {Barclay}, {Barish}, {Barker}, {Barkett}, {Barone}, {Barr}, {Barsotti}, {Barsuglia}, {Barta}, {Barthelmy}, {Bartlett}, {Bartos}, {Bassiri}, {Basti}, {Batch}, {Bawaj}, {Bayley}, {Bazzan}, {B{\'e}csy}, {Beer}, {Bejger}, {Belahcene}, {Bell}, {Berger}, {Bergmann}, {Bernuzzi}, {Bero}, {Berry}, {Bersanetti}, {Bertolini}, {Betzwieser}, {Bhagwat}, {Bhandare}, {Bilenko}, {Billingsley}, {Billman},
  {Birch}, {Birney}, {Birnholtz}, {Biscans}, {Biscoveanu}, {Bisht}, {Bitossi}, {Biwer}, {Bizouard}, {Blackburn}, {Blackman}, {Blair}, {Blair}, {Blair}, {Bloemen}, {Bock}, {Bode}, {Boer}, {Bogaert}, {Bohe}, {Bondu}, {Bonilla}, {Bonnand}, {Boom}, {Bork}, {Boschi}, {Bose}, {Bossie}, {Bouffanais}, {Bozzi}, {Bradaschia}, {Brady}, {Branchesi}, {Brau}, {Briant}, {Brillet}, {Brinkmann}, {Brisson}, {Brockill}, {Broida}, {Brooks}, {Brown}, {Brown}, {Brunett}, {Buchanan}, {Buikema}, {Bulik}, {Bulten}, {Buonanno}, {Buskulic}, {Buy}, {Byer}, {Cabero}, {Cadonati}, {Cagnoli}, {Cahillane}, {Calder{\'o}n Bustillo}, {Callister}, {Calloni}, {Camp}, {Canepa}, {Canizares}, {Cannon}, {Cao}, {Cao}, {Capano}, {Capocasa}, {Carbognani}, {Caride}, {Carney}, {Carullo}, {Casanueva Diaz}, {Casentini}, {Caudill}, {Cavagli{\`a}}, {Cavalier}, {Cavalieri}, {Cella}, {Cepeda}, {Cerd{\'a}-Dur{\'a}n}, {Cerretani}, {Cesarini}, {Chamberlin}, {Chan}, {Chao}, {Charlton}, {Chase}, {Chassande-Mottin}, {Chatterjee}, {Chatziioannou}, {Cheeseboro},
  {Chen}, {Chen}, {Chen}, {Cheng}, {Chia}, {Chincarini}, {Chiummo}, {Chmiel}, {Cho}, {Cho}, {Chow}, {Christensen}, {Chu}, {Chua}, {Chua}, {Chung}, {Chung}, {Ciani}, {Ciolfi}, {Cirelli}, {Cirone}, {Clara}, {Clark}, {Clearwater}, {Cleva}, {Cocchieri}, {Coccia}, {Cohadon}, {Cohen}, {Colla}, {Collette}, {Cominsky}, {Constancio}, {Conti}, {Cooper}, {Corban}, {Corbitt}, {Cordero-Carri{\'o}n}, {Corley}, {Cornish}, {Corsi}, {Cortese}, {Costa}, {Coughlin}, {Coughlin}, {Coulon}, {Countryman}, {Couvares}, {Covas}, {Cowan}, {Coward}, {Cowart}, {Coyne}, {Coyne}, {Creighton}, {Creighton}, {Cripe}, {Crowder}, {Cullen}, {Cumming}, {Cunningham}, {Cuoco}, {Dal Canton}, {D{\'a}lya}, {Danilishin}, {D'Antonio}, {Danzmann}, {Dasgupta}, {Da Silva Costa}, {Dattilo}, {Dave}, {Davier}, {Davis}, {Daw}, {Day}, {De}, {DeBra}, {Degallaix}, {De Laurentis}, {Del{\'e}glise}, {Del Pozzo}, {Demos}, {Denker}, {Dent}, {De Pietri}, {Dergachev}, {De Rosa}, {DeRosa}, {De Rossi}, {DeSalvo}, {de Varona}, {Devenson}, {Dhurandhar}, {D{\'\i}az},
  {Dietrich}, {Di Fiore}, {Di Giovanni}, {Di Girolamo}, {Di Lieto}, {Di Pace}, {Di Palma}, {Di Renzo}, {Doctor}, {Dolique}, {Donovan}, {Dooley}, {Doravari}, {Dorrington}, {Douglas}, {Dovale {\'A}lvarez}, {Downes}, {Drago}, {Dreissigacker}, {Driggers}, {Du}, {Ducrot}, {Dudi}, {Dupej}, {Dwyer}, {Edo}, {Edwards}, {Effler}, {Eggenstein}, {Ehrens}, {Eichholz}, {Eikenberry}, {Eisenstein}, {Essick}, {Estevez}, {Etienne}, {Etzel}, {Evans}, {Evans}, {Factourovich}, {Fafone}, {Fair}, {Fairhurst}, {Fan}, {Farinon}, {Farr}, {Farr}, {Fauchon-Jones}, {Favata}, {Fays}, {Fee}, {Fehrmann}, {Feicht}, {Fejer}, {Fernandez-Galiana}, {Ferrante}, {Ferreira}, {Ferrini}, {Fidecaro}, {Finstad}, {Fiori}, {Fiorucci}, {Fishbach}, {Fisher}, {Fitz-Axen}, {Flaminio}, {Fletcher}, {Fong}, {Font}, {Forsyth}, {Forsyth}, {Fournier}, {Frasca}, {Frasconi}, {Frei}, {Freise}, {Frey}, {Frey}, {Fries}, {Fritschel}, {Frolov}, {Fulda}, {Fyffe}, {Gabbard}, {Gadre}, {Gaebel}, {Gair}, {Gammaitoni}, {Ganija}, {Gaonkar}, {Garcia-Quiros}, {Garufi}, {Gateley},
  {Gaudio}, {Gaur}, {Gayathri}, {Gehrels}, {Gemme}, {Genin}, {Gennai}, {George}, {George}, {Gergely}, {Germain}, {Ghonge}, {Ghosh}, {Ghosh}, {Ghosh}, {Giaime}, {Giardina}, {Giazotto}, {Gill}, {Glover}, {Goetz}, {Goetz}, {Gomes}, {Goncharov}, {Gonz{\'a}lez}, {Gonzalez Castro}, {Gopakumar}, {Gorodetsky}, {Gossan}, {Gosselin}, {Gouaty}, {Grado}, {Graef}, {Granata}, {Grant}, {Gras}, {Gray}, {Greco}, {Green}, {Gretarsson}, {Groot}, {Grote}, {Grunewald}, {Gruning}, {Guidi}, {Guo}, {Gupta}, {Gupta}, {Gushwa}, {Gustafson}, {Gustafson}, {Halim}, {Hall}, {Hall}, {Hamilton}, {Hammond}, {Haney}, {Hanke}, {Hanks}, {Hanna}, {Hannam}, {Hannuksela}, {Hanson}, {Hardwick}, {Harms}, {Harry}, {Harry}, {Hart}, {Haster}, {Haughian}, {Healy}, {Heidmann}, {Heintze}, {Heitmann}, {Hello}, {Hemming}, {Hendry}, {Heng}, {Hennig}, {Heptonstall}, {Heurs}, {Hild}, {Hinderer}, {Ho}, {Hoak}, {Hofman}, {Holt}, {Holz}, {Hopkins}, {Horst}, {Hough}, {Houston}, {Howell}, {Hreibi}, {Hu}, {Huerta}, {Huet}, {Hughey}, {Husa}, {Huttner}, {Huynh-Dinh},
  {Indik}, {Inta}, {Intini}, {Isa}, {Isac}, {Isi}, {Iyer}, {Izumi}, {Jacqmin}, {Jani}, {Jaranowski}, {Jawahar}, {Jim{\'e}nez-Forteza}, {Johnson}, {Johnson-McDaniel}, {Jones}, {Jones}, {Jonker}, {Ju}, {Junker}, {Kalaghatgi}, {Kalogera}, {Kamai}, {Kandhasamy}, {Kang}, {Kanner}, {Kapadia}, {Karki}, {Karvinen}, {Kasprzack}, {Kastaun}, {Katolik}, {Katsavounidis}, {Katzman}, {Kaufer}, {Kawabe}, {K{\'e}f{\'e}lian}, {Keitel}, {Kemball}, {Kennedy}, {Kent}, {Key}, {Khalili}, {Khan}, {Khan}, {Khan}, {Khazanov}, {Kijbunchoo}, {Kim}, {Kim}, {Kim}, {Kim}, {Kim}, {Kim}, {Kimbrell}, {King}, {King}, {Kinley-Hanlon}, {Kirchhoff}, {Kissel}, {Kleybolte}, {Klimenko}, {Knowles}, {Koch}, {Koehlenbeck}, {Koley}, {Kondrashov}, {Kontos}, {Korobko}, {Korth}, {Kowalska}, {Kozak}, {Kr{\"a}mer}, {Kringel}, {Krishnan}, {Kr{\'o}lak}, {Kuehn}, {Kumar}, {Kumar}, {Kumar}, {Kuo}, {Kutynia}, {Kwang}, {Lackey}, {Lai}, {Landry}, {Lang}, {Lange}, {Lantz}, {Lanza}, {Larson}, {Lartaux-Vollard}, {Lasky}, {Laxen}, {Lazzarini}, {Lazzaro}, {Leaci},
  {Leavey}, {Lee}, {Lee}, {Lee}, {Lee}, {Lee}, {Lehmann}, {Lenon}, {Leon}, {Leonardi}, {Leroy}, {Letendre}, {Levin}, {Li}, {Linker}, {Littenberg}, {Liu}, {Liu}, {Lo}, {Lockerbie}, {London}, {Lord}, {Lorenzini}, {Loriette}, {Lormand}, {Losurdo}, {Lough}, {Lousto}, {Lovelace}, {L{\"u}ck}, {Lumaca}, {Lundgren}, {Lynch}, {Ma}, {Macas}, {Macfoy}, {Machenschalk}, {MacInnis}, {Macleod}, {Maga{\~n}a Hernandez}, {Maga{\~n}a-Sandoval}, {Maga{\~n}a Zertuche}, {Magee}, {Majorana}, {Maksimovic}, {Man}, {Mandic}, {Mangano}, {Mansell}, {Manske}, {Mantovani}, {Marchesoni}, {Marion}, {M{\'a}rka}, {M{\'a}rka}, {Markakis}, {Markosyan}, {Markowitz}, {Maros}, {Marquina}, {Marsh}, {Martelli}, {Martellini}, {Martin}, {Martin}, {Martynov}, {Marx}, {Mason}, {Massera}, {Masserot}, {Massinger}, {Masso-Reid}, {Mastrogiovanni}, {Matas}, {Matichard}, {Matone}, {Mavalvala}, {Mazumder}, {McCarthy}, {McClelland}, {McCormick}, {McCuller}, {McGuire}, {McIntyre}, {McIver}, {McManus}, {McNeill}, {McRae}, {McWilliams}, {Meacher}, {Meadors},
  {Mehmet}, {Meidam}, {Mejuto-Villa}, {Melatos}, {Mendell}, {Mercer}, {Merilh}, {Merzougui}, {Meshkov}, {Messenger}, {Messick}, {Metzdorff}, {Meyers}, {Miao}, {Michel}, {Middleton}, {Mikhailov}, {Milano}, {Miller}, {Miller}, {Miller}, {Millhouse}, {Milovich-Goff}, {Minazzoli}, {Minenkov}, {Ming}, {Mishra}, {Mitra}, {Mitrofanov}, {Mitselmakher}, {Mittleman}, {Moffa}, {Moggi}, {Mogushi}, {Mohan}, {Mohapatra}, {Molina}, {Montani}, {Moore}, {Moraru}, {Moreno}, {Morisaki}, {Morriss}, {Mours}, {Mow-Lowry}, {Mueller}, {Muir}, {Mukherjee}, {Mukherjee}, {Mukherjee}, {Mukund}, {Mullavey}, {Munch}, {Mu{\~n}iz}, {Muratore}, {Murray}, {Nagar}, {Napier}, {Nardecchia}, {Naticchioni}, {Nayak}, {Neilson}, {Nelemans}, {Nelson}, {Nery}, {Neunzert}, {Nevin}, {Newport}, {Newton}, {Ng}, {Nguyen}, {Nguyen}, {Nichols}, {Nielsen}, {Nissanke}, {Nitz}, {Noack}, {Nocera}, {Nolting}, {North}, {Nuttall}, {Oberling}, {O'Dea}, {Ogin}, {Oh}, {Oh}, {Ohme}, {Okada}, {Oliver}, {Oppermann}, {Oram}, {O'Reilly}, {Ormiston}, {Ortega},
  {O'Shaughnessy}, {Ossokine}, {Ottaway}, {Overmier}, {Owen}, {Pace}, {Page}, {Page}, {Pai}, {Pai}, {Palamos}, {Palashov}, {Palomba}, {Pal-Singh}, {Pan}, {Pan}, {Pang}, {Pang}, {Pankow}, {Pannarale}, {Pant}, {Paoletti}, {Paoli}, {Papa}, {Parida}, {Parker}, {Pascucci}, {Pasqualetti}, {Passaquieti}, {Passuello}, {Patil}, {Patricelli}, {Pearlstone}, {Pedraza}, {Pedurand}, {Pekowsky}, {Pele}, {Penn}, {Perez}, {Perreca}, {Perri}, {Pfeiffer}, {Phelps}, {Piccinni}, {Pichot}, {Piergiovanni}, {Pierro}, {Pillant}, {Pinard}, {Pinto}, {Pirello}, {Pitkin}, {Poe}, {Poggiani}, {Popolizio}, {Porter}, {Post}, {Powell}, {Prasad}, {Pratt}, {Pratten}, {Predoi}, {Prestegard}, {Prijatelj}, {Principe}, {Privitera}, {Prix}, {Prodi}, {Prokhorov}, {Puncken}, {Punturo}, {Puppo}, {P{\"u}rrer}, {Qi}, {Quetschke}, {Quintero}, {Quitzow-James}, {Raab}, {Rabeling}, {Radkins}, {Raffai}, {Raja}, {Rajan}, {Rajbhandari}, {Rakhmanov}, {Ramirez}, {Ramos-Buades}, {Rapagnani}, {Raymond}, {Razzano}, {Read}, {Regimbau}, {Rei}, {Reid}, {Reitze}, {Ren},
  {Reyes}, {Ricci}, {Ricker}, {Rieger}, {Riles}, {Rizzo}, {Robertson}, {Robie}, {Robinet}, {Rocchi}, {Rolland}, {Rollins}, {Roma}, {Romano}, {Romano}, {Romel}, {Romie}, {Rosi{\'n}ska}, {Ross}, {Rowan}, {R{\"u}diger}, {Ruggi}, {Rutins}, {Ryan}, {Sachdev}, {Sadecki}, {Sadeghian}, {Sakellariadou}, {Salconi}, {Saleem}, {Salemi}, {Samajdar}, {Sammut}, {Sampson}, {Sanchez}, {Sanchez}, {Sanchis-Gual}, {Sandberg}, {Sanders}, {Sassolas}, {Sathyaprakash}, {Saulson}, {Sauter}, {Savage}, {Sawadsky}, {Schale}, {Scheel}, {Scheuer}, {Schmidt}, {Schmidt}, {Schnabel}, {Schofield}, {Sch{\"o}nbeck}, {Schreiber}, {Schuette}, {Schulte}, {Schutz}, {Schwalbe}, {Scott}, {Scott}, {Seidel}, {Sellers}, {Sengupta}, {Sentenac}, {Sequino}, {Sergeev}, {Shaddock}, {Shaffer}, {Shah}, {Shahriar}, {Shaner}, {Shao}, {Shapiro}, {Shawhan}, {Sheperd}, {Shoemaker}, {Shoemaker}, {Siellez}, {Siemens}, {Sieniawska}, {Sigg}, {Silva}, {Singer}, {Singh}, {Singhal}, {Sintes}, {Slagmolen}, {Smith}, {Smith}, {Smith}, {Somala}, {Son}, {Sonnenberg}, {Sorazu},
  {Sorrentino}, {Souradeep}, {Spencer}, {Srivastava}, {Staats}, {Staley}, {Steinke}, {Steinlechner}, {Steinlechner}, {Steinmeyer}, {Stevenson}, {Stone}, {Stops}, {Strain}, {Stratta}, {Strigin}, {Strunk}, {Sturani}, {Stuver}, {Summerscales}, {Sun}, {Sunil}, {Suresh}, {Sutton}, {Swinkels}, {Szczepa{\'n}czyk}, {Tacca}, {Tait}, {Talbot}, {Talukder}, {Tanner}, {T{\'a}pai}, {Taracchini}, {Tasson}, {Taylor}, {Taylor}, {Tewari}, {Theeg}, {Thies}, {Thomas}, {Thomas}, {Thomas}, {Thorne}, {Thorne}, {Thrane}, {Tiwari}, {Tiwari}, {Tokmakov}, {Toland}, {Tonelli}, {Tornasi}, {Torres-Forn{\'e}}, {Torrie}, {T{\"o}yr{\"a}}, {Travasso}, {Traylor}, {Trinastic}, {Tringali}, {Trozzo}, {Tsang}, {Tse}, {Tso}, {Tsukada}, {Tsuna}, {Tuyenbayev}, {Ueno}, {Ugolini}, {Unnikrishnan}, {Urban}, {Usman}, {Vahlbruch}, {Vajente}, {Valdes}, {Vallisneri}, {van Bakel}, {van Beuzekom}, {van den Brand}, {Van Den Broeck}, {Vander-Hyde}, {van der Schaaf}, {van Heijningen}, {van Veggel}, {Vardaro}, {Varma}, {Vass}, {Vas{\'u}th}, {Vecchio}, {Vedovato},
  {Veitch}, {Veitch}, {Venkateswara}, {Venugopalan}, {Verkindt}, {Vetrano}, {Vicer{\'e}}, {Viets}, {Vinciguerra}, {Vine}, {Vinet}, {Vitale}, {Vo}, {Vocca}, {Vorvick}, {Vyatchanin}, {Wade}, {Wade}, {Wade}, {Walet}, {Walker}, {Wallace}, {Walsh}, {Wang}, {Wang}, {Wang}, {Wang}, {Wang}, {Ward}, {Warner}, {Was}, {Watchi}, {Weaver}, {Wei}, {Weinert}, {Weinstein}, {Weiss}, {Wen}, {Wessel}, {We{\ss}els}, {Westerweck}, {Westphal}, {Wette}, {Whelan}, {Whitcomb}, {Whiting}, {Whittle}, {Wilken}, {Williams}, {Williams}, {Williamson}, {Willis}, {Willke}, {Wimmer}, {Winkler}, {Wipf}, {Wittel}, {Woan}, {Woehler}, {Wofford}, {Wong}, {Worden}, {Wright}, {Wu}, {Wysocki}, {Xiao}, {Yamamoto}, {Yancey}, {Yang}, {Yap}, {Yazback}, {Yu}, {Yu}, {Yvert}, {Zadro{\.Z}ny}, {Zanolin}, {Zelenova}, {Zendri}, {Zevin}, {Zhang}, {Zhang}, {Zhang}, {Zhang}, {Zhao}, {Zhou}, {Zhou}, {Zhu}, {Zhu}, {Zimmerman}, {Zucker}, {Zweizig}, {LIGO Scientific Collaboration}, \& {Virgo Collaboration}}]{GW170817}
---. 2017, \prl, 119, 161101

\bibitem[{{Abbott} {et~al.}(2020){Abbott}, {Abbott}, {Abraham}, {Acernese}, {Ackley}, {Adams}, {Adhikari}, {Adya}, {Affeldt}, {Agathos}, {Agatsuma}, {Aggarwal}, {Aguiar}, {Aich}, {Aiello}, {Ain}, {Ajith}, {Akcay}, {Allen}, {Allocca}, {Altin}, {Amato}, {Anand}, {Ananyeva}, {Anderson}, {Anderson}, {Angelova}, {Ansoldi}, {Antier}, {LIGO Scientific Collaboration}, \& {Virgo Collaboration}}]{GW190521}
{Abbott}, R., {Abbott}, T.~D., {et~al.} 2020, \prl, 125, 101102

\bibitem[{{Abbott} {et~al.}(2023){Abbott}, {Abbott}, {Acernese}, {Ackley}, {Adams}, {Adhikari}, {Adhikari}, {Adya}, {Affeldt}, {Agarwal}, {Agathos}, {Agatsuma}, {Aggarwal}, {Aguiar}, {Aiello}, {Ain}, {Ajith}, {Akutsu}, {de Alarc{\'o}n}, {Akcay}, {Albanesi}, {Allocca}, {Altin}, {Amato}, {Anand}, {Anand}, {Ananyeva}, {Anderson}, {Anderson}, {Ando}, {Andrade}, {Andres}, {Andri{\'c}}, {Angelova}, {Ansoldi}, {Antelis}, {Antier}, {Antonini}, {Appert}, {Arai}, {Arai}, {Arai}, {Araki}, {Araya}, {Araya}, {Areeda}, {Ar{\`e}ne}, {Aritomi}, {Arnaud}, {Arogeti}, {Aronson}, {Arun}, {Asada}, {Asali}, {Ashton}, {Aso}, {Assiduo}, {Aston}, {Astone}, {Aubin}, {Austin}, {Babak}, {Badaracco}, {Bader}, {Badger}, {Bae}, {Bae}, {Baer}, {Bagnasco}, {Bai}, {Baiotti}, {Baird}, {Bajpai}, {Ball}, {Ballardin}, {Ballmer}, {Balsamo}, {Baltus}, {Banagiri}, {Bankar}, {Barayoga}, {Barbieri}, {Barish}, {Barker}, {Barneo}, {Barone}, {Barr}, {Barsotti}, {Barsuglia}, {Barta}, {Bartlett}, {Barton}, {Bartos}, {Bassiri}, {Basti}, {Bawaj}, {Bayley},
  {Baylor}, {Bazzan}, {B{\'e}csy}, {Bedakihale}, {Bejger}, {Belahcene}, {Benedetto}, {Beniwal}, {Bennett}, {Bentley}, {Benyaala}, {Bergamin}, {Berger}, {Bernuzzi}, {Berry}, {Bersanetti}, {Bertolini}, {Betzwieser}, {Beveridge}, {Bhandare}, {Bhardwaj}, {Bhattacharjee}, {Bhaumik}, {Bilenko}, {Billingsley}, {Bini}, {Birney}, {Birnholtz}, {Biscans}, {Bischi}, {Biscoveanu}, {Bisht}, {Biswas}, {Bitossi}, {Bizouard}, {Blackburn}, {Blair}, {Blair}, {Blair}, {Bobba}, {Bode}, {Boer}, {Bogaert}, {Boldrini}, {Bonavena}, {Bondu}, {Bonilla}, {Bonnand}, {Booker}, {Boom}, {Bork}, {Boschi}, {Bose}, {Bose}, {Bossilkov}, {Boudart}, {Bouffanais}, {Bozzi}, {Bradaschia}, {Brady}, {Bramley}, {Branch}, {Branchesi}, {Brandt}, {Brau}, {Breschi}, {Briant}, {Briggs}, {Brillet}, {Brinkmann}, {Brockill}, {Brooks}, {Brooks}, {Brown}, {Brunett}, {Bruno}, {Bruntz}, {Bryant}, {Bulik}, {Bulten}, {Buonanno}, {Buscicchio}, {Buskulic}, {Buy}, {Byer}, {Cadonati}, {Cagnoli}, {Cahillane}, {Bustillo}, {Callaghan}, {Callister}, {Calloni}, {Cameron},
  {Camp}, {Canepa}, {Canevarolo}, {Cannavacciuolo}, {Cannon}, {Cao}, {Cao}, {Capocasa}, {Capote}, {Carapella}, {Carbognani}, {Carlin}, {Carney}, {Carpinelli}, {Carrillo}, {Carullo}, {Carver}, {Diaz}, {Casentini}, {Castaldi}, {Caudill}, {Cavagli{\`a}}, {Cavalier}, {Cavalieri}, {Ceasar}, {Cella}, {Cerd{\'a}-Dur{\'a}n}, {Cesarini}, {Chaibi}, {Chakravarti}, {Subrahmanya}, {Champion}, {Chan}, {Chan}, {Chan}, {Chan}, {Chan}, {Chandra}, {Chanial}, {Chao}, {Chapman-Bird}, {Charlton}, {Chase}, {Chassande-Mottin}, {Chatterjee}, {Chatterjee}, {Chatterjee}, {Chaturvedi}, {Chaty}, {Chatziioannou}, {Chen}, {Chen}, {Chen}, {Chen}, {Chen}, {Chen}, {Chen}, {Chen}, {Cheng}, {Cheong}, {Cheung}, {Chia}, {Chiadini}, {Chiang}, {Chiarini}, {Chierici}, {Chincarini}, {Chiofalo}, {Chiummo}, {Cho}, {Cho}, {Choudhary}, {Choudhary}, {Christensen}, {Chu}, {Chu}, {Chu}, {Chua}, {Chung}, {Ciani}, {Ciecielag}, {Cie{\'s}lar}, {Cifaldi}, {Ciobanu}, {Ciolfi}, {Cipriano}, {Cirone}, {Clara}, {Clark}, {Clark}, {Clarke}, {Clearwater}, {Clesse},
  {Cleva}, {Coccia}, {Codazzo}, {Cohadon}, {Cohen}, {Cohen}, {Colleoni}, {Collette}, {Colombo}, {Colpi}, {Compton}, {Constancio}, {Conti}, {Cooper}, {Corban}, {Corbitt}, {Cordero-Carri{\'o}n}, {Corezzi}, {Corley}, {Cornish}, {Corre}, {Corsi}, {Cortese}, {Costa}, {Cotesta}, {Coughlin}, {Coulon}, {Countryman}, {Cousins}, {Couvares}, {Coward}, {Cowart}, {Coyne}, {Coyne}, {Creighton}, {Creighton}, {Criswell}, {Croquette}, {Crowder}, {Cudell}, {Cullen}, {Cumming}, {Cummings}, {Cunningham}, {Cuoco}, {Cury{\l}o}, {Dabadie}, {Canton}, {Dall'Osso}, {D{\'a}lya}, {Dana}, {Daneshgaranbajastani}, {D'Angelo}, {Danila}, {Danilishin}, {D'Antonio}, {Danzmann}, {Darsow-Fromm}, {Dasgupta}, {Datrier}, {Datta}, {Dattilo}, {Dave}, {Davier}, {Davies}, {Davis}, {Davis}, {Daw}, {Dean}, {Debra}, {Deenadayalan}, {Degallaix}, {de Laurentis}, {Del{\'e}glise}, {Del Favero}, {de Lillo}, {de Lillo}, {Del Pozzo}, {Demarchi}, {de Matteis}, {D'Emilio}, {Demos}, {Dent}, {Depasse}, {de Pietri}, {De Rosa}, {de Rossi}, {Desalvo}, {de Simone},
  {Dhurandhar}, {D{\'\i}az}, {Diaz-Ortiz}, {Didio}, {Dietrich}, {di Fiore}, {di Fronzo}, {di Giorgio}, {di Giovanni}, {di Giovanni}, {di Girolamo}, {di Lieto}, {Ding}, {di Pace}, {di Palma}, {di Renzo}, {Divakarla}, {Dmitriev}, {Doctor}, {D'Onofrio}, {Donovan}, {Dooley}, {Doravari}, {Dorrington}, {Drago}, {Driggers}, {Drori}, {Ducoin}, {Dupej}, {Durante}, {D'Urso}, {Duverne}, {Dwyer}, {Eassa}, {Easter}, {Ebersold}, {Eckhardt}, {Eddolls}, {Edelman}, {Edo}, {Edy}, {Effler}, {Eguchi}, {Eichholz}, {Eikenberry}, {Eisenmann}, {Eisenstein}, {Ejlli}, {Engelby}, {Enomoto}, {Errico}, {Essick}, {Estell{\'e}s}, {Estevez}, {Etienne}, {Etzel}, {Evans}, {Evans}, {Ewing}, {Fafone}, {Fair}, {Fairhurst}, {Farah}, {Farinon}, {Farr}, {Farr}, {Farrow}, {Fauchon-Jones}, {Favaro}, {Favata}, {Fays}, {Fazio}, {Feicht}, {Fejer}, {Fenyvesi}, {Ferguson}, {Fernandez-Galiana}, {Ferrante}, {Ferreira}, {Fidecaro}, {Figura}, {Fiori}, {Fishbach}, {Fisher}, {Fittipaldi}, {Fiumara}, {Flaminio}, {Floden}, {Fong}, {Font}, {Fornal}, {Forsyth},
  {Franke}, {Frasca}, {Frasconi}, {Frederick}, {Freed}, {Frei}, {Freise}, {Frey}, {Fritschel}, {Frolov}, {Fronz{\'e}}, {Fujii}, {Fujikawa}, {Fukunaga}, {Fukushima}, {Fulda}, {Fyffe}, {Gabbard}, {Gadre}, {Gair}, {Gais}, {Galaudage}, {Gamba}, {Ganapathy}, {Ganguly}, {Gao}, {Gaonkar}, {Garaventa}, {Garc{\'\i}a}, {Garc{\'\i}a-N{\'u}{\~n}ez}, {Garc{\'\i}a-Quir{\'o}s}, {Garufi}, {Gateley}, {Gaudio}, {Gayathri}, {Ge}, {Gemme}, {Gennai}, {George}, {George}, {Gerberding}, {Gergely}, {Gewecke}, {Ghonge}, {Ghosh}, {Ghosh}, {Ghosh}, {Ghosh}, {Giacomazzo}, {Giacoppo}, {Giaime}, {Giardina}, {Gibson}, {Gier}, {Giesler}, {Giri}, {Gissi}, {Glanzer}, {Gleckl}, {Godwin}, {Golomb}, {Goetz}, {Goetz}, {Gohlke}, {Goncharov}, {Gonz{\'a}lez}, {Gopakumar}, {Gosselin}, {Gouaty}, {Gould}, {Grace}, {Grado}, {Granata}, {Granata}, {Grant}, {Gras}, {Grassia}, {Gray}, {Gray}, {Greco}, {Green}, {Green}, {Gretarsson}, {Gretarsson}, {Griffith}, {Griffiths}, {Griggs}, {Grignani}, {Grimaldi}, {Grimm}, {Grote}, {Grunewald}, {Gruning}, {Guerra},
  {Guidi}, {Guimaraes}, {Guix{\'e}}, {Gulati}, {Guo}, {Guo}, {Gupta}, {Gupta}, {Gupta}, {Gustafson}, {Gustafson}, {Guzman}, {Ha}, {Haegel}, {Hagiwara}, {Haino}, {Halim}, {Hall}, {Hamilton}, {Hammond}, {Han}, {Haney}, {Hanks}, {Hanna}, {Hannam}, {Hannuksela}, {Hansen}, {Hansen}, {Hanson}, {Harder}, {Hardwick}, {Haris}, {Harms}, {Harry}, {Harry}, {Hartwig}, {Hasegawa}, {Haskell}, {Hasskew}, {Haster}, {Hattori}, {Haughian}, {Hayakawa}, {Hayama}, {Hayes}, {Healy}, {Heidmann}, {Heidt}, {Heintze}, {Heinze}, {Heinzel}, {Heitmann}, {Hellman}, {Hello}, {Helmling-Cornell}, {Hemming}, {Hendry}, {Heng}, {Hennes}, {Hennig}, {Hennig}, {Hernandez}, {Vivanco}, {Heurs}, {Hild}, {Hill}, {Himemoto}, {Hines}, {Hiranuma}, {Hirata}, {Hirose}, {Hochheim}, {Hofman}, {Hohmann}, {Holcomb}, {Holland}, {Hollows}, {Holmes}, {Holt}, {Holz}, {Hong}, {Hopkins}, {Hough}, {Hourihane}, {Howell}, {Hoy}, {Hoyland}, {Hreibi}, {Hsieh}, {Hsu}, {Huang}, {Huang}, {Huang}, {Huang}, {Huang}, {Huang}, {H{\"u}bner}, {Huddart}, {Hughey}, {Hui}, {Hui},
  {Husa}, {Huttner}, {Huxford}, {Huynh-Dinh}, {Ide}, {Idzkowski}, {Iess}, {Ikenoue}, {Imam}, {Inayoshi}, {Ingram}, {Inoue}, {Ioka}, {Isi}, {Isleif}, {Ito}, {Itoh}, {Iyer}, {Izumi}, {Jaberianhamedan}, {Jacqmin}, {Jadhav}, {Jadhav}, {James}, {Jan}, {Jani}, {Janquart}, {Janssens}, {Janthalur}, {Jaranowski}, {Jariwala}, {Jaume}, {Jenkins}, {Jenner}, {Jeon}, {Jeunon}, {Jia}, {Jin}, {Johns}, {Jones}, {Jones}, {Jones}, {Jones}, {Jones}, {Jonker}, {Ju}, {Jung}, {Jung}, {Junker}, {Juste}, {Kaihotsu}, {Kajita}, {Kakizaki}, {Kalaghatgi}, {Kalogera}, {Kamai}, {Kamiizumi}, {Kanda}, {Kandhasamy}, {Kang}, {Kanner}, {Kao}, {Kapadia}, {Kapasi}, {Karat}, {Karathanasis}, {Karki}, {Kashyap}, {Kasprzack}, {Kastaun}, {Katsanevas}, {Katsavounidis}, {Katzman}, {Kaur}, {Kawabe}, {Kawaguchi}, {Kawai}, {Kawasaki}, {K{\'e}f{\'e}lian}, {Keitel}, {Key}, {Khadka}, {Khalili}, {Khan}, {Khazanov}, {Khetan}, {Khursheed}, {Kijbunchoo}, {Kim}, {Kim}, {Kim}, {Kim}, {Kim}, {Kim}, {Kimball}, {Kimura}, {Kinley-Hanlon}, {Kirchhoff}, {Kissel}, {Kita},
  {Kitazawa}, {Kleybolte}, {Klimenko}, {Knee}, {Knowles}, {Knyazev}, {Koch}, {Koekoek}, {Kojima}, {Kokeyama}, {Koley}, {Kolitsidou}, {Kolstein}, {Komori}, {Kondrashov}, {Kong}, {Kontos}, {Koper}, {Korobko}, {Kotake}, {Kovalam}, {Kozak}, {Kozakai}, {Kozu}, {Kringel}, {Krishnendu}, {Kr{\'o}lak}, {Kuehn}, {Kuei}, {Kuijer}, {Kulkarni}, {Kumar}, {Kumar}, {Kumar}, {Kumar}, {Kume}, {Kuns}, {Kuo}, {Kuo}, {Kuromiya}, {Kuroyanagi}, {Kusayanagi}, {Kuwahara}, {Kwak}, {Lagabbe}, {Laghi}, {Lalande}, {Lam}, {Lamberts}, {Landry}, {Landry}, {Lane}, {Lang}, {Lange}, {Lantz}, {La Rosa}, {Lartaux-Vollard}, {Lasky}, {Laxen}, {Lazzarini}, {Lazzaro}, {Leaci}, {Leavey}, {Lecoeuche}, {Lee}, {Lee}, {Lee}, {Lee}, {Lee}, {Lee}, {Lehmann}, {Lema{\^\i}tre}, {Leonardi}, {Leroy}, {Letendre}, {Levesque}, {Levin}, {Leviton}, {Leyde}, {Li}, {Li}, {Li}, {Li}, {Li}, {Li}, {Lin}, {Lin}, {Lin}, {Lin}, {Lin}, {Linde}, {Linker}, {Linley}, {Littenberg}, {Liu}, {Liu}, {Liu}, {Liu}, {Llamas}, {Llorens-Monteagudo}, {Lo}, {Lockwood}, {Loh}, {London},
  {Longo}, {Lopez}, {Portilla}, {Lorenzini}, {Loriette}, {Lormand}, {Losurdo}, {Lott}, {Lough}, {Lousto}, {Lovelace}, {Lucaccioni}, {L{\"u}ck}, {Lumaca}, {Lundgren}, {Luo}, {Lynam}, {Macas}, {Macinnis}, {MacLeod}, {MacMillan}, {Macquet}, {Hernandez}, {Magazz{\`u}}, {Magee}, {Maggiore}, {Magnozzi}, {Mahesh}, {Majorana}, {Makarem}, {Maksimovic}, {Maliakal}, {Malik}, {Man}, {Mandic}, {Mangano}, {Mango}, {Mansell}, {Manske}, {Mantovani}, {Mapelli}, {Marchesoni}, {Marchio}, {Marion}, {Mark}, {M{\'a}rka}, {M{\'a}rka}, {Markakis}, {Markosyan}, {Markowitz}, {Maros}, {Marquina}, {Marsat}, {Martelli}, {Martin}, {Martin}, {Martinez}, {Martinez}, {Martinez}, {Martinovic}, {Martynov}, {Marx}, {Masalehdan}, {Mason}, {Massera}, {Masserot}, {Massinger}, {Masso-Reid}, {Mastrogiovanni}, {Matas}, {Mateu-Lucena}, {Matichard}, {Matiushechkina}, {Mavalvala}, {McCann}, {McCarthy}, {McClelland}, {McClincy}, {McCormick}, {McCuller}, {McGhee}, {McGuire}, {McIsaac}, {McIver}, {McRae}, {McWilliams}, {Meacher}, {Mehmet}, {Mehta},
  {Meijer}, {Melatos}, {Melchor}, {Mendell}, {Menendez-Vazquez}, {Menoni}, {Mercer}, {Mereni}, {Merfeld}, {Merilh}, {Merritt}, {Merzougui}, {Meshkov}, {Messenger}, {Messick}, {Meyers}, {Meylahn}, {Mhaske}, {Miani}, {Miao}, {Michaloliakos}, {Michel}, {Michimura}, {Middleton}, {Milano}, {Miller}, {Miller}, {Miller}, {Miller}, {Millhouse}, {Mills}, {Milotti}, {Minazzoli}, {Minenkov}, {Mio}, {Mir}, {Miravet-Ten{\'e}s}, {Mishra}, {Mishra}, {Mistry}, {Mitra}, {Mitrofanov}, {Mitselmakher}, {Mittleman}, {Miyakawa}, {Miyamoto}, {Miyazaki}, {Miyo}, {Miyoki}, {Mo}, {Modafferi}, {Moguel}, {Mogushi}, {Mohapatra}, {Mohite}, {Molina}, {Molina-Ruiz}, {Mondin}, {Montani}, {Moore}, {Moraru}, {Morawski}, {More}, {Moreno}, {Moreno}, {Mori}, {Morisaki}, {Moriwaki}, {Morr{\'a}s}, {Mours}, {Mow-Lowry}, {Mozzon}, {Muciaccia}, {Mukherjee}, {Mukherjee}, {Mukherjee}, {Mukherjee}, {Mukherjee}, {Mukund}, {Mullavey}, {Munch}, {Mu{\~n}iz}, {Murray}, {Musenich}, {Muusse}, {Nadji}, {Nagano}, {Nagano}, {Nagar}, {Nakamura}, {Nakano}, {Nakano},
  {Nakashima}, {Nakayama}, {Napolano}, {Nardecchia}, {Narikawa}, {Naticchioni}, {Nayak}, {Nayak}, {Negishi}, {Neil}, {Neilson}, {Nelemans}, {Nelson}, {Nery}, {Neubauer}, {Neunzert}, {Ng}, {Ng}, {Nguyen}, {Nguyen}, {Nguyen}, {Quynh}, {Ni}, {Nichols}, {Nishizawa}, {Nissanke}, {Nitoglia}, {Nocera}, {Norman}, {North}, {Nozaki}, {Siles}, {Nuttall}, {Oberling}, {O'Brien}, {Obuchi}, {O'Dell}, {Oelker}, {Ogaki}, {Oganesyan}, {Oh}, {Oh}, {Oh}, {Ohashi}, {Ohishi}, {Ohkawa}, {Ohme}, {Ohta}, {Okada}, {Okutani}, {Okutomi}, {Olivetto}, {Oohara}, {Ooi}, {Oram}, {O'Reilly}, {Ormiston}, {Ormsby}, {Ortega}, {O'Shaughnessy}, {O'Shea}, {Oshino}, {Ossokine}, {Osthelder}, {Otabe}, {Ottaway}, {Overmier}, {Pace}, {Pagano}, {Page}, {Pagliaroli}, {Pai}, {Pai}, {Palamos}, {Palashov}, {Palomba}, {Pan}, {Pan}, {Panda}, {Pang}, {Pang}, {Pankow}, {Pannarale}, {Pant}, {Panther}, {Paoletti}, {Paoli}, {Paolone}, {Parisi}, {Park}, {Park}, {Parker}, {Pascucci}, {Pasqualetti}, {Passaquieti}, {Passuello}, {Patel}, {Pathak}, {Patricelli},
  {Patron}, {Paul}, {Payne}, {Pedraza}, {Pegoraro}, {Pele}, {Arellano}, {Penn}, {Perego}, {Pereira}, {Pereira}, {Perez}, {P{\'e}rigois}, {Perkins}, {Perreca}, {Perri{\`e}s}, {Petermann}, {Petterson}, {Pfeiffer}, {Pham}, {Phukon}, {Piccinni}, {Pichot}, {Piendibene}, {Piergiovanni}, {Pierini}, {Pierro}, {Pillant}, {Pillas}, {Pilo}, {Pinard}, {Pinto}, {Pinto}, {Piotrzkowski}, {Piotrzkowski}, {Pirello}, {Pitkin}, {Placidi}, {Planas}, {Plastino}, {Pluchar}, {Poggiani}, {Polini}, {Pong}, {Ponrathnam}, {Popolizio}, {Porter}, {Poulton}, {Powell}, {Pracchia}, {Pradier}, {Prajapati}, {Prasai}, {Prasanna}, {Pratten}, {Principe}, {Prodi}, {Prokhorov}, {Prosposito}, {Prudenzi}, {Puecher}, {Punturo}, {Puosi}, {Puppo}, {P{\"u}rrer}, {Qi}, {Quetschke}, {Quitzow-James}, {Raab}, {Raaijmakers}, {Radkins}, {Radulesco}, {Raffai}, {Rail}, {Raja}, {Rajan}, {Ramirez}, {Ramirez}, {Ramos-Buades}, {Rana}, {Rapagnani}, {Rapol}, {Ray}, {Raymond}, {Raza}, {Razzano}, {Read}, {Rees}, {Regimbau}, {Rei}, {Reid}, {Reid}, {Reitze}, {Relton},
  {Renzini}, {Rettegno}, {Reza}, {Rezac}, {Ricci}, {Richards}, {Richardson}, {Richardson}, {Riemenschneider}, {Riles}, {Rinaldi}, {Rink}, {Rizzo}, {Robertson}, {Robie}, {Robinet}, {Rocchi}, {Rodriguez}, {Rolland}, {Rollins}, {Romanelli}, {Romano}, {Romel}, {Romero-Rodr{\'\i}guez}, {Romero-Shaw}, {Romie}, {Ronchini}, {Rosa}, {Rose}, {Rosi{\'n}ska}, {Ross}, {Rowan}, {Rowlinson}, {Roy}, {Roy}, {Roy}, {Rozza}, {Ruggi}, {Ryan}, {Sachdev}, {Sadecki}, {Sadiq}, {Sago}, {Saito}, {Saito}, {Sakai}, {Sakai}, {Sakellariadou}, {Sakuno}, {Salafia}, {Salconi}, {Saleem}, {Salemi}, {Samajdar}, {Sanchez}, {Sanchez}, {Sanchez}, {Sanchis-Gual}, {Sanders}, {Sanuy}, {Saravanan}, {Sarin}, {Sassolas}, {Satari}, {Sathyaprakash}, {Sato}, {Sato}, {Sauter}, {Savage}, {Sawada}, {Sawant}, {Sawant}, {Sayah}, {Schaetzl}, {Scheel}, {Scheuer}, {Schiworski}, {Schmidt}, {Schmidt}, {Schnabel}, {Schneewind}, {Schofield}, {Sch{\"o}nbeck}, {Schulte}, {Schutz}, {Schwartz}, {Scott}, {Scott}, {Seglar-Arroyo}, {Sekiguchi}, {Sekiguchi}, {Sellers},
  {Sengupta}, {Sentenac}, {Seo}, {Sequino}, {Sergeev}, {Setyawati}, {Shaffer}, {Shahriar}, {Shams}, {Shao}, {Sharma}, {Sharma}, {Shawhan}, {Shcheblanov}, {Shibagaki}, {Shikauchi}, {Shimizu}, {Shimoda}, {Shimode}, {Shinkai}, {Shishido}, {Shoda}, {Shoemaker}, {Shoemaker}, {Shyamsundar}, {Sieniawska}, {Sigg}, {Singer}, {Singh}, {Singh}, {Singha}, {Sintes}, {Sipala}, {Skliris}, {Slagmolen}, {Slaven-Blair}, {Smetana}, {Smith}, {Smith}, {Soldateschi}, {Somala}, {Somiya}, {Son}, {Soni}, {Soni}, {Sordini}, {Sorrentino}, {Sorrentino}, {Sotani}, {Soulard}, {Souradeep}, {Sowell}, {Spagnuolo}, {Spencer}, {Spera}, {Srinivasan}, {Srivastava}, {Srivastava}, {Staats}, {Stachie}, {Steer}, {Steinhoff}, {Steinlechner}, {Steinlechner}, {Stevenson}, {Stops}, {Stover}, {Strain}, {Strang}, {Stratta}, {Strunk}, {Sturani}, {Stuver}, {Sudhagar}, {Sudhir}, {Sugimoto}, {Suh}, {Sullivan}, {Summerscales}, {Sun}, {Sun}, {Sunil}, {Sur}, {Suresh}, {Sutton}, {Suzuki}, {Suzuki}, {Swinkels}, {Szczepa{\'n}czyk}, {Szewczyk}, {Tacca}, {Tagoshi},
  {Tait}, {Takahashi}, {Takahashi}, {Takamori}, {Takano}, {Takeda}, {Takeda}, {Talbot}, {Talbot}, {Tanaka}, {Tanaka}, {Tanaka}, {Tanaka}, {Tanaka}, {Tanasijczuk}, {Tanioka}, {Tanner}, {Tao}, {Tao}, {Mart{\'\i}n}, {Taranto}, {Tasson}, {Telada}, {Tenorio}, {Terhune}, {Terkowski}, {Thirugnanasambandam}, {Thomas}, {Thomas}, {Thomas}, {Thompson}, {Thondapu}, {Thorne}, {Thrane}, {Tiwari}, {Tiwari}, {Tiwari}, {Toivonen}, {Toland}, {Tolley}, {Tomaru}, {Tomigami}, {Tomura}, {Tonelli}, {Torres-Forn{\'e}}, {Torrie}, {E Melo}, {T{\"o}yr{\"a}}, {Trapananti}, {Travasso}, {Traylor}, {Trevor}, {Tringali}, {Tripathee}, {Troiano}, {Trovato}, {Trozzo}, {Trudeau}, {Tsai}, {Tsai}, {Tsang}, {Tsang}, {Tsao}, {Tse}, {Tso}, {Tsubono}, {Tsuchida}, {Tsukada}, {Tsuna}, {Tsutsui}, {Tsuzuki}, {Turbang}, {Turconi}, {Tuyenbayev}, {Ubhi}, {Uchikata}, {Uchiyama}, {Udall}, {Ueda}, {Uehara}, {Ueno}, {Ueshima}, {Unnikrishnan}, {Uraguchi}, {Urban}, {Ushiba}, {Utina}, {Vahlbruch}, {Vajente}, {Vajpeyi}, {Valdes}, {Valentini}, {Valsan}, {van Bakel},
  {van Beuzekom}, {van den Brand}, {van den Broeck}, {Vander-Hyde}, {van der Schaaf}, {van Heijningen}, {Vanosky}, {van Putten}, {van Remortel}, {Vardaro}, {Vargas}, {Varma}, {Vas{\'u}th}, {Vecchio}, {Vedovato}, {Veitch}, {Veitch}, {Venneberg}, {Venugopalan}, {Verkindt}, {Verma}, {Verma}, {Veske}, {Vetrano}, {Vicer{\'e}}, {Vidyant}, {Viets}, {Vijaykumar}, {Villa-Ortega}, {Vinet}, {Virtuoso}, {Vitale}, {Vo}, {Vocca}, {von Reis}, {von Wrangel}, {Vorvick}, {Vyatchanin}, {Wade}, {Wade}, {Wagner}, {Walet}, {Walker}, {Wallace}, {Wallace}, {Walsh}, {Wang}, {Wang}, {Wang}, {Ward}, {Warner}, {Was}, {Washimi}, {Washington}, {Watchi}, {Weaver}, {Webster}, {Weinert}, {Weinstein}, {Weiss}, {Weller}, {Wellmann}, {Wen}, {We{\ss}els}, {Wette}, {Whelan}, {White}, {Whiting}, {Whittle}, {Wilken}, {Williams}, {Williams}, {Williamson}, {Willis}, {Willke}, {Wilson}, {Winkler}, {Wipf}, {Wlodarczyk}, {Woan}, {Woehler}, {Wofford}, {Wong}, {Wu}, {Wu}, {Wu}, {Wu}, {Wysocki}, {Xiao}, {Xu}, {Yamada}, {Yamamoto}, {Yamamoto}, {Yamamoto},
  {Yamamoto}, {Yamashita}, {Yamazaki}, {Yang}, {Yang}, {Yang}, {Yang}, {Yang}, {Yap}, {Yeeles}, {Yelikar}, {Ying}, {Yokogawa}, {Yokoyama}, {Yokozawa}, {Yoo}, {Yoshioka}, {Yu}, {Yu}, {Yuzurihara}, {Zadro{\.z}ny}, {Zanolin}, {Zeidler}, {Zelenova}, {Zendri}, {Zevin}, {Zhan}, {Zhang}, {Zhang}, {Zhang}, {Zhang}, {Zhang}, {Zhao}, {Zhao}, {Zhao}, {Zhao}, {Zheng}, {Zhou}, {Zhou}, {Zhu}, {Zhu}, {Zimmerman}, {Zlochower}, {Zucker}, {Zweizig}, {LIGO Scientific Collaboration}, {VIRGO Collaboration}, \& {KAGRA Collaboration}}]{LIGO2023}
---. 2023, Physical Review X, 13, 011048

\bibitem[{{Agazie} {et~al.}(2023){Agazie}, {Anumarlapudi}, {Archibald}, {Arzoumanian}, {Baker}, {B{\'e}csy}, {Blecha}, {Brazier}, {Brook}, {Burke-Spolaor}, {Burnette}, {Case}, {Charisi}, {Chatterjee}, {Chatziioannou}, {Cheeseboro}, {Chen}, {Cohen}, {Cordes}, {Cornish}, {Crawford}, {Cromartie}, {Crowter}, {Cutler}, {Decesar}, {Degan}, {Demorest}, {Deng}, {Dolch}, {Drachler}, {Ellis}, {Ferrara}, {Fiore}, {Fonseca}, {Freedman}, {Garver-Daniels}, {Gentile}, {Gersbach}, {Glaser}, {Good}, {G{\"u}ltekin}, {Hazboun}, {Hourihane}, {Islo}, {Jennings}, {Johnson}, {Jones}, {Kaiser}, {Kaplan}, {Kelley}, {Kerr}, {Key}, {Klein}, {Laal}, {Lam}, {Lamb}, {Lazio}, {Lewandowska}, {Littenberg}, {Liu}, {Lommen}, {Lorimer}, {Luo}, {Lynch}, {Ma}, {Madison}, {Mattson}, {McEwen}, {McKee}, {McLaughlin}, {McMann}, {Meyers}, {Meyers}, {Mingarelli}, {Mitridate}, {Natarajan}, {Ng}, {Nice}, {Ocker}, {Olum}, {Pennucci}, {Perera}, {Petrov}, {Pol}, {Radovan}, {Ransom}, {Ray}, {Romano}, {Sardesai}, {Schmiedekamp}, {Schmiedekamp}, {Schmitz},
  {Schult}, {Shapiro-Albert}, {Siemens}, {Simon}, {Siwek}, {Stairs}, {Stinebring}, {Stovall}, {Sun}, {Susobhanan}, {Swiggum}, {Taylor}, {Taylor}, {Turner}, {Unal}, {Vallisneri}, {van Haasteren}, {Vigeland}, {Wahl}, {Wang}, {Witt}, {Young}, \& {Nanograv Collaboration}}]{NANOGrav2023}
{Agazie}, G., {Anumarlapudi}, A., {et~al.} 2023, \apjl, 951, L8

\bibitem[{{Amaro-Seoane} {et~al.}(2023){Amaro-Seoane}, {Andrews}, {Arca Sedda}, {Askar}, {Baghi}, {Balasov}, {Bartos}, {Bavera}, {Bellovary}, {Berry}, {Berti}, {Bianchi}, {Blecha}, {Blondin}, {Bogdanovi{\'c}}, {Boissier}, {Bonetti}, {Bonoli}, {Bortolas}, {Breivik}, {Capelo}, {Caramete}, {Cattorini}, {Charisi}, {Chaty}, {Chen}, {Chru{\'s}li{\'n}ska}, {Chua}, {Church}, {Colpi}, {D'Orazio}, {Danielski}, {Davies}, {Dayal}, {De Rosa}, {Derdzinski}, {Destounis}, {Dotti}, {Dutan}, {Dvorkin}, {Fabj}, {Foglizzo}, {Ford}, {Fouvry}, {Franchini}, {Fragos}, {Fryer}, {Gaspari}, {Gerosa}, {Graziani}, {Groot}, {Habouzit}, {Haggard}, {Haiman}, {Han}, {Istrate}, {Johansson}, {Khan}, {Kimpson}, {Kokkotas}, {Kong}, {Korol}, {Kremer}, {Kupfer}, {Lamberts}, {Larson}, {Lau}, {Liu}, {Lloyd-Ronning}, {Lodato}, {Lupi}, {Ma}, {Maccarone}, {Mandel}, {Mangiagli}, {Mapelli}, {Mathis}, {Mayer}, {McGee}, {McKernan}, {Miller}, {Mota}, {Mumpower}, {Nasim}, {Nelemans}, {Noble}, {Pacucci}, {Panessa}, {Paschalidis}, {Pfister}, {Porquet},
  {Quenby}, {Ricarte}, {R{\"o}pke}, {Regan}, {Rosswog}, {Ruiter}, {Ruiz}, {Runnoe}, {Schneider}, {Schnittman}, {Secunda}, {Sesana}, {Seto}, {Shao}, {Shapiro}, {Sopuerta}, {Stone}, {Suvorov}, {Tamanini}, {Tamfal}, {Tauris}, {Temmink}, {Tomsick}, {Toonen}, {Torres-Orjuela}, {Toscani}, {Tsokaros}, {Unal}, {V{\'a}zquez-Aceves}, {Valiante}, {van Putten}, {van Roestel}, {Vignali}, {Volonteri}, {Wu}, {Younsi}, {Yu}, {Zane}, {Zwick}, {Antonini}, {Baibhav}, {Barausse}, {Bonilla Rivera}, {Branchesi}, {Branduardi-Raymont}, {Burdge}, {Chakraborty}, {Cuadra}, {Dage}, {Davis}, {de Mink}, {Decarli}, {Doneva}, {Escoffier}, {Gandhi}, {Haardt}, {Lousto}, {Nissanke}, {Nordhaus}, {O'Shaughnessy}, {Portegies Zwart}, {Pound}, {Schussler}, {Sergijenko}, {Spallicci}, {Vernieri}, \& {Vigna-G{\'o}mez}}]{LISA2023}
{Amaro-Seoane}, P., {Andrews}, J., {et~al.} 2023, Living Reviews in Relativity, 26, 2

\bibitem[{{Anderson} {et~al.}(1990){Anderson}, {Gorham}, {Kulkarni}, {Prince}, \& {Wolszczan}}]{Anderson1990}
{Anderson}, S.~B., {Gorham}, P.~W., {et~al.} 1990, \nat, 346, 42

\bibitem[{{Andrews} \& {Kalogera}(2022)}]{AndrewsKalogera2022}
{Andrews}, J.~J., \& {Kalogera}, V. 2022, \apj, 930, 159

\bibitem[{{Antonini} \& {Perets}(2012)}]{AntoniniPerets2012}
{Antonini}, F., \& {Perets}, H.~B. 2012, \apj, 757, 27

\bibitem[{{Antonini} \& {Rasio}(2016{\natexlab{a}})}]{Antonini2016}
{Antonini}, F., \& {Rasio}, F.~A. 2016{\natexlab{a}}, \apj, 831, 187

\bibitem[{{Antonini} \& {Rasio}(2016{\natexlab{b}})}]{AntoniniRasio2016}
---. 2016{\natexlab{b}}, The Astrophysical Journal, 831, 187

\bibitem[{{Antonov}(1962)}]{Antonov1962}
{Antonov}, V.~A. 1962, {Solution of the problem of stability of stellar system Emden's density law and the spherical distribution of velocities}

\bibitem[{{Arca Sedda} {et~al.}(2018){Arca Sedda}, {Askar}, \& {Giersz}}]{ArcaSedda2018}
{Arca Sedda}, M., {Askar}, A., \& {Giersz}, M. 2018, \mnras, 479, 4652

\bibitem[{{Atallah} {et~al.}(2024){Atallah}, {Weatherford}, {Trani}, \& {Rasio}}]{Atallah2024}
{Atallah}, D., {Weatherford}, N.~C., {et~al.} 2024, arXiv e-prints, arXiv:2402.12429

\bibitem[{{Bacon} {et~al.}(2014){Bacon}, {Vernet}, {Borisova}, {Bouch{\'e}}, {Brinchmann}, {Carollo}, {Carton}, {Caruana}, {Cerda}, {Contini}, {Franx}, {Girard}, {Guerou}, {Haddad}, {Hau}, {Herenz}, {Herrera}, {Husemann}, {Husser}, {Jarno}, {Kamann}, {Krajnovic}, {Lilly}, {Mainieri}, {Martinsson}, {Palsa}, {Patricio}, {P{\'e}contal}, {Pello}, {Piqueras}, {Richard}, {Sandin}, {Schroetter}, {Selman}, {Shirazi}, {Smette}, {Soto}, {Streicher}, {Urrutia}, {Weilbacher}, {Wisotzki}, \& {Zins}}]{Bacon2014}
{Bacon}, R., {Vernet}, J., {et~al.} 2014, The Messenger, 157, 13

\bibitem[{{Bagchi} {et~al.}(2011){Bagchi}, {Lorimer}, \& {Chennamangalam}}]{Bagchi2011}
{Bagchi}, M., {Lorimer}, D.~R., \& {Chennamangalam}, J. 2011, \mnras, 418, 477

\bibitem[{{Bahramian} {et~al.}(2013){Bahramian}, {Heinke}, {Sivakoff}, \& {Gladstone}}]{Bahramian2013}
{Bahramian}, A., {Heinke}, C.~O., {et~al.} 2013, \apj, 766, 136

\bibitem[{{Bahramian} {et~al.}(2017){Bahramian}, {Heinke}, {Tudor}, {Miller-Jones}, {Bogdanov}, {Maccarone}, {Knigge}, {Sivakoff}, {Chomiuk}, {Strader}, {Garcia}, \& {Kallman}}]{Bahramian2017}
---. 2017, \mnras, 467, 2199

\bibitem[{{Baibhav} {et~al.}(2019){Baibhav}, {Berti}, {Gerosa}, {Mapelli}, {Giacobbo}, {Bouffanais}, \& {Di Carlo}}]{Baibhav2019}
{Baibhav}, V., {Berti}, E., {et~al.} 2019, \prd, 100, 064060

\bibitem[{{Bailyn}(1995)}]{Bailyn1995}
{Bailyn}, C.~D. 1995, \araa, 33, 133

\bibitem[{{Balakrishnan} {et~al.}(2023){Balakrishnan}, {Freire}, {Ransom}, {Ridolfi}, {Barr}, {Chen}, {Krishnan}, {Champion}, {Kramer}, {Gautam}, {Padmanabh}, {Men}, {Abbate}, {Stappers}, {Stairs}, {Keane}, \& {Possenti}}]{Balakrishnan2023}
{Balakrishnan}, V., {Freire}, P. C.~C., {et~al.} 2023, \apjl, 942, L35

\bibitem[{{Balbinot} {et~al.}(2024){Balbinot}, {Dodd}, {Matsuno}, {Lardo}, {Helmi}, {Panuzzo}, {Mazeh}, {Holl}, {Caffau}, {Jorissen}, {Babusiaux}, {Gavras}, {Wyrzykowski}, {Eyer}, {Leclerc}, {Bombrun}, {Mowlavi}, {Seabroke}, {Cabrera-Ziri}, {Callingham}, {Ruiz-Lara}, \& {Starkenburg}}]{Balbinot2024}
{Balbinot}, E., {Dodd}, E., {et~al.} 2024, \aap, 687, L3

\bibitem[{{Banerjee}(2017)}]{Banerjee2017}
{Banerjee}, S. 2017, \mnras, 467, 524

\bibitem[{{Barr} {et~al.}(2024){Barr}, {Dutta}, {Freire}, {Cadelano}, {Gautam}, {Kramer}, {Pallanca}, {Ransom}, {Ridolfi}, {Stappers}, {Tauris}, {Venkatraman Krishnan}, {Wex}, {Bailes}, {Behrend}, {Buchner}, {Burgay}, {Chen}, {Champion}, {Chen}, {Corongiu}, {Geyer}, {Men}, {Padmanabh}, \& {Possenti}}]{Barr2024}
{Barr}, E.~D., {Dutta}, A., {et~al.} 2024, Science, 383, 275

\bibitem[{Belczynski {et~al.}(2016)Belczynski, Holz, Bulik, \& O’Shaughnessy}]{Belczynski2016a}
Belczynski, K., Holz, D.~E., {et~al.} 2016, Nature, 534, 512

\bibitem[{{Binney} \& {Tremaine}(2008)}]{BinneyTremaine2008}
{Binney}, J., \& {Tremaine}, S. 2008, {Galactic Dynamics: Second Edition}

\bibitem[{{Bird} {et~al.}(2016){Bird}, {Cholis}, {Mu{\~n}oz}, {Ali-Ha{\"i}moud}, {Kamionkowski}, {Kovetz}, {Raccanelli}, \& {Riess}}]{Bird2016}
{Bird}, S., {Cholis}, I., {et~al.} 2016, Physical Review Letters, 116, 201301

\bibitem[{{Bochenek} {et~al.}(2020){Bochenek}, {Ravi}, {Belov}, {Hallinan}, {Kocz}, {Kulkarni}, \& {McKenna}}]{Bochenek2020}
{Bochenek}, C.~D., {Ravi}, V., {et~al.} 2020, \nat, 587, 59

\bibitem[{{Boyles} {et~al.}(2011){Boyles}, {Lorimer}, {Turk}, {Mnatsakanov}, {Lynch}, {Ransom}, {Freire}, \& {Belczynski}}]{Boyles2011}
{Boyles}, J., {Lorimer}, D.~R., {et~al.} 2011, \apj, 742, 51

\bibitem[{{Breen} \& {Heggie}(2013)}]{BreenHeggie2013}
{Breen}, P.~G., \& {Heggie}, D.~C. 2013, \mnras, 432, 2779

\bibitem[{{Bregman} {et~al.}(2024){Bregman}, {Gnedin}, {Seitzer}, \& {Qu}}]{Bregman2024}
{Bregman}, J.~N., {Gnedin}, O.~Y., {et~al.} 2024, \apjl, 968, L6

\bibitem[{{Breivik} {et~al.}(2016){Breivik}, {Rodriguez}, {Larson}, {Kalogera}, \& {Rasio}}]{Breivik2016}
{Breivik}, K., {Rodriguez}, C.~L., {et~al.} 2016, \apjl, 830, L18

\bibitem[{{Brodie} \& {Strader}(2006)}]{BrodieStrader2006}
{Brodie}, J.~P., \& {Strader}, J. 2006, \araa, 44, 193

\bibitem[{{Chomiuk} {et~al.}(2013){Chomiuk}, {Strader}, {Maccarone}, {Miller-Jones}, {Heinke}, {Noyola}, {Seth}, \& {Ransom}}]{Chomiuk2013}
{Chomiuk}, L., {Strader}, J., {et~al.} 2013, \apj, 777, 69

\bibitem[{Clark(1975)}]{Clark1975}
Clark, G. 1975, \apj, 199, L143

\bibitem[{{Corral-Santana} {et~al.}(2016){Corral-Santana}, {Casares}, {Mu{\~n}oz-Darias}, {Bauer}, {Mart{\'\i}nez-Pais}, \& {Russell}}]{Corral-Santana2016}
{Corral-Santana}, J.~M., {Casares}, J., {et~al.} 2016, \aap, 587, A61

\bibitem[{{Damour}(2001)}]{Damour2001}
{Damour}, T. 2001, \prd, 64, 124013

\bibitem[{{Di Carlo} {et~al.}(2024){Di Carlo}, {Agrawal}, {Rodriguez}, \& {Breivik}}]{DiCarlo2024}
{Di Carlo}, U.~N., {Agrawal}, P., {et~al.} 2024, \apj, 965, 22

\bibitem[{{Di Carlo} {et~al.}(2019){Di Carlo}, {Giacobbo}, {Mapelli}, {Pasquato}, {Spera}, {Wang}, \& {Haardt}}]{DiCarlo2019}
{Di Carlo}, U.~N., {Giacobbo}, N., {et~al.} 2019, arXiv e-prints, arXiv:1901.00863

\bibitem[{{D'Orazio} \& {Samsing}(2018)}]{D'OrazioSamsing2018}
{D'Orazio}, D.~J., \& {Samsing}, J. 2018, \mnras, 481, 4775

\bibitem[{{El-Badry} {et~al.}(2023{\natexlab{a}}){El-Badry}, {Rix}, {Cendes}, {Rodriguez}, {Conroy}, {Quataert}, {Hawkins}, {Zari}, {Hobson}, {Breivik}, {Rau}, {Berger}, {Shahaf}, {Seeburger}, {Burdge}, {Latham}, {Buchhave}, {Bieryla}, {Bashi}, {Mazeh}, \& {Faigler}}]{El-Badry2023b}
{El-Badry}, K., {Rix}, H.-W., {et~al.} 2023{\natexlab{a}}, \mnras, 521, 4323

\bibitem[{{El-Badry} {et~al.}(2023{\natexlab{b}}){El-Badry}, {Rix}, {Quataert}, {Howard}, {Isaacson}, {Fuller}, {Hawkins}, {Breivik}, {Wong}, {Rodriguez}, {Conroy}, {Shahaf}, {Mazeh}, {Arenou}, {Burdge}, {Bashi}, {Faigler}, {Weisz}, {Seeburger}, {Almada Monter}, \& {Wojno}}]{El-Badry2023a}
---. 2023{\natexlab{b}}, \mnras, 518, 1057

\bibitem[{{Fabian} {et~al.}(1975){Fabian}, {Pringle}, \& {Rees}}]{Fabian1975}
{Fabian}, A.~C., {Pringle}, J.~E., \& {Rees}, M.~J. 1975, \mnras, 172, 15p

\bibitem[{{Farmer} {et~al.}(2019){Farmer}, {Renzo}, {de Mink}, {Marchant}, \& {Justham}}]{Farmer2019}
{Farmer}, R., {Renzo}, M., {et~al.} 2019, \apj, 887, 53

\bibitem[{{Favata} {et~al.}(2004){Favata}, {Hughes}, \& {Holz}}]{Favata2004}
{Favata}, M., {Hughes}, S.~A., \& {Holz}, D.~E. 2004, \apjl, 607, L5

\bibitem[{{Feldmeier} {et~al.}(2013){Feldmeier}, {L{\"u}tzgendorf}, {Neumayer}, {Kissler-Patig}, {Gebhardt}, {Baumgardt}, {Noyola}, {de Zeeuw}, \& {Jalali}}]{Feldmeier2013}
{Feldmeier}, A., {L{\"u}tzgendorf}, N., {et~al.} 2013, \aap, 554, A63

\bibitem[{{Fellhauer} {et~al.}(2003){Fellhauer}, {Lin}, {Bolte}, {Aarseth}, \& {Williams}}]{Fellhauer2003}
{Fellhauer}, M., {Lin}, D.~N.~C., {et~al.} 2003, \apjl, 595, L53

\bibitem[{{Ferrarese} \& {Merritt}(2000)}]{FerrareseMerritt2000}
{Ferrarese}, L., \& {Merritt}, D. 2000, \apjl, 539, L9

\bibitem[{{Fishbach} {et~al.}(2022){Fishbach}, {Kimball}, \& {Kalogera}}]{Fishbach2022}
{Fishbach}, M., {Kimball}, C., \& {Kalogera}, V. 2022, \apjl, 935, L26

\bibitem[{{Fowler} \& {Hoyle}(1964)}]{Fowler1964}
{Fowler}, W.~A., \& {Hoyle}, F. 1964, \apjs, 9, 201

\bibitem[{{Freire} {et~al.}(2004){Freire}, {Gupta}, {Ransom}, \& {Ishwara-Chandra}}]{Freire2004}
{Freire}, P.~C., {Gupta}, Y., {et~al.} 2004, \apjl, 606, L53

\bibitem[{{Fryer} {et~al.}(2012){Fryer}, {Belczynski}, {Wiktorowicz}, {Dominik}, {Kalogera}, \& {Holz}}]{Fryer2012}
{Fryer}, C.~L., {Belczynski}, K., {et~al.} 2012, \apj, 749, 91

\bibitem[{{Gaia Collaboration} {et~al.}(2023){Gaia Collaboration}, {Vallenari}, {Brown}, {Prusti}, {de Bruijne}, {Arenou}, {Babusiaux}, {Biermann}, {Creevey}, {Ducourant}, {Evans}, {Eyer}, {Guerra}, {Hutton}, {Jordi}, {Klioner}, {Lammers}, {Lindegren}, {Luri}, {Mignard}, {Panem}, {Pourbaix}, {Randich}, {Sartoretti}, {Soubiran}, {Tanga}, {Walton}, {Bailer-Jones}, {Bastian}, {Drimmel}, {Jansen}, {Katz}, {Lattanzi}, {van Leeuwen}, {Bakker}, {Cacciari}, {Casta{\~n}eda}, {De Angeli}, {Fabricius}, {Fouesneau}, {Fr{\'e}mat}, {Galluccio}, {Guerrier}, {Heiter}, {Masana}, {Messineo}, {Mowlavi}, {Nicolas}, {Nienartowicz}, {Pailler}, {Panuzzo}, {Riclet}, {Roux}, {Seabroke}, {Sordo}, {Th{\'e}venin}, {Gracia-Abril}, {Portell}, {Teyssier}, {Altmann}, {Andrae}, {Audard}, {Bellas-Velidis}, {Benson}, {Berthier}, {Blomme}, {Burgess}, {Busonero}, {Busso}, {C{\'a}novas}, {Carry}, {Cellino}, {Cheek}, {Clementini}, {Damerdji}, {Davidson}, {de Teodoro}, {Nu{\~n}ez Campos}, {Delchambre}, {Dell'Oro}, {Esquej},
  {Fern{\'a}ndez-Hern{\'a}ndez}, {Fraile}, {Garabato}, {Garc{\'\i}a-Lario}, {Gosset}, {Haigron}, {Halbwachs}, {Hambly}, {Harrison}, {Hern{\'a}ndez}, {Hestroffer}, {Hodgkin}, {Holl}, {Jan{\ss}en}, {Jevardat de Fombelle}, {Jordan}, {Krone-Martins}, {Lanzafame}, {L{\"o}ffler}, {Marchal}, {Marrese}, {Moitinho}, {Muinonen}, {Osborne}, {Pancino}, {Pauwels}, {Recio-Blanco}, {Reyl{\'e}}, {Riello}, {Rimoldini}, {Roegiers}, {Rybizki}, {Sarro}, {Siopis}, {Smith}, {Sozzetti}, {Utrilla}, {van Leeuwen}, {Abbas}, {{\'A}brah{\'a}m}, {Abreu Aramburu}, {Aerts}, {Aguado}, {Ajaj}, {Aldea-Montero}, {Altavilla}, {{\'A}lvarez}, {Alves}, {Anders}, {Anderson}, {Anglada Varela}, {Antoja}, {Baines}, {Baker}, {Balaguer-N{\'u}{\~n}ez}, {Balbinot}, {Balog}, {Barache}, {Barbato}, {Barros}, {Barstow}, {Bartolom{\'e}}, {Bassilana}, {Bauchet}, {Becciani}, {Bellazzini}, {Berihuete}, {Bernet}, {Bertone}, {Bianchi}, {Binnenfeld}, {Blanco-Cuaresma}, {Blazere}, {Boch}, {Bombrun}, {Bossini}, {Bouquillon}, {Bragaglia}, {Bramante}, {Breedt},
  {Bressan}, {Brouillet}, {Brugaletta}, {Bucciarelli}, {Burlacu}, {Butkevich}, {Buzzi}, {Caffau}, {Cancelliere}, {Cantat-Gaudin}, {Carballo}, {Carlucci}, {Carnerero}, {Carrasco}, {Casamiquela}, {Castellani}, {Castro-Ginard}, {Chaoul}, {Charlot}, {Chemin}, {Chiaramida}, {Chiavassa}, {Chornay}, {Comoretto}, {Contursi}, {Cooper}, {Cornez}, {Cowell}, {Crifo}, {Cropper}, {Crosta}, {Crowley}, {Dafonte}, {Dapergolas}, {David}, {David}, {de Laverny}, {De Luise}, {De March}, {De Ridder}, {de Souza}, {de Torres}, {del Peloso}, {del Pozo}, {Delbo}, {Delgado}, {Delisle}, {Demouchy}, {Dharmawardena}, {Di Matteo}, {Diakite}, {Diener}, {Distefano}, {Dolding}, {Edvardsson}, {Enke}, {Fabre}, {Fabrizio}, {Faigler}, {Fedorets}, {Fernique}, {Fienga}, {Figueras}, {Fournier}, {Fouron}, {Fragkoudi}, {Gai}, {Garcia-Gutierrez}, {Garcia-Reinaldos}, {Garc{\'\i}a-Torres}, {Garofalo}, {Gavel}, {Gavras}, {Gerlach}, {Geyer}, {Giacobbe}, {Gilmore}, {Girona}, {Giuffrida}, {Gomel}, {Gomez}, {Gonz{\'a}lez-N{\'u}{\~n}ez},
  {Gonz{\'a}lez-Santamar{\'\i}a}, {Gonz{\'a}lez-Vidal}, {Granvik}, {Guillout}, {Guiraud}, {Guti{\'e}rrez-S{\'a}nchez}, {Guy}, {Hatzidimitriou}, {Hauser}, {Haywood}, {Helmer}, {Helmi}, {Sarmiento}, {Hidalgo}, {Hilger}, {H{\l}adczuk}, {Hobbs}, {Holland}, {Huckle}, {Jardine}, {Jasniewicz}, {Jean-Antoine Piccolo}, {Jim{\'e}nez-Arranz}, {Jorissen}, {Juaristi Campillo}, {Julbe}, {Karbevska}, {Kervella}, {Khanna}, {Kontizas}, {Kordopatis}, {Korn}, {K{\'o}sp{\'a}l}, {Kostrzewa-Rutkowska}, {Kruszy{\'n}ska}, {Kun}, {Laizeau}, {Lambert}, {Lanza}, {Lasne}, {Le Campion}, {Lebreton}, {Lebzelter}, {Leccia}, {Leclerc}, {Lecoeur-Taibi}, {Liao}, {Licata}, {Lindstr{\o}m}, {Lister}, {Livanou}, {Lobel}, {Lorca}, {Loup}, {Madrero Pardo}, {Magdaleno Romeo}, {Managau}, {Mann}, {Manteiga}, {Marchant}, {Marconi}, {Marcos}, {Marcos Santos}, {Mar{\'\i}n Pina}, {Marinoni}, {Marocco}, {Marshall}, {Martin Polo}, {Mart{\'\i}n-Fleitas}, {Marton}, {Mary}, {Masip}, {Massari}, {Mastrobuono-Battisti}, {Mazeh}, {McMillan}, {Messina}, {Michalik},
  {Millar}, {Mints}, {Molina}, {Molinaro}, {Moln{\'a}r}, {Monari}, {Mongui{\'o}}, {Montegriffo}, {Montero}, {Mor}, {Mora}, {Morbidelli}, {Morel}, {Morris}, {Muraveva}, {Murphy}, {Musella}, {Nagy}, {Noval}, {Oca{\~n}a}, {Ogden}, {Ordenovic}, {Osinde}, {Pagani}, {Pagano}, {Palaversa}, {Palicio}, {Pallas-Quintela}, {Panahi}, {Payne-Wardenaar}, {Pe{\~n}alosa Esteller}, {Penttil{\"a}}, {Pichon}, {Piersimoni}, {Pineau}, {Plachy}, {Plum}, {Poggio}, {Pr{\v{s}}a}, {Pulone}, {Racero}, {Ragaini}, {Rainer}, {Raiteri}, {Rambaux}, {Ramos}, {Ramos-Lerate}, {Re Fiorentin}, {Regibo}, {Richards}, {Rios Diaz}, {Ripepi}, {Riva}, {Rix}, {Rixon}, {Robichon}, {Robin}, {Robin}, {Roelens}, {Rogues}, {Rohrbasser}, {Romero-G{\'o}mez}, {Rowell}, {Royer}, {Ruz Mieres}, {Rybicki}, {Sadowski}, {S{\'a}ez N{\'u}{\~n}ez}, {Sagrist{\`a} Sell{\'e}s}, {Sahlmann}, {Salguero}, {Samaras}, {Sanchez Gimenez}, {Sanna}, {Santove{\~n}a}, {Sarasso}, {Schultheis}, {Sciacca}, {Segol}, {Segovia}, {S{\'e}gransan}, {Semeux}, {Shahaf}, {Siddiqui}, {Siebert},
  {Siltala}, {Silvelo}, {Slezak}, {Slezak}, {Smart}, {Snaith}, {Solano}, {Solitro}, {Souami}, {Souchay}, {Spagna}, {Spina}, {Spoto}, {Steele}, {Steidelm{\"u}ller}, {Stephenson}, {S{\"u}veges}, {Surdej}, {Szabados}, {Szegedi-Elek}, {Taris}, {Taylor}, {Teixeira}, {Tolomei}, {Tonello}, {Torra}, {Torra}, {Torralba Elipe}, {Trabucchi}, {Tsounis}, {Turon}, {Ulla}, {Unger}, {Vaillant}, {van Dillen}, {van Reeven}, {Vanel}, {Vecchiato}, {Viala}, {Vicente}, {Voutsinas}, {Weiler}, {Wevers}, {Wyrzykowski}, {Yoldas}, {Yvard}, {Zhao}, {Zorec}, {Zucker}, \& {Zwitter}}]{Gaia2023}
{Gaia Collaboration}, {Vallenari}, A., {et~al.} 2023, \aap, 674, A1

\bibitem[{{Gaia Collaboration} {et~al.}(2024){Gaia Collaboration}, {Panuzzo}, {Mazeh}, {Arenou}, {Holl}, {Caffau}, {Jorissen}, {Babusiaux}, {Gavras}, {Sahlmann}, {Bastian}, {Wyrzykowski}, {Eyer}, {Leclerc}, {Bauchet}, {Bombrun}, {Mowlavi}, {Seabroke}, {Teyssier}, {Balbinot}, {Helmi}, {Brown}, {Vallenari}, {Prusti}, {de Bruijne}, {Barbier}, {Biermann}, {Creevey}, {Ducourant}, {Evans}, {Guerra}, {Hutton}, {Jordi}, {Klioner}, {Lammers}, {Lindegren}, {Luri}, {Mignard}, {Nicolas}, {Randich}, {Sartoretti}, {Smiljanic}, {Tanga}, {Walton}, {Aerts}, {Bailer-Jones}, {Cropper}, {Drimmel}, {Jansen}, {Katz}, {Lattanzi}, {Soubiran}, {Th{\'e}venin}, {van Leeuwen}, {Andrae}, {Audard}, {Bakker}, {Blomme}, {Casta{\~n}eda}, {De Angeli}, {Fabricius}, {Fouesneau}, {Fr{\'e}mat}, {Galluccio}, {Guerrier}, {Heiter}, {Masana}, {Messineo}, {Nienartowicz}, {Pailler}, {Riclet}, {Roux}, {Sordo}, {Gracia-Abril}, {Portell}, {Altmann}, {Benson}, {Berthier}, {Burgess}, {Busonero}, {Busso}, {Cacciari}, {C{\'a}novas}, {Carrasco}, {Carry},
  {Cellino}, {Cheek}, {Clementini}, {Damerdji}, {Davidson}, {de Teodoro}, {Delchambre}, {Dell'Oro}, {Fraile Garcia}, {Garabato}, {Garc{\'\i}a-Lario}, {Haigron}, {Hambly}, {Harrison}, {Hatzidimitriou}, {Hern{\'a}ndez}, {Hestroffer}, {Hodgkin}, {Jamal}, {Jevardat de Fombelle}, {Jordan}, {Krone-Martins}, {Lanzafame}, {L{\"o}ffler}, {Lorca}, {Marchal}, {Marrese}, {Moitinho}, {Muinonen}, {Nu{\~n}ez Campos}, {Oreshina-Slezak}, {Osborne}, {Pancino}, {Pauwels}, {Recio-Blanco}, {Riello}, {Rimoldini}, {Robin}, {Roegiers}, {Sarro}, {Schultheis}, {Smith}, {Sozzetti}, {Utrilla}, {van Leeuwen}, {Weingrill}, {Abbas}, {{\'A}brah{\'a}m}, {Abreu Aramburu}, {Ahmed}, {Altavilla}, {{\'A}lvarez}, {Anders}, {Anderson}, {Anglada Varela}, {Antoja}, {Baig}, {Baines}, {Baker}, {Balaguer-N{\'u}{\~n}ez}, {Balog}, {Barache}, {Barros}, {Barstow}, {Bartolom{\'e}}, {Bashi}, {Bassilana}, {Baudeau}, {Becciani}, {Bedin}, {Bellas-Velidis}, {Bellazzini}, {Beordo}, {Bernet}, {Bertolotto}, {Bertone}, {Bianchi}, {Binnenfeld}, {Blanco-Cuaresma},
  {Bland-Hawthorn}, {Blazere}, {Boch}, {Bossini}, {Bouquillon}, {Bragaglia}, {Braine}, {Bratsolis}, {Breedt}, {Bressan}, {Brouillet}, {Brugaletta}, {Bucciarelli}, {Butkevich}, {Buzzi}, {Camut}, {Cancelliere}, {Cantat-Gaudin}, {Capilla Guilarte}, {Carballo}, {Carlucci}, {Carnerero}, {Carretero}, {Carton}, {Casamiquela}, {Casey}, {Castellani}, {Castro-Ginard}, {Ceraj}, {Cesare}, {Charlot}, {Chaudet}, {Chemin}, {Chiavassa}, {Chornay}, {Chosson}, {Cooper}, {Cornez}, {Cowell}, {Crosta}, {Crowley}, {Cruz Reyes}, {Dafonte}, {Dal Ponte}, {David}, {de Laverny}, {De Luise}, {De March}, {de Torres}, {del Peloso}, {Delbo}, {Delgado}, {Delisle}, {Demouchy}, {Denis}, {Dharmawardena}, {Di Giacomo}, {Diener}, {Distefano}, {Dolding}, {Dsilva}, {Enke}, {Fabre}, {Fabrizio}, {Faigler}, {Fatovi{\'c}}, {Fedorets}, {Fern{\'a}ndez-Hern{\'a}ndez}, {Fernique}, {Figueras}, {Fouron}, {Fragkoudi}, {Gai}, {Galinier}, {Garcia-Serrano}, {Garc{\'\i}a-Torres}, {Garofalo}, {Gerlach}, {Geyer}, {Giacobbe}, {Gilmore}, {Girona}, {Giuffrida},
  {Gomboc}, {Gomez}, {Gonz{\'a}lez-Santamar{\'\i}a}, {Gosset}, {Granvik}, {Gregori Barrera}, {Guti{\'e}rrez-S{\'a}nchez}, {Haywood}, {Helmer}, {Hidalgo}, {Hilger}, {Hobbs}, {Hottier}, {Huckle}, {Jim{\'e}nez-Arranz}, {Juaristi Campillo}, {Kaczmarek}, {Kervella}, {Khanna}, {Kontizas}, {Kordopatis}, {Korn}, {K{\'o}sp{\'a}l}, {Kostrzewa-Rutkowska}, {Kruszy{\'n}ska}, {Kun}, {Lambert}, {Lanza}, {Lebreton}, {Lebzelter}, {Leccia}, {Lecoutre}, {Liao}, {Liberato}, {Licata}, {Livanou}, {Lobel}, {L{\'o}pez-Miralles}, {Loup}, {Madar{\'a}sz}, {Mahy}, {Mann}, {Manteiga}, {Marcellino}, {Marchant}, {Marconi}, {Mar{\'\i}n Pina}, {Marinoni}, {Marshall}, {Mart{\'\i}n Lozano}, {Martin Polo}, {Mart{\'\i}n-Fleitas}, {Marton}, {Mascarenhas}, {Masip}, {Mastrobuono-Battisti}, {McMillan}, {Meichsner}, {Merc}, {Messina}, {Millar}, {Mints}, {Mohamed}, {Molina}, {Molinaro}, {Moln{\'a}r}, {Mongui{\'o}}, {Montegriffo}, {Monti}, {Mora}, {Morbidelli}, {Morris}, {Mudimadugula}, {Muraveva}, {Musella}, {Nagy}, {Nardetto}, {Navarrete}, {Oh},
  {Ordenovic}, {Orenstein}, {Pagani}, {Pagano}, {Palaversa}, {Palicio}, {Pallas-Quintela}, {Pawlak}, {Penttil{\"a}}, {Pesciullesi}, {Pinamonti}, {Plachy}, {Planquart}, {Plum}, {Poggio}, {Pourbaix}, {Price-Whelan}, {Pulone}, {Rabin}, {Rainer}, {Raiteri}, {Ramos}, {Ramos-Lerate}, {Ratajczak}, {Re Fiorentin}, {Regibo}, {Reyl{\'e}}, {Ripepi}, {Riva}, {Rix}, {Rixon}, {Robert}, {Robichon}, {Robin}, {Romero-G{\'o}mez}, {Rowell}, {Ruz Mieres}, {Rybicki}, {Sadowski}, {Sagrist{\`a} Sell{\'e}s}, {Sanna}, {Santove{\~n}a}, {Sarasso}, {Sarmiento}, {Sarrate Riera}, {Sciacca}, {S{\'e}gransan}, {Semczuk}, {Shahaf}, {Siebert}, {Slezak}, {Smart}, {Snaith}, {Solano}, {Solitro}, {Souami}, {Souchay}, {Spitoni}, {Spoto}, {Squillante}, {Steele}, {Steidelm{\"u}ller}, {Surdej}, {Szabados}, {Taris}, {Taylor}, {Teixeira}, {Tepper-Garcia}, {Thuillot}, {Tolomei}, {Tonello}, {Torra}, {Torralba Elipe}, {Trabucchi}, {Trentin}, {Tsantaki}, {Turon}, {Ulla}, {Unger}, {Valtchanov}, {Vanel}, {Vecchiato}, {Vicente}, {Villar}, {Weiler}, {Zhao},
  {Zorec}, {Zucker}, {{\v{Z}}upi{\'c}}, \& {Zwitter}}]{GaiaBH3_2024}
{Gaia Collaboration}, {Panuzzo}, P., {et~al.} 2024, \aap, 686, L2

\bibitem[{{Gerosa} \& {Berti}(2017)}]{GerosaBerti2017}
{Gerosa}, D., \& {Berti}, E. 2017, \prd, 95, 124046

\bibitem[{{Ghez} {et~al.}(2008){Ghez}, {Salim}, {Weinberg}, {Lu}, {Do}, {Dunn}, {Matthews}, {Morris}, {Yelda}, {Becklin}, {Kremenek}, {Milosavljevic}, \& {Naiman}}]{Ghez2008}
{Ghez}, A.~M., {Salim}, S., {et~al.} 2008, \apj, 689, 1044

\bibitem[{{Gieles} {et~al.}(2018){Gieles}, {Balbinot}, {Yaaqib}, {H{\'e}nault-Brunet}, {Zocchi}, {Peuten}, \& {Jonker}}]{Gieles2018}
{Gieles}, M., {Balbinot}, E., {et~al.} 2018, \mnras, 473, 4832

\bibitem[{{Gieles} {et~al.}(2021){Gieles}, {Erkal}, {Antonini}, {Balbinot}, \& {Pe{\~n}arrubia}}]{Gieles2021}
{Gieles}, M., {Erkal}, D., {et~al.} 2021, Nature Astronomy, 5, 957

\bibitem[{{Giesers} {et~al.}(2018){Giesers}, {Dreizler}, {Husser}, {Kamann}, {Anglada Escud{\'e}}, {Brinchmann}, {Carollo}, {Roth}, {Weilbacher}, \& {Wisotzki}}]{Giesers2018}
{Giesers}, B., {Dreizler}, S., {et~al.} 2018, \mnras, 475, L15

\bibitem[{{Giesers} {et~al.}(2019){Giesers}, {Kamann}, {Dreizler}, {Husser}, {Askar}, {G{\"o}ttgens}, {Brinchmann}, {Latour}, {Weilbacher}, {Wendt}, \& {Roth}}]{Giesers2019}
{Giesers}, B., {Kamann}, S., {et~al.} 2019, arXiv e-prints, arXiv:1909.04050

\bibitem[{{Gilliland} {et~al.}(2000){Gilliland}, {Brown}, {Guhathakurta}, {Sarajedini}, {Milone}, {Albrow}, {Baliber}, {Bruntt}, {Burrows}, {Charbonneau}, {Choi}, {Cochran}, {Edmonds}, {Frandsen}, {Howell}, {Lin}, {Marcy}, {Mayor}, {Naef}, {Sigurdsson}, {Stagg}, {Vandenberg}, {Vogt}, \& {Williams}}]{Gilliland2000}
{Gilliland}, R.~L., {Brown}, T.~M., {et~al.} 2000, \apjl, 545, L47

\bibitem[{{Gonz{\'a}lez} {et~al.}(2021){Gonz{\'a}lez}, {Kremer}, {Chatterjee}, {Fragione}, {Rodriguez}, {Weatherford}, {Ye}, \& {Rasio}}]{Gonzalez2021}
{Gonz{\'a}lez}, E., {Kremer}, K., {et~al.} 2021, \apjl, 908, L29

\bibitem[{{Gordon} {et~al.}(2023){Gordon}, {Fong}, {Kilpatrick}, {Eftekhari}, {Leja}, {Prochaska}, {Nugent}, {Bhandari}, {Blanchard}, {Caleb}, {Day}, {Deller}, {Dong}, {Glowacki}, {Gourdji}, {Mannings}, {Mahoney}, {Marnoch}, {Miller}, {Paterson}, {Rastinejad}, {Ryder}, {Sadler}, {Scott}, {Sears}, {Shannon}, {Simha}, {Stappers}, \& {Tejos}}]{Gordon2023}
{Gordon}, A.~C., {Fong}, W.-f., {et~al.} 2023, \apj, 954, 80

\bibitem[{{Greene} {et~al.}(2020){Greene}, {Strader}, \& {Ho}}]{Greene2020}
{Greene}, J.~E., {Strader}, J., \& {Ho}, L.~C. 2020, \araa, 58, 257

\bibitem[{{Grindlay} {et~al.}(2001){Grindlay}, {Heinke}, {Edmonds}, \& {Murray}}]{Grindlay2001}
{Grindlay}, J.~E., {Heinke}, C., {et~al.} 2001, Science, 292, 2290

\bibitem[{{H{\"a}berle} {et~al.}(2024){H{\"a}berle}, {Neumayer}, {Seth}, {Bellini}, {Libralato}, {Baumgardt}, {Whitaker}, {Dumont}, {Alfaro Cuello}, {Anderson}, {Clontz}, {Kacharov}, {Kamann}, {Feldmeier-Krause}, {Milone}, {Nitschai}, {Pechetti}, \& {van de Ven}}]{Haberle2024}
{H{\"a}berle}, M., {Neumayer}, N., {et~al.} 2024, arXiv e-prints, arXiv:2405.06015

\bibitem[{{Harris}(1996)}]{Harris1996}
{Harris}, W.~E. 1996, \aj, 112, 1487

\bibitem[{{Heggie} \& {Hut}(2003)}]{HeggieHut2003}
{Heggie}, D., \& {Hut}, P. 2003, {The Gravitational Million-Body Problem: A Multidisciplinary Approach to Star Cluster Dynamics}

\bibitem[{Heggie(1975)}]{Heggie1975}
Heggie, D.~C. 1975, Mon.~Not.~R.~Astron.~Soc, 173, 729

\bibitem[{{Heinke} {et~al.}(2003){Heinke}, {Grindlay}, {Lugger}, {Cohn}, {Edmonds}, {Lloyd}, \& {Cool}}]{Heinke2003}
{Heinke}, C.~O., {Grindlay}, J.~E., {et~al.} 2003, \apj, 598, 501

\bibitem[{{H{\'e}non}(1961)}]{Henon1961}
{H{\'e}non}, M. 1961, Annales d'Astrophysique, 24, 369

\bibitem[{{Hills}(1975)}]{Hills1975}
{Hills}, J.~G. 1975, \aj, 80, 809

\bibitem[{Hobbs {et~al.}(2005)Hobbs, Lorimer, Lyne, \& Kramer}]{Hobbs2005}
Hobbs, G., Lorimer, D.~R., {et~al.} 2005, Monthly Notices of the Royal Astronomical Society, 360, 974

\bibitem[{{Hut} {et~al.}(1992){Hut}, {McMillan}, {Goodman}, {Mateo}, {Phinney}, {Pryor}, {Richer}, {Verbunt}, \& {Weinberg}}]{Hut1992}
{Hut}, P., {McMillan}, S., {et~al.} 1992, \pasp, 104, 981

\bibitem[{{Ivezi{\'c}} {et~al.}(2019){Ivezi{\'c}}, {Kahn}, {Tyson}, {Abel}, {Acosta}, {Allsman}, {Alonso}, {AlSayyad}, {Anderson}, {Andrew}, {Angel}, {Angeli}, {Ansari}, {Antilogus}, {Araujo}, {Armstrong}, {Arndt}, {Astier}, {Aubourg}, {Auza}, {Axelrod}, {Bard}, {Barr}, {Barrau}, {Bartlett}, {Bauer}, {Bauman}, {Baumont}, {Bechtol}, {Bechtol}, {Becker}, {Becla}, {Beldica}, {Bellavia}, {Bianco}, {Biswas}, {Blanc}, {Blazek}, {Blandford}, {Bloom}, {Bogart}, {Bond}, {Booth}, {Borgland}, {Borne}, {Bosch}, {Boutigny}, {Brackett}, {Bradshaw}, {Brandt}, {Brown}, {Bullock}, {Burchat}, {Burke}, {Cagnoli}, {Calabrese}, {Callahan}, {Callen}, {Carlin}, {Carlson}, {Chandrasekharan}, {Charles-Emerson}, {Chesley}, {Cheu}, {Chiang}, {Chiang}, {Chirino}, {Chow}, {Ciardi}, {Claver}, {Cohen-Tanugi}, {Cockrum}, {Coles}, {Connolly}, {Cook}, {Cooray}, {Covey}, {Cribbs}, {Cui}, {Cutri}, {Daly}, {Daniel}, {Daruich}, {Daubard}, {Daues}, {Dawson}, {Delgado}, {Dellapenna}, {de Peyster}, {de Val-Borro}, {Digel}, {Doherty}, {Dubois},
  {Dubois-Felsmann}, {Durech}, {Economou}, {Eifler}, {Eracleous}, {Emmons}, {Fausti Neto}, {Ferguson}, {Figueroa}, {Fisher-Levine}, {Focke}, {Foss}, {Frank}, {Freemon}, {Gangler}, {Gawiser}, {Geary}, {Gee}, {Geha}, {Gessner}, {Gibson}, {Gilmore}, {Glanzman}, {Glick}, {Goldina}, {Goldstein}, {Goodenow}, {Graham}, {Gressler}, {Gris}, {Guy}, {Guyonnet}, {Haller}, {Harris}, {Hascall}, {Haupt}, {Hernandez}, {Herrmann}, {Hileman}, {Hoblitt}, {Hodgson}, {Hogan}, {Howard}, {Huang}, {Huffer}, {Ingraham}, {Innes}, {Jacoby}, {Jain}, {Jammes}, {Jee}, {Jenness}, {Jernigan}, {Jevremovi{\'c}}, {Johns}, {Johnson}, {Johnson}, {Jones}, {Juramy-Gilles}, {Juri{\'c}}, {Kalirai}, {Kallivayalil}, {Kalmbach}, {Kantor}, {Karst}, {Kasliwal}, {Kelly}, {Kessler}, {Kinnison}, {Kirkby}, {Knox}, {Kotov}, {Krabbendam}, {Krughoff}, {Kub{\'a}nek}, {Kuczewski}, {Kulkarni}, {Ku}, {Kurita}, {Lage}, {Lambert}, {Lange}, {Langton}, {Le Guillou}, {Levine}, {Liang}, {Lim}, {Lintott}, {Long}, {Lopez}, {Lotz}, {Lupton}, {Lust}, {MacArthur}, {Mahabal},
  {Mandelbaum}, {Markiewicz}, {Marsh}, {Marshall}, {Marshall}, {May}, {McKercher}, {McQueen}, {Meyers}, {Migliore}, {Miller}, {Mills}, {Miraval}, {Moeyens}, {Moolekamp}, {Monet}, {Moniez}, {Monkewitz}, {Montgomery}, {Morrison}, {Mueller}, {Muller}, {Mu{\~n}oz Arancibia}, {Neill}, {Newbry}, {Nief}, {Nomerotski}, {Nordby}, {O'Connor}, {Oliver}, {Olivier}, {Olsen}, {O'Mullane}, {Ortiz}, {Osier}, {Owen}, {Pain}, {Palecek}, {Parejko}, {Parsons}, {Pease}, {Peterson}, {Peterson}, {Petravick}, {Libby Petrick}, {Petry}, {Pierfederici}, {Pietrowicz}, {Pike}, {Pinto}, {Plante}, {Plate}, {Plutchak}, {Price}, {Prouza}, {Radeka}, {Rajagopal}, {Rasmussen}, {Regnault}, {Reil}, {Reiss}, {Reuter}, {Ridgway}, {Riot}, {Ritz}, {Robinson}, {Roby}, {Roodman}, {Rosing}, {Roucelle}, {Rumore}, {Russo}, {Saha}, {Sassolas}, {Schalk}, {Schellart}, {Schindler}, {Schmidt}, {Schneider}, {Schneider}, {Schoening}, {Schumacher}, {Schwamb}, {Sebag}, {Selvy}, {Sembroski}, {Seppala}, {Serio}, {Serrano}, {Shaw}, {Shipsey}, {Sick}, {Silvestri},
  {Slater}, {Smith}, {Smith}, {Sobhani}, {Soldahl}, {Storrie-Lombardi}, {Stover}, {Strauss}, {Street}, {Stubbs}, {Sullivan}, {Sweeney}, {Swinbank}, {Szalay}, {Takacs}, {Tether}, {Thaler}, {Thayer}, {Thomas}, {Thornton}, {Thukral}, {Tice}, {Trilling}, {Turri}, {Van Berg}, {Vanden Berk}, {Vetter}, {Virieux}, {Vucina}, {Wahl}, {Walkowicz}, {Walsh}, {Walter}, {Wang}, {Wang}, {Warner}, {Wiecha}, {Willman}, {Winters}, {Wittman}, {Wolff}, {Wood-Vasey}, {Wu}, {Xin}, {Yoachim}, \& {Zhan}}]{LSST2019}
{Ivezi{\'c}}, {\v{Z}}., {Kahn}, S.~M., {et~al.} 2019, \apj, 873, 111

\bibitem[{{Jacoby} {et~al.}(2006){Jacoby}, {Cameron}, {Jenet}, {Anderson}, {Murty}, \& {Kulkarni}}]{Jacoby2006}
{Jacoby}, B.~A., {Cameron}, P.~B., {et~al.} 2006, \apjl, 644, L113

\bibitem[{{Jonker} \& {Nelemans}(2004)}]{JonkerNelemans2004}
{Jonker}, P.~G., \& {Nelemans}, G. 2004, \mnras, 354, 355

\bibitem[{{Kalirai} {et~al.}(2012){Kalirai}, {Richer}, {Anderson}, {Dotter}, {Fahlman}, {Hansen}, {Hurley}, {King}, {Reitzel}, {Rich}, {Shara}, {Stetson}, \& {Woodley}}]{Kalirai2012}
{Kalirai}, J.~S., {Richer}, H.~B., {et~al.} 2012, \aj, 143, 11

\bibitem[{{Kamann} {et~al.}(2016){Kamann}, {Husser}, {Brinchmann}, {Emsellem}, {Weilbacher}, {Wisotzki}, {Wendt}, {Krajnovi{\'c}}, {Roth}, {Bacon}, \& {Dreizler}}]{Kamann2016}
{Kamann}, S., {Husser}, T.~O., {et~al.} 2016, \aap, 588, A149

\bibitem[{{Katz}(1975)}]{Katz1975}
{Katz}, J.~I. 1975, \nat, 253, 698

\bibitem[{{Kirsten} {et~al.}(2022){Kirsten}, {Marcote}, {Nimmo}, {Hessels}, {Bhardwaj}, {Tendulkar}, {Keimpema}, {Yang}, {Snelders}, {Scholz}, {Pearlman}, {Law}, {Peters}, {Giroletti}, {Paragi}, {Bassa}, {Hewitt}, {Bach}, {Bezrukovs}, {Burgay}, {Buttaccio}, {Conway}, {Corongiu}, {Feiler}, {Forss{\'e}n}, {Gawro{\'n}ski}, {Karuppusamy}, {Kharinov}, {Lindqvist}, {Maccaferri}, {Melnikov}, {Ould-Boukattine}, {Possenti}, {Surcis}, {Wang}, {Yuan}, {Aggarwal}, {Anna-Thomas}, {Bower}, {Blaauw}, {Burke-Spolaor}, {Cassanelli}, {Clarke}, {Fonseca}, {Gaensler}, {Gopinath}, {Kaspi}, {Kassim}, {Lazio}, {Leung}, {Li}, {Lin}, {Masui}, {Mckinven}, {Michilli}, {Mikhailov}, {Ng}, {Orbidans}, {Pen}, {Petroff}, {Rahman}, {Ransom}, {Shin}, {Smith}, {Stairs}, \& {Vlemmings}}]{Kirsten2022}
{Kirsten}, F., {Marcote}, B., {et~al.} 2022, \nat, 602, 585

\bibitem[{{Kremer} {et~al.}(2019{\natexlab{a}}){Kremer}, {Chatterjee}, {Ye}, {Rodriguez}, \& {Rasio}}]{Kremer2019a}
{Kremer}, K., {Chatterjee}, S., {et~al.} 2019{\natexlab{a}}, \apj, 871, 38

\bibitem[{{Kremer} {et~al.}(2023{\natexlab{a}}){Kremer}, {Fuller}, {Piro}, \& {Ransom}}]{Kremer2023_psr}
{Kremer}, K., {Fuller}, J., {et~al.} 2023{\natexlab{a}}, \mnras, 525, L22

\bibitem[{{Kremer} {et~al.}(2023{\natexlab{b}}){Kremer}, {Li}, {Lu}, {Piro}, \& {Zhang}}]{Kremer2023_frb}
{Kremer}, K., {Li}, D., {et~al.} 2023{\natexlab{b}}, \apj, 944, 6

\bibitem[{{Kremer} {et~al.}(2022){Kremer}, {Lombardi}, {Lu}, {Piro}, \& {Rasio}}]{Kremer2022_sph}
{Kremer}, K., {Lombardi}, J.~C., {et~al.} 2022, \apj, 933, 203

\bibitem[{{Kremer} {et~al.}(2021{\natexlab{a}}){Kremer}, {Piro}, \& {Li}}]{Kremer2021_frb}
{Kremer}, K., {Piro}, A.~L., \& {Li}, D. 2021{\natexlab{a}}, \apjl, 917, L11

\bibitem[{{Kremer} {et~al.}(2021{\natexlab{b}}){Kremer}, {Rui}, {Weatherford}, {Chatterjee}, {Fragione}, {Rasio}, {Rodriguez}, \& {Ye}}]{Kremer2021_wd}
{Kremer}, K., {Rui}, N.~Z., {et~al.} 2021{\natexlab{b}}, \apj, 917, 28

\bibitem[{{Kremer} {et~al.}(2018){Kremer}, {Ye}, {Chatterjee}, {Rodriguez}, \& {Rasio}}]{Kremer2018b}
{Kremer}, K., {Ye}, C.~S., {et~al.} 2018, \apjl, 855, L15

\bibitem[{{Kremer} {et~al.}(2020{\natexlab{a}}){Kremer}, {Ye}, {Chatterjee}, {Rodriguez}, \& {Rasio}}]{Kremer2020_bhburning}
{Kremer}, K., {Ye}, C.~S., {et~al.} 2020{\natexlab{a}}, in Star Clusters: From the Milky Way to the Early Universe, ed. A.~{Bragaglia}, M.~{Davies}, A.~{Sills}, \& E.~{Vesperini}, Vol. 351, 357--366

\bibitem[{{Kremer} {et~al.}(2019{\natexlab{b}}){Kremer}, {Rodriguez}, {Amaro-Seoane}, {Breivik}, {Chatterjee}, {Katz}, {Larson}, {Rasio}, {Samsing}, {Ye}, \& {Zevin}}]{Kremer2019b}
{Kremer}, K., {Rodriguez}, C.~L., {et~al.} 2019{\natexlab{b}}, \prd, 99, 063003

\bibitem[{{Kremer} {et~al.}(2020{\natexlab{b}}){Kremer}, {Ye}, {Rui}, {Weatherford}, {Chatterjee}, {Fragione}, {Rodriguez}, {Spera}, \& {Rasio}}]{Kremer2020_catalog}
{Kremer}, K., {Ye}, C.~S., {et~al.} 2020{\natexlab{b}}, \apjs, 247, 48

\bibitem[{{Kremer} {et~al.}(2020{\natexlab{c}}){Kremer}, {Spera}, {Becker}, {Chatterjee}, {Di Carlo}, {Fragione}, {Rodriguez}, {Ye}, \& {Rasio}}]{Kremer2020_imbh}
{Kremer}, K., {Spera}, M., {et~al.} 2020{\natexlab{c}}, \apj, 903, 45

\bibitem[{{Kroupa}(2001)}]{Kroupa2001}
{Kroupa}, P. 2001, \mnras, 322, 231

\bibitem[{{Kulkarni} {et~al.}(1993){Kulkarni}, {Hut}, \& {McMillan}}]{Kulkarni1993}
{Kulkarni}, S.~R., {Hut}, P., \& {McMillan}, S. 1993, \nat, 364, 421

\bibitem[{{Lada} \& {Lada}(2003)}]{LadaLada2003}
{Lada}, C.~J., \& {Lada}, E.~A. 2003, \araa, 41, 57

\bibitem[{{Law} {et~al.}(2024){Law}, {Sharma}, {Ravi}, {Chen}, {Catha}, {Connor}, {Faber}, {Hallinan}, {Harnach}, {Hellbourg}, {Hobbs}, {Hodge}, {Hodges}, {Lamb}, {Rasmussen}, {Sherman}, {Shi}, {Simard}, {Squillace}, {Weinreb}, {Woody}, \& {Yurk}}]{Law2024}
{Law}, C.~J., {Sharma}, K., {et~al.} 2024, \apj, 967, 29

\bibitem[{{Lu} {et~al.}(2022){Lu}, {Beniamini}, \& {Kumar}}]{Lu2022}
{Lu}, W., {Beniamini}, P., \& {Kumar}, P. 2022, \mnras, 510, 1867

\bibitem[{{Lugger} {et~al.}(2007){Lugger}, {Cohn}, {Heinke}, {Grindlay}, \& {Edmonds}}]{Lugger2007}
{Lugger}, P.~M., {Cohn}, H.~N., {et~al.} 2007, \apj, 657, 286

\bibitem[{{L{\"u}tzgendorf} {et~al.}(2011){L{\"u}tzgendorf}, {Kissler-Patig}, {Noyola}, {Jalali}, {de Zeeuw}, {Gebhardt}, \& {Baumgardt}}]{Lutz2011}
{L{\"u}tzgendorf}, N., {Kissler-Patig}, M., {et~al.} 2011, \aap, 533, A36

\bibitem[{{Lynch} {et~al.}(2012){Lynch}, {Freire}, {Ransom}, \& {Jacoby}}]{Lynch2012}
{Lynch}, R.~S., {Freire}, P. C.~C., {et~al.} 2012, \apj, 745, 109

\bibitem[{{Lynden-Bell} \& {Wood}(1968)}]{Lynden-Bell1968}
{Lynden-Bell}, D., \& {Wood}, R. 1968, \mnras, 138, 495

\bibitem[{Lyne {et~al.}(1987)Lyne, Brinklow, Middleditch, Kulkarni, Backer, \& Clifton}]{Lyne1987}
Lyne, A., Brinklow, A., {et~al.} 1987, \nat, 328, 399

\bibitem[{{Lyne} {et~al.}(1988){Lyne}, {Biggs}, {Brinklow}, {Ashworth}, \& {McKenna}}]{Lyne1988}
{Lyne}, A.~G., {Biggs}, J.~D., {et~al.} 1988, \nat, 332, 45

\bibitem[{{Maccarone} {et~al.}(2007){Maccarone}, {Kundu}, {Zepf}, \& {Rhode}}]{Maccarone2007}
{Maccarone}, T.~J., {Kundu}, A., {et~al.} 2007, \nat, 445, 183

\bibitem[{{Mackey} {et~al.}(2008){Mackey}, {Wilkinson}, {Davies}, \& {Gilmore}}]{Mackey2008}
{Mackey}, A.~D., {Wilkinson}, M.~I., {et~al.} 2008, \mnras, 386, 65

\bibitem[{{Manchester} {et~al.}(2005){Manchester}, {Hobbs}, {Teoh}, \& {Hobbs}}]{Manchester2005}
{Manchester}, R.~N., {Hobbs}, G.~B., {et~al.} 2005, \aj, 129, 1993

\bibitem[{{Manchester} {et~al.}(1991){Manchester}, {Lyne}, {Robinson}, {D'Amico}, {Bailes}, \& {Lim}}]{Manchester1991}
{Manchester}, R.~N., {Lyne}, A.~G., {et~al.} 1991, \nat, 352, 219

\bibitem[{{Mar{\'\i}n Pina} {et~al.}(2024){Mar{\'\i}n Pina}, {Rastello}, {Gieles}, {Kremer}, {Fitzgerald}, \& {Rando}}]{MarinPina2024}
{Mar{\'\i}n Pina}, D., {Rastello}, S., {et~al.} 2024, arXiv e-prints, arXiv:2404.13036

\bibitem[{{McKernan} {et~al.}(2018){McKernan}, {Ford}, {Bellovary}, {Leigh}, {Haiman}, {Kocsis}, {Lyra}, {Mac Low}, {Metzger}, {O'Dowd}, {Endlich}, \& {Rosen}}]{McKernan2018}
{McKernan}, B., {Ford}, K.~E.~S., {et~al.} 2018, \apj, 866, 66

\bibitem[{{Metzger} \& {Pejcha}(2017)}]{MetzgerPejcha2017}
{Metzger}, B.~D., \& {Pejcha}, O. 2017, \mnras, 471, 3200

\bibitem[{{Migliari} \& {Fender}(2006)}]{MigliariFender2006}
{Migliari}, S., \& {Fender}, R.~P. 2006, \mnras, 366, 79

\bibitem[{{Miller} \& {Hamilton}(2002)}]{MillerHamilton2002}
{Miller}, M.~C., \& {Hamilton}, D.~P. 2002, \mnras, 330, 232

\bibitem[{{Miller-Jones} {et~al.}(2015){Miller-Jones}, {Strader}, {Heinke}, {Maccarone}, {van den Berg}, {Knigge}, {Chomiuk}, {Noyola}, {Russell}, {Seth}, \& {Sivakoff}}]{Miller-Jones2015}
{Miller-Jones}, J.~C.~A., {Strader}, J., {et~al.} 2015, \mnras, 453, 3918

\bibitem[{{Mirabel} {et~al.}(2001){Mirabel}, {Dhawan}, {Mignani}, {Rodrigues}, \& {Guglielmetti}}]{Mirabel2001}
{Mirabel}, I.~F., {Dhawan}, V., {et~al.} 2001, \nat, 413, 139

\bibitem[{Morscher {et~al.}(2015)Morscher, Pattabiraman, Rodriguez, Rasio, \& Umbreit}]{Morscher2015}
Morscher, M., Pattabiraman, B., {et~al.} 2015, The Astrophysical Journal, 800, 9

\bibitem[{{Neumayer} {et~al.}(2020){Neumayer}, {Seth}, \& {B{\"o}ker}}]{Neumayer2020}
{Neumayer}, N., {Seth}, A., \& {B{\"o}ker}, T. 2020, \aapr, 28, 4

\bibitem[{{Noyola} {et~al.}(2010){Noyola}, {Gebhardt}, {Kissler-Patig}, {L{\"u}tzgendorf}, {Jalali}, {de Zeeuw}, \& {Baumgardt}}]{Noyola2010}
{Noyola}, E., {Gebhardt}, K., {et~al.} 2010, \apjl, 719, L60

\bibitem[{{O'Leary} {et~al.}(2009){O'Leary}, {Kocsis}, \& {Loeb}}]{O'Leary2009}
{O'Leary}, R.~M., {Kocsis}, B., \& {Loeb}, A. 2009, \mnras, 395, 2127

\bibitem[{{Pan} {et~al.}(2021){Pan}, {Qian}, {Ma}, {Liu}, {Wang}, {Luo}, {Yan}, {Ransom}, {Lorimer}, {Li}, \& {Jiang}}]{FAST2021}
{Pan}, Z., {Qian}, L., {et~al.} 2021, \apjl, 915, L28

\bibitem[{{Payne} {et~al.}(2024){Payne}, {Kremer}, \& {Zevin}}]{Payne2024}
{Payne}, E., {Kremer}, K., \& {Zevin}, M. 2024, \apjl, 966, L16

\bibitem[{{Perera} {et~al.}(2017){Perera}, {Stappers}, {Lyne}, {Bassa}, {Cognard}, {Guillemot}, {Kramer}, {Theureau}, \& {Desvignes}}]{Perera2017}
{Perera}, B.~B.~P., {Stappers}, B.~W., {et~al.} 2017, Monthly Notices of the Royal Astronomical Society, 468, 2114

\bibitem[{{Perets} {et~al.}(2016){Perets}, {Li}, {Lombardi}, \& {Milcarek}}]{Perets2016}
{Perets}, H.~B., {Li}, Z., {et~al.} 2016, \apj, 823, 113

\bibitem[{Peters(1964)}]{Peters1964}
Peters, P. 1964, Physical Review, 136, B1224

\bibitem[{{Phinney} \& {Kulkarni}(1994)}]{PhinneyKulkarni1994}
{Phinney}, E.~S., \& {Kulkarni}, S.~R. 1994, \araa, 32, 591

\bibitem[{{Pooley} {et~al.}(2002){Pooley}, {Lewin}, {Homer}, {Verbunt}, {Anderson}, {Gaensler}, {Margon}, {Miller}, {Fox}, {Kaspi}, \& {van der Klis}}]{Pooley2002}
{Pooley}, D., {Lewin}, W. H.~G., {et~al.} 2002, \apj, 569, 405

\bibitem[{{Pooley} {et~al.}(2003){Pooley}, {Lewin}, {Anderson}, {Baumgardt}, {Filippenko}, {Gaensler}, {Homer}, {Hut}, {Kaspi}, {Makino}, {Margon}, {McMillan}, {Portegies Zwart}, {van der Klis}, \& {Verbunt}}]{Pooley2003}
---. 2003, \apjl, 591, L131

\bibitem[{{Portegies Zwart} {et~al.}(2004){Portegies Zwart}, {Baumgardt}, {Hut}, {Makino}, \& {McMillan}}]{PortegiesZwart2004}
{Portegies Zwart}, S.~F., {Baumgardt}, H., {et~al.} 2004, \nat, 428, 724

\bibitem[{{Postnov} \& {Yungelson}(2014)}]{Postnov2014}
{Postnov}, K.~A., \& {Yungelson}, L.~R. 2014, Living Reviews in Relativity, 17, 3

\bibitem[{{Quinlan} \& {Shapiro}(1987)}]{QuinlanShapiro1987}
{Quinlan}, G.~D., \& {Shapiro}, S.~L. 1987, \apj, 321, 199

\bibitem[{{Ransom} {et~al.}(2004){Ransom}, {Stairs}, {Backer}, {Greenhill}, {Bassa}, {Hessels}, \& {Kaspi}}]{Ransom2004}
{Ransom}, S.~M., {Stairs}, I.~H., {et~al.} 2004, \apj, 604, 328

\bibitem[{{Rastello} {et~al.}(2023){Rastello}, {Iorio}, {Mapelli}, {Arca-Sedda}, {Di Carlo}, {Escobar}, {Shenar}, \& {Torniamenti}}]{Rastello2023}
{Rastello}, S., {Iorio}, G., {et~al.} 2023, \mnras, 526, 740

\bibitem[{{Remillard} \& {McClintock}(2006)}]{RemillardMcClintock2006}
{Remillard}, R.~A., \& {McClintock}, J.~E. 2006, \araa, 44, 49

\bibitem[{{Ridolfi} {et~al.}(2019){Ridolfi}, {Freire}, {Gupta}, \& {Ransom}}]{Ridolfi2019}
{Ridolfi}, A., {Freire}, P.~C.~C., {et~al.} 2019, \mnras, 490, 3860

\bibitem[{{Ridolfi} {et~al.}(2021){Ridolfi}, {Gautam}, {Freire}, {Ransom}, {Buchner}, {Possenti}, {Venkatraman Krishnan}, {Bailes}, {Kramer}, {Stappers}, {Abbate}, {Barr}, {Burgay}, {Camilo}, {Corongiu}, {Jameson}, {Padmanabh}, {Vleeschower}, {Champion}, {Chen}, {Geyer}, {Karastergiou}, {Karuppusamy}, {Parthasarathy}, {Reardon}, {Serylak}, {Shannon}, \& {Spiewak}}]{TRAPUM2021}
{Ridolfi}, A., {Gautam}, T., {et~al.} 2021, \mnras, 504, 1407

\bibitem[{{Rodriguez} {et~al.}(2018{\natexlab{a}}){Rodriguez}, {Amaro-Seoane}, {Chatterjee}, {Kremer}, {Rasio}, {Samsing}, {Ye}, \& {Zevin}}]{Rodriguez2018b}
{Rodriguez}, C.~L., {Amaro-Seoane}, P., {et~al.} 2018{\natexlab{a}}, \prd, 98, 123005

\bibitem[{{Rodriguez} {et~al.}(2018{\natexlab{b}}){Rodriguez}, {Amaro-Seoane}, {Chatterjee}, \& {Rasio}}]{Rodriguez2018a}
---. 2018{\natexlab{b}}, Physical Review Letters, 120, 151101

\bibitem[{Rodriguez {et~al.}(2016)Rodriguez, Chatterjee, \& Rasio}]{Rodriguez2016a}
Rodriguez, C.~L., Chatterjee, S., \& Rasio, F.~A. 2016, Physical Review D, 93, 084029

\bibitem[{{Rodriguez} {et~al.}(2021){Rodriguez}, {Kremer}, {Chatterjee}, {Fragione}, {Loeb}, {Rasio}, {Weatherford}, \& {Ye}}]{Rodriguez2021_RNAAS}
{Rodriguez}, C.~L., {Kremer}, K., {et~al.} 2021, Research Notes of the American Astronomical Society, 5, 19

\bibitem[{Rodriguez {et~al.}(2015)Rodriguez, Morscher, Pattabiraman, Chatterjee, Haster, \& Rasio}]{Rodriguez2015a}
Rodriguez, C.~L., Morscher, M., {et~al.} 2015, Physical Review Letters, 115, 051101

\bibitem[{{Rodriguez} {et~al.}(2019){Rodriguez}, {Zevin}, {Amaro-Seoane}, {Chatterjee}, {Kremer}, {Rasio}, \& {Ye}}]{Rodriguez2019}
{Rodriguez}, C.~L., {Zevin}, M., {et~al.} 2019, \prd, 100, 043027

\bibitem[{{Rodriguez} {et~al.}(2016){Rodriguez}, {Zevin}, {Pankow}, {Kalogera}, \& {Rasio}}]{Rodriguez2016c}
---. 2016, \apjl, 832, L2

\bibitem[{{Ruderman} \& {Sutherland}(1975)}]{RudermanSutherland1975}
{Ruderman}, M.~A., \& {Sutherland}, P.~G. 1975, \apj, 196, 51

\bibitem[{{Rutledge} {et~al.}(2002){Rutledge}, {Bildsten}, {Brown}, {Pavlov}, \& {Zavlin}}]{Rutledge2002}
{Rutledge}, R.~E., {Bildsten}, L., {et~al.} 2002, \apj, 578, 405

\bibitem[{{Samsing} {et~al.}(2020){Samsing}, {D'Orazio}, {Kremer}, {Rodriguez}, \& {Askar}}]{Samsing2020}
{Samsing}, J., {D'Orazio}, D.~J., {et~al.} 2020, \prd, 101, 123010

\bibitem[{{Samsing} {et~al.}(2014){Samsing}, {MacLeod}, \& {Ramirez-Ruiz}}]{Samsing2014}
{Samsing}, J., {MacLeod}, M., \& {Ramirez-Ruiz}, E. 2014, \apj, 784, 71

\bibitem[{{Sana} {et~al.}(2009){Sana}, {Gosset}, \& {Evans}}]{Sana2009}
{Sana}, H., {Gosset}, E., \& {Evans}, C.~J. 2009, \mnras, 400, 1479

\bibitem[{{Sana} {et~al.}(2012){Sana}, {de Mink}, {de Koter}, {Langer}, {Evans}, {Gieles}, {Gosset}, {Izzard}, {Le Bouquin}, \& {Schneider}}]{Sana2012}
{Sana}, H., {de Mink}, S.~E., {et~al.} 2012, Science, 337, 444

\bibitem[{{Schmidt} {et~al.}(2015){Schmidt}, {Ohme}, \& {Hannam}}]{Schmidt2015}
{Schmidt}, P., {Ohme}, F., \& {Hannam}, M. 2015, \prd, 91, 024043

\bibitem[{{Shapiro} \& {Teukolsky}(1983)}]{ShapiroTeukolsky1983}
{Shapiro}, S.~L., \& {Teukolsky}, S.~A. 1983, {Black holes, white dwarfs and neutron stars. The physics of compact objects}, doi:10.1002/9783527617661

\bibitem[{{Shen} {et~al.}(2019){Shen}, {Quataert}, \& {Pakmor}}]{Shen2019}
{Shen}, K.~J., {Quataert}, E., \& {Pakmor}, R. 2019, arXiv e-prints, arXiv:1908.08056

\bibitem[{{Shi} {et~al.}(2023){Shi}, {Kremer}, {Grudi{\'c}}, {Gerling-Dunsmore}, \& {Hopkins}}]{Shi2023}
{Shi}, Y., {Kremer}, K., {et~al.} 2023, \mnras, 518, 3606

\bibitem[{{Shishkovsky} {et~al.}(2018){Shishkovsky}, {Strader}, {Chomiuk}, {Bahramian}, {Tremou}, {Li}, {Salinas}, {Tudor}, {Miller-Jones}, {Maccarone}, {Heinke}, \& {Sivakoff}}]{Shishkovsky2018}
{Shishkovsky}, L., {Strader}, J., {et~al.} 2018, \apj, 855, 55

\bibitem[{{Sigurdsson} {et~al.}(2003){Sigurdsson}, {Richer}, {Hansen}, {Stairs}, \& {Thorsett}}]{Sigurdsson2003}
{Sigurdsson}, S., {Richer}, H.~B., {et~al.} 2003, Science, 301, 193

\bibitem[{{Silsbee} \& {Tremaine}(2017)}]{SilsbeeTremaine2017}
{Silsbee}, K., \& {Tremaine}, S. 2017, \apj, 836, 39

\bibitem[{Silsbee \& Tremaine(2017)}]{Silsbee2017}
Silsbee, K., \& Tremaine, S. 2017, The Astrophysical Journal, 836, 39

\bibitem[{{Smartt}(2009)}]{Smartt2009}
{Smartt}, S.~J. 2009, \araa, 47, 63

\bibitem[{{Spitzer}(1969)}]{Spitzer1969}
{Spitzer}, Lyman, J. 1969, \apjl, 158, L139

\bibitem[{{Sridhar} {et~al.}(2021){Sridhar}, {Metzger}, {Beniamini}, {Margalit}, {Renzo}, {Sironi}, \& {Kovlakas}}]{Sridhar2021}
{Sridhar}, N., {Metzger}, B.~D., {et~al.} 2021, \apj, 917, 13

\bibitem[{{Strader} {et~al.}(2012){Strader}, {Chomiuk}, {Maccarone}, {Miller-Jones}, \& {Seth}}]{Strader2012}
{Strader}, J., {Chomiuk}, L., {et~al.} 2012, \nat, 490, 71

\bibitem[{{Tauris} {et~al.}(2013){Tauris}, {Sanyal}, {Yoon}, \& {Langer}}]{Tauris2013}
{Tauris}, T.~M., {Sanyal}, D., {et~al.} 2013, \aap, 558, A39

\bibitem[{{Thorsett} {et~al.}(1993){Thorsett}, {Arzoumanian}, \& {Taylor}}]{Thorsett1993}
{Thorsett}, S.~E., {Arzoumanian}, Z., \& {Taylor}, J.~H. 1993, \apjl, 412, L33

\bibitem[{{Tremou} {et~al.}(2018){Tremou}, {Strader}, {Chomiuk}, {Shishkovsky}, {Maccarone}, {Miller-Jones}, {Tudor}, {Heinke}, {Sivakoff}, {Seth}, \& {Noyola}}]{Tremou2018}
{Tremou}, E., {Strader}, J., {et~al.} 2018, \apj, 862, 16

\bibitem[{{Vink} {et~al.}(2001){Vink}, {de Koter}, \& {Lamers}}]{Vink2001}
{Vink}, J.~S., {de Koter}, A., \& {Lamers}, H.~J.~G.~L.~M. 2001, \aap, 369, 574

\bibitem[{{Vitral} {et~al.}(2022){Vitral}, {Kremer}, {Libralato}, {Mamon}, \& {Bellini}}]{Vitral2022}
{Vitral}, E., {Kremer}, K., {et~al.} 2022, \mnras, 514, 806

\bibitem[{{Vitral} {et~al.}(2023){Vitral}, {Libralato}, {Kremer}, {Mamon}, {Bellini}, {Bedin}, \& {Anderson}}]{Vitral2023}
{Vitral}, E., {Libralato}, M., {et~al.} 2023, \mnras, 522, 5740

\bibitem[{{Wang} {et~al.}(2016){Wang}, {Spurzem}, {Aarseth}, {Giersz}, {Askar}, {Berczik}, {Naab}, {Schadow}, \& {Kouwenhoven}}]{Wang2016}
{Wang}, L., {Spurzem}, R., {et~al.} 2016, \mnras, 458, 1450

\bibitem[{{Weatherford} {et~al.}(2020){Weatherford}, {Chatterjee}, {Kremer}, \& {Rasio}}]{Weatherford2020}
{Weatherford}, N.~C., {Chatterjee}, S., {et~al.} 2020, \apj, 898, 162

\bibitem[{{Woosley}(2017)}]{Woosley2017}
{Woosley}, S.~E. 2017, \apj, 836, 244

\bibitem[{{Woosley} {et~al.}(2002){Woosley}, {Heger}, \& {Weaver}}]{Woosley2002}
{Woosley}, S.~E., {Heger}, A., \& {Weaver}, T.~A. 2002, Reviews of Modern Physics, 74, 1015

\bibitem[{{Ye} {et~al.}(2020){Ye}, {Fong}, {Kremer}, {Rodriguez}, {Chatterjee}, {Fragione}, \& {Rasio}}]{Ye2020_BNS}
{Ye}, C.~S., {Fong}, W.-f., {et~al.} 2020, \apjl, 888, L10

\bibitem[{{Ye} {et~al.}(2019){Ye}, {Kremer}, {Chatterjee}, {Rodriguez}, \& {Rasio}}]{Ye2018}
{Ye}, C.~S., {Kremer}, K., {et~al.} 2019, \apj, 877, 122

\bibitem[{{Zevin} {et~al.}(2021{\natexlab{a}}){Zevin}, {Romero-Shaw}, {Kremer}, {Thrane}, \& {Lasky}}]{Zevin2021}
{Zevin}, M., {Romero-Shaw}, I.~M., {et~al.} 2021{\natexlab{a}}, \apjl, 921, L43

\bibitem[{{Zevin} {et~al.}(2021{\natexlab{b}}){Zevin}, {Romero-Shaw}, {Kremer}, {Thrane}, \& {Lasky}}]{Zevin2021_ecc}
---. 2021{\natexlab{b}}, \apjl, 921, L43

\bibitem[{{Zhou} {et~al.}(2024){Zhou}, {Wang}, {Li}, {Fang}, {Miao}, {Freire}, {Zhang}, {Zhang}, {Chen}, {Feng}, {Xiao}, {Xie}, {Zhang}, {Jin}, {Wang}, {Ke}, {Guo}, {Zhao}, {Niu}, {Zhu}, {Xue}, {Wang}, {Wu}, {Gan}, {Sun}, {Wang}, {Zhang}, {Zhang}, {Cao}, \& {Lu}}]{Zhou2024}
{Zhou}, D., {Wang}, P., {et~al.} 2024, Science China Physics, Mechanics, and Astronomy, 67, 269512

\bibitem[{{Zocchi} {et~al.}(2019){Zocchi}, {Gieles}, \& {H{\'e}nault-Brunet}}]{Zocchi2019}
{Zocchi}, A., {Gieles}, M., \& {H{\'e}nault-Brunet}, V. 2019, \mnras, 482, 4713

\end{thebibliography}
\end{multicols}

\end{document}